\documentclass[preprint2]{aastex}
\usepackage[caption=false]{subfig}
\nonstopmode

\begin{document}

\title{The Luminous Convolution Model}

\author{S.~Cisneros\altaffilmark{1}, N.\,S.~Oblath, J.\,A.~Formaggio\altaffilmark{1}, G.~Goedecke\altaffilmark{2}, D.~Chester\altaffilmark{1,3}, R.~Ott\altaffilmark{1,4}, A.~Ashley\altaffilmark{1}, and A.~Rodriguez\altaffilmark{1}}

\altaffiltext{1}{Laboratory for Nuclear Science, Massachusetts Institute of Technology, Cambridge, MA~02139, USA}
\altaffiltext{2}{Department of Physics, New Mexico State University, Las Cruces, NM~88003, USA}
\altaffiltext{3}{Present address: University of California at Los Angeles}
\altaffiltext{4}{Present address: University of California at Davis}
\slugcomment{Submitted to ApJ, with thanks to Ed Bertschinger,  Janet Conrad, Peter Fisher   \&  V.\,P. Nair.}

\begin{abstract}
 We present a heuristic model for predicting the  rotation curves of spiral galaxies.
 The Luminous Convolution Model (LCM) utilizes  Lorentz-type transformations 
 of very small changes in  photon frequencies from curved space-times to construct 
 a  model predictive of galaxy rotation profile observations.
 These frequency changes are derived from  the Schwarzschild red-shift result or 
 the analogous  result from a Kerr wave equation. 
 The LCM  maps the small curvatures of the emitter galactic frame onto those of the  receiver galactic frame, and then
returns the map to the associated flat frames  where measurements are made. 
 This treatment   rests upon estimates of the luminous matter in both the emitter   and receiver galaxies to 
determine these small curvatures.   
 The LCM is tested on a sample of 23   galaxies, represented in 35 different data sets. 
 LCM fits are  compared  to those of the Navarro, Frenk  and White (NFW) Dark Matter Model, and/or to the Modified Newtonian Dynamics (MOND)  model when possible. The high degree of sensitivity of the LCM to the initial assumption 
 of a luminous mass-to-light ratio ($M_L/L$)   is shown.  We demonstrate that the LCM is successful across a wide range of spiral galaxies for predicting the observed rotation curves.
\end{abstract}

\keywords{cosmology: dark matter, theory; galaxies: distances and redshifts}

\section{Introduction}
\label{sec:INTRO}

Since the 1960's, much work has been done to identify the source of what is sometimes known as the flat-rotation curve problem; that is, the discrepancy between the luminous and dynamical models of matter distributions within observed galaxies.  Investigations into this problem have mainly followed one of two paths: either an alteration of the laws of physics governing gravitation or a new source of matter.  The most popular among the new physics models is the Modified Newtonian Dynamics (MOfD)~\citep{Milgrom}, in which the gravitational constant changes gradually over the large distance scales of galaxies and clusters of galaxies.  The concept of new sources of matter is more commonly referred to as dark matter, with the most popular dark matter model to date  proposed by  Navarro, Frenk \& White (NFW) \citep{NFW}.  Both approaches successfully explain a great variety of cosmological observations to date beyond spiral galaxies (for a full review of these two approaches, see  \citet{SanMcGa} and \citet{Gianfranco}).  Indications of dark matter may have been observed in the DAMA \citep{DAMA}, CoGeNT \citep{CoGeNT}, and CDMS \citep{CDMS} Experiments, although these results are in some conflict with the limits of XENON100 \citep{XENON}.   Thus, despite their successes, it remains true that neither phenomena --deviations from General Relativity (GR) predictions nor direct detection of dark matter-- has been observed in decisive terrestrial experiments. New clues and new approaches from particle physics and astrophysics are needed to understand the effect.

Luminous mass modeling for galaxies dates back to as early as 1912 with the advent of the Hertzsprung-Russell (HR) diagram \citep{HRD}.  The HR diagram allows for  identification of mass associated with  observed light. The photometric identification of the individual  masses of stars is then  extrapolated to  population synthesis models which
 estimate   the total luminous stellar mass of  galaxies.  Stellar masses are then added to estimated gas masses and  reported as   luminous mass-to-light $( M_L/L)$ ratios  for a given galaxy, where $M_L$ is the total luminous mass and $L$ is the total luminosity.\\  

As a constraint on the luminous masses,  a second, dynamical (hence orthogonal) measurement of the mass distribution was introduced. However,  this dynamical measure, $v_{obs}$, reported independently by Oort  and   Zwicky in the 1930s \citep{Zwick,Bergh}, and  later confirmed by \citet{Rubin} and  \citet{Bosma}, clearly demonstrated flat-rotation speeds at high radii,   a strong deviation  from the luminous Keplerian predictions.  The   mismatch between the dynamical ($M'$) and luminous ($M_L$) mass distributions  within galaxies  has become known as the flat-rotation curve problem.  The result of the discrepancy was two fold; a `missing mass' component,  the dark matter  ($M_{DM}$), and an underconstrained luminous matter modeling problem \citep{Conroy}.   \\

We propose a new model to rectify the differences between galactic rotation curves and luminous mass models.  This predictive model, which we will refer to as the Luminous Convolution Model (LCM), provides an alternative to the dark matter and MOND hypotheses: it represents a new approach to  rectifying the observed galaxy masses $M'$ with the luminous galaxy masses $M_L$, and is  constructed entirely from the luminous matter $M_L$.    The LCM relies upon the GR evaluation of small galactic curvatures, mapped from an emitter galaxy onto the receiver galaxy, to 
 compare the relative gravitational potentials.  All evaluations of galaxies are done under idealized geometries and assumptions consistent 
 with the current Newtonian treatment of the dark-matter problem.  External factors such as the Hubble flow are  assumed to be incorporated into the results of the two free parameters of the model.  As this is a heuristic model, physical interpretation of the two parameters will be left for future work.\\
 
The LCM  borrows concepts from both models of modified gravity and dark matter to be consistent with the formalisms developed to date.    Similar to MOND we will consider the concept of small changes to gravitational acceleration at large distances, though we will interpret the changes in acceleration as small changes in curvature.  The changes in  curvature will be evaluated by an appropriate homogeneous wave equation  as gravitationally shifted  frequencies. We will use the convenient parametrization 
of  the rotation curve velocity prediction, $v_{tot}$, that is typically used in dark-matter models:
\begin{equation}
v_{tot}^2=  v_{lum}^2+  v_{DM}^2, 
\label{eq:zonte1}
\end{equation}
where  $v_{DM}$ is the model specific dark matter contribution, and $v_{lum}$ is 
the Keplerian contribution to the rotations from the luminous component, $M_L$.  The velocity prediction will be fit to 
the observed rotation velocities $v_{obs}$.\\

 The paper is divided as follows. Section~\ref{sec:station} briefly describes the formalism adopted  in comparing the  mass distributions of galaxies.  Section~\ref{sec:introLCM} introduces the LCM formalism, and  Section~\ref{sec:LCM} reports the LCM results for   23  spiral galaxies, represented in 35 different data sets.  Some general conclusions are presented in Section~\ref{sec:conclusion}.  \\
 
\section{ Mass Distributions in Spiral Galaxies}
\label{sec:station}
\subsection{Spherical Symmetry}
The quadratic velocity sum   in  
 Eq.~(\ref{eq:zonte1}) comes from  the  sum of 
 the mass elements, 
\begin{equation}
    M'=M_{L}+M_{ DM },
     \label{eq:mass1}
 \end{equation}
 where $M'$ is the total gravitational mass, composed of the luminous $M_L$ and dark $M_{DM}$ masses.  
Each mass   in Eq.~(\ref{eq:mass1})   is  the mass enclosed up to a radius $r$,  
$M(r)$,  and  is related to an  associated orbital  velocity   $v(r)$ by    Newton's second law. In the case of
spherical symmetry, that relation is simply:
 \begin{equation}
F(r)=- \frac{mv(r)^2}{r}=-\frac{mM(r)G}{r^2},
 \label{eq:dynamical}
 \end{equation} 
where $m$  is the mass of a  test particle   orbiting at a radial distance
$r$   from the center of the spherical 
 mass distribution $M(r)$, $G$ the gravitational constant, and  $F(r)$ the  Newtonian gravitational force. Hence  the resulting quadratic 
 sum of velocities in Eq.~(\ref{eq:zonte1}) implies like geometry for the components summed.   \\

 The most commonly employed geometry is  spherical, as the matter distribution  is very diffuse.   Deviations from spherical symmetry are higher order corrections to the potential and thereby the force \citep{Binney}.
Newtonian  gravity potentials
 \begin{equation}
\Phi(r)=-\frac{1}{m}\int F(r) dr+\Phi_o,
\label{eq:potentialgeneral}
\end{equation}
are related to  forces $F(r)$   in the traditional way.    The  integration constant $\Phi_o$ is generally fixed such that $\Phi(r) \to 0$ as $r\to \infty$.   The  gauge choice for a system of two galaxies  will be discussed in  Sect.\ref{sec:gauge}. \\
  
  The gravity potential $\Phi(r)$ used in the LCM will be  evaluated for the luminous matter $M_L$ alone,  and hence  will be called $\Phi_L$.   The LCM is applied  in the weak field limit of spiral galaxies, in which case the 
 GR metric $g_{\mu \nu}$  can  be  parametrized with the Newtonian potentials   $\Phi_L$.  
   The  small curvature contributions convolved in the LCM function, $v_{LC}^2$, defined in Eq.~(\ref{eq:convolutionFunc}), will  come from scalar wave equations
(Sect.~\ref{sec:Schw} and~\ref{sec:kerr}).   It is   ``the metric $g_{\mu \nu}$ which forms the components of the gravitational potential'' \citep[see][p.204]{Weyl} .   Each data set reported the observed velocities in the context of comparison to the luminous mass model chosen.   The stellar $M_L/L$ ratios  used for each galaxy are   reported in  Table.\ref{tab:referenceDATAs}.  \\
 
 \subsection{Luminous Galaxy Mass}
 \label{sec:lumGal}
Estimates of the total luminous mass in spiral galaxies vary widely, as the modeling process is underconstrained.  Models for the 
total light,  interpreted for distance and extinction corrections,  are done in 
specific   wavelength bands.  These  associated bands  and  the $M_L/L$ ratios reported in each source are reported  in Table.\ref{tab:referenceDATAs}.  These $M_L/L$ ratios are generally chosen in the context of a specific model, which are also reported in 
Table.\ref{tab:referenceDATAs}.  \\

Luminous  masses $M_L$ are  reported as the Keplerian rotation velocities $v_{lum}$ implied by  Eq.~(\ref{eq:dynamical}).   Luminous matter 
models  involve assumptions regarding  metallicities, types of stars, and geometries of the   individual components of the galaxies; the  thin and thick stellar disks, stellar bulge,  
and gas. In general, these components are modeled individually using different observational techniques \citep{Binney}.  The individual  geometries  
  used to  calculate the appropriate  potentials, and hence  force relations,   are not considered when 
  the final velocity sum is calculated.   The  total Keplerian rotation velocity  $v_{lum}$
 is  taken to be the  quadratic 
  sum of the components \citep{Maria},  as was done for Eq.~(\ref{eq:zonte1}), 
 \begin{equation}
 v^2_{lum}=v^2_{bulge}+v^2_{disk1}+v^2_{disk2}+v^2_{gas}. 
   \label{eq:quadrature}
 \end{equation}
This sum is an implicit assumption of like geometry for 
the components, as it arises from the mass sum,
 \begin{equation}
 M_{lum}=M_{bulge}+M_{disk1}+M_{disk2}+M_{gas}.
 \label{eq:lumMasses}
 \end{equation}
The   error introduced in  assuming  like geometry  (in Eqs.~(\ref{eq:zonte1}) or  ~(\ref{eq:quadrature})) for 
disk and spherical mass distributions
is an underestimate in the magnitude  of the Newtonian gravitational potential $\Phi(r)$~\citep{Chatterjee} ,   not   in 
the functional shape.  As such, the introduction of such 
an error can not be responsible for the dark matter problem, which is a functional difference in the potential at large
  $r$.     \\  
 
The  spherical assumption in Eq.~(\ref{eq:dynamical}) is commonly employed  and   offers the additional  benefit of 
 Newton's shell theorem, where  the  gravitational  field   at  each point $r$ 
   is  composed  of contributions from only those  mass elements \emph{interior} to $r$ \citep{Fowles}. The shell 
   theorem will be  advantageous  in construction of the LCM, allowing use of the   exterior metric in the 
   plane of the galactic disk.
   
\subsection{Doppler Shifts}
\label{sec:dynamical}
The LCM will rely upon three different types of  frequencies that are related to the general  Lorentz Doppler shift formula.\\

The general Lorentz Doppler shift formula is applicable in a variety of   contexts, from earth-bound measurements to cosmological 
distance estimates. A general definition of the  Lorentz Doppler shift formula (LDSF), Eq.~\ref{eq:LorentzDefine}, 
 \begin{equation}
\frac{v}{c}=
\frac{  \frac{\omega_s}{\omega_o}- \frac{\omega_o}{\omega_s}}{ 
\frac{\omega_s}{\omega_o}+ \frac{\omega_o}{\omega_s}},
\label{eq:LorentzDefine}
\end{equation} 
relates  the   characteristic  frequency  $\omega_o$  to the received shifted frequency $\omega_s$ for some photon. The relative velocity parameter $v$ describes the motion of the source with respect to the receiver directly towards or away from each other. The characteristic frequency  $\omega_o$ 
is   typically identified for some well known atomic transition, as measured on earth. This frequency is assumed to be the same in any flat frame in any system based upon the constancy of the speed of light.\\

In the context of astrophysics,   when $\omega_s>\omega_o$ the light is considered to be blue-shifted, and when $\omega_s<\omega_o$, the light is considered to be red-shifted.  Cosmological  distance indicators, translational effects (e.g. relative motion), and gravitational effects are all characterized by the dimensionless  quantity  $z$: 
\begin{equation}
1+z=\frac{\omega_o}{\omega_s}.
\label{eq:zParam}
\end{equation}
  For red-shifts  $z$  is positive and  
 for blue shifts  $z$ is negative \citep{Hartle}.\\

The three frequencies of interest are based upon the  two  fundamental 
observables in the flat-rotation curve problem.   The first observable is  the  shifted frequency $\omega'$, measured as a 
function of radius for a given galaxy.       The second  observable is the 
total light, which when interpreted through a population synthesis model gives the 
 luminous mass $M_L$.  This luminous  mass   implies the  Keplerian rotation term $v_{lum}$ in Eq.~(\ref{eq:zonte1}), by Eq.~(\ref{eq:dynamical}).  \\
 
The first frequency of interest is the observed frequency    $\omega'$.  This frequency    
 yields the observed  flat-rotation curve velocity parameter  $v_{obs}$ by  a LDSF,
\begin{equation}
\frac{v_{obs}}{c}=
\frac{  \frac{\omega'}{\omega_o}- \frac{\omega_o}{\omega'}}{ 
\frac{\omega'}{\omega_o}+ \frac{\omega_o}{\omega'}}.
\label{eq:dataLorentz}
\end{equation}
The shifted-frequency  $\omega'$ is  assumed to be measured for a photon  emitted along the line of sight from a   test particle   in a stable, circular orbit.   The characteristic frequency  $\omega_o$ remains defined as in Eq.~(\ref{eq:LorentzDefine}).  The   shifted frequency  $\omega'$  implies the total gravitational mass $M'$  in Eq.~(\ref{eq:mass1}) by Newton's second law, Eq.~(\ref{eq:dynamical}).  \\

The second frequency of interest is that frequency which would have been measured if the luminous matter alone were responsible for the observed rotations.  The Keplerian rotation velocity $v_{lum}$   implies a frequency $\omega_l$ by the LDSF relation:
   \begin{equation}
\frac{v_{lum}}{c}=   \frac{  \frac{\omega_{l}}{\omega_{o}}  -   \frac{\omega_{o}}{\omega_{l}}  }{  \frac{\omega_{l}}{\omega_{o}} +   \frac{\omega_{l}}{\omega_{o}} }. 
\label{eq:primeLUM}
\end{equation} 
The frequency $\omega_l$  will be used to   characterize our knowledge of  the local  frames  in Sect.\ref{sec:convolution}.   The characteristic frequency  $\omega_o$ remains defined as in Eq.~(\ref{eq:LorentzDefine}).   \\

The third type of  frequency of interest  are those from  gravitational effects, often known as gravitational redshifts.   For these gravitationally shifted frequencies, the  associated  curvatures of the space-time will be derived in  Sect.~\ref{sec:Schw} and~\ref{sec:kerr} for the luminous matter $M_L$ alone.  These frequencies reflect metric curvature of the space-time and  
will be  denoted as $\omega_{gal}$ for  the   emitter-galaxy and $\omega_{mw}$ for the  receiver galaxy. These two gravitationally shifted frequencies  will be convolved  by an application of  the LDSF as a mapping in Sect.~\ref{sec:convolution}. \\

\section{ The Luminous Convolution Model} 
\label{sec:introLCM}

Constructed in analogy to Eq.~(\ref{eq:zonte1}), 
the   LCM rotation curve prediction   $v_{Lmod}$   is:
\begin{equation}
v_{Lmod}^2=   \zeta v_{lum}^2+  \alpha v_{LC}^2,
\label{eq:zople1}
\end{equation}
where  $ \zeta$   and $\alpha$ are the fitting parameters of the model,    $v_{lum}^2$ is   the square of the  Keplerian rotation velocity, and $v_{LC}^2$ is  the convolution term,  to be defined in Sect.\ref{sec:convolution}.  We will fit $v_{Lmod}$ to the observed rotation curve data $v_{obs}$.  The parameter $\zeta$ scales the luminous mass distribution. The parameter $\alpha$  relates the LCM function $v_{LC}^2$  to the dark matter term $v_{DM}^2$ in Sect.\ref{sec:zeta}.  The $v_{LC}^2$ requires two inputs: the luminous matter $M_L$ profiles of both the emitter and the receiver galaxies. \\

The parametrization in Eq.~(\ref{eq:zople1})   is convenient,  as it allows the LCM to isolate  effects on the photons from 
translation and acceleration.  As is known to  occur in nature,  effects on  electric and magnetic fields (hence light)   separate cleanly into two separate terms
based upon the  translation and acceleration of the source of the fields \citep[see][Eqs.~(14.13) \& (14.14)]{Jack}.   Translation  effects   will be encapsulated in  the term $v_{lum}^2$ and curvature (e.g. acceleration) effects will be encapsulated in 
 the term $v_{LC}^2$.     As we transition from Newtonian to General Relativity, we readily replace the word acceleration with curvature.  \\

\subsection{Convolution Term $v_{LC}^2$}
\label{sec:convolution}

The convolution term  $ v_{LC}^2$ in Eq.~(\ref{eq:zonte1})
  is composed of three terms:
\begin{equation}
v_{LC}^2=   \kappa  v_{1}  v_{2}, 
\label{eq:convolutionFunc}
\end{equation}
where  $\kappa$ is the ratio of galactic curvatures, $v_1$ is  a mapping of the galactic frames,  and  $v_2$ is  a  
 mapping back to the associated flat tangent frames where physical measurements are made.  \\

The first term in Eq.~(\ref{eq:convolutionFunc}),   $\kappa$,   is a measure  of the deviation from   flat space-time for  a given pair of emitter-receiver galaxies:    
\begin{equation}
\kappa=\frac{\Delta c_{gal}}{ \Delta c_{mw}},
\label{eq:kappa}
\end{equation}
where curvatures are measured based upon the difference of the coordinate  light speed $\tilde{c}_{i}$ from $c$ (see Sect.~\ref{sec:curvature}), 
\begin{equation}
\Delta c_{i}=  c-\tilde{c}_{i}  
\label{eq:equiv}
\end{equation}
for the emitter-galaxy  $i={gal}$ and the receiver galaxy  $i={mw}$.   The quantity $\Delta c_{i}$ is   sensitive to small curvatures.   Physically, $\kappa=1$ when the two galaxies have equal deviations from flatness (i.e. the luminous galaxy masses are approximately the same as a function of radius); $\kappa<1$ when the emitter-galaxy  is less massive than the MW, and vice versa  for $\kappa>1$.  The $\kappa$ ratio acts to normalize the two galactic frames to the same `level,' so we can  apply   LDSF in  a series of two mappings.    As $\kappa$ is undefined in the limit   $\Delta c_{i}\to 0$,   the LCM is only  applicable in circumstances for which   both emitter and receiver frames have  non-zero curvature.   However, as the curvatures for spiral galaxies are exceedingly small, this constraint does not exclude work in very diffuse space-times. \\

\begin{figure}
 \epsscale{0.9}
	\plotone{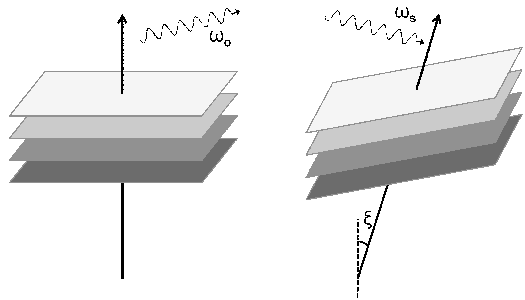} 
	\caption{A  graphic representation of a  Lorentz transformation as a  mapping from emitter to receiver frame.   The mapping of frames will be 
	  defined by the frequencies.\label{fig:noah}}
\end{figure}

The  second term in Eq.~(\ref{eq:convolutionFunc}), $v_1$, is a mapping of the galactic frames.  In order to construct this mapping, 
we  first investigate the geometric interpretation of the LDSF.   We can rewrite  Eq.~(\ref{eq:LorentzDefine}) as a   hyperbolic rotation 
 \begin{equation}
\frac{v}{c}= \tanh \xi= \frac{e^\xi - e^{-\xi}}{e^\xi + e^{-\xi}},
\label{eq:map}
\end{equation}
through the  rapidity angle $\xi$, as shown in~Fig.~\ref{fig:noah}.  
 
The angle $\xi$  is defined as a  positive rotation   away from the vertical time axis within the Special Relativistic light-cone~\citep[p67]{MTW}.   As Special Relativity is symmetric between the two frames (i.e. it  is meaningless to assign absolute motion to 
either frame),  this rotation angle must always be positive.  However, as we transition to  curved frames, it is  important to consistently associate the emitted and received frequencies with  specific frames.  In Fig.~\ref{fig:noah},  we associate the  
characteristic  frequency $\omega_o$ with the emitter's frame, as seen by an observer at the point of emission, and the    shifted-frequency, $\omega_s$ with the receiver's frame.\\

 By comparing Eq.~(\ref{eq:LorentzDefine}) to Eq.~(\ref{eq:map}), 
we define what we will call a  fundamental convolution  of frequencies (FCF):
\begin{equation}
e^\xi=\frac{\omega_{s}}{\omega_{o}}.  
\label{eq:definite}
\end{equation}
which  will be used  to map frames based upon    ratios of frequencies emitted 
and received.  The use of light   to define  the  mapping of frames   is apt, as light  defines the  metric  and thereby the curvatures on any space-time.\\

The FCF for the $v_1$ mapping, $e^{\xi_1}$, is defined by the gravitationally shifted 
frequencies for the emitter and receiver galaxies, $\omega_{gal}$ and $\omega_{mw}$,  defined in Sect.~\ref{sec:Schw} and~\ref{sec:kerr}.  The galactic  FCF
 \begin{equation}
e^{\xi_1}=\frac{\omega_{mw}}{\omega_{gal}},  
\label{eq:specific}
\end{equation}
yields, by the general LDSF, 
\begin{equation}
\frac{v_1}{c}=\frac{  \frac{\omega_{mw}}{\omega_{gal}}- \frac{\omega_{gal}}{\omega_{mw}}}{ \frac{\omega_{mw}}{\omega_{gal}}+ \frac{\omega_{gal}}{\omega_{mw}}}.
\label{eq:prime1}
\end{equation}
The \emph{parameter}  $v_1$   describes the mapping, not a physical speed. 
The use of the   LDSF for slightly 
curved frames  is  justified by the weak field assumptions,  which are in common  use  for  evaluation 
of the  flat-rotation curve problem.    \\

The  second term in Eq.~(\ref{eq:convolutionFunc}), $v_2$,  is a mapping back to the   flat frames where 
  physical measurements are made.  The FCF for the  $v_2$ mapping, $e^{\xi_2}$, is defined by four frames: the two
  curved frames of the emitter/receiver galaxies $e^{\xi_{curved}}$ and the two
  flat tangent frames $e^{\xi_{flat}}$ of the emitter/receiver galaxies.  This FCF is defined
 \begin{equation}
 e^{2\xi_2}= e^{\xi_{flat}}/ e^{\xi_{curved}}.
 \label{eq:FCFtwo}
\end{equation}
where the curved FCF is $e^{\xi_{curved}}=e^{\xi_1}$, from   Eq.~(\ref{eq:specific}).
  The FCF  for  the flat frames is defined by our expectation of  the  frequencies in  Eq.~(\ref{eq:primeLUM}):
  \begin{equation}
  e^{\xi_{flat}} =  \left( \frac{\omega_{l} }{\omega_{o}}\right);
  \label{eq:flatflat}
  \end{equation}
which describe our understanding of the luminous matter $M_L$ within 
  the Newtonian limit, and thereby  the associated 
  flat frames of the emitter and receiver.  \\

 In order to convolve the FCF of Eq.~(\ref{eq:FCFtwo}) into the $v_2$ LDSF, we must assess the quality of a 
 transformation from a curved to a flat frame.   In Eqs.~(\ref{eq:LorentzDefine}) and~(\ref{eq:map})  boosts  are always 
  defined as positive rotations away from the time axis in Fig.~\ref{fig:noah},  since the `rest'  frame is arbitrary.    However,  since 
  it  is the rest frame  which we associate  with the  flat frames in Eq.~(\ref{eq:flatflat}), we need to construct a 
  reverse boost.\\
  
A reverse boost does not exist in Special Relativity, but we propose the   
  the reciprocal of  Eq.~(\ref{eq:map}): 
 \begin{equation}
\frac{v_2 }{c}= \coth \xi_2= \frac{e^{\xi_2 } +e^{-\xi_2 }}{e^{\xi_2 } - e^{-\xi_2 }}, 
\label{eq:hyperbolicreference}
\end{equation}  
 Since the $v_2$  is a mapping of four frames it is  convenient to rewrite   Eq.~(\ref{eq:hyperbolicreference}) in 
 the form:
 \begin{equation}
\frac{v_2 }{c}=  \frac{e^{ 2\xi_2  }+1}{e^{ 2\xi_2 } - 1}, 
\label{eq:hyperbolico}
\end{equation}  
such that the last mapping, $v_2$,  is   
  \begin{equation}
 \frac{v_2}{c}=
 \frac{
 \frac{\omega_{l} }{\omega_{o}}
 + \frac{\omega_{mw}}{\omega_{gal}}}{ 
  \frac{\omega_{l}}{\omega_{o}}
 - \frac{\omega_{mw}}{\omega_{gal}}} .
 \label{eq:MAPTWO}
\end{equation}
Again,  $v_2$ is a \emph{parameter} describing the mapping, 
 not a physical velocity.  
\subsection{Parameters $\zeta$ and  $\alpha$}
\label{sec:zeta}
\begin{figure*}
\centering
\includegraphics[width=.40\textwidth]{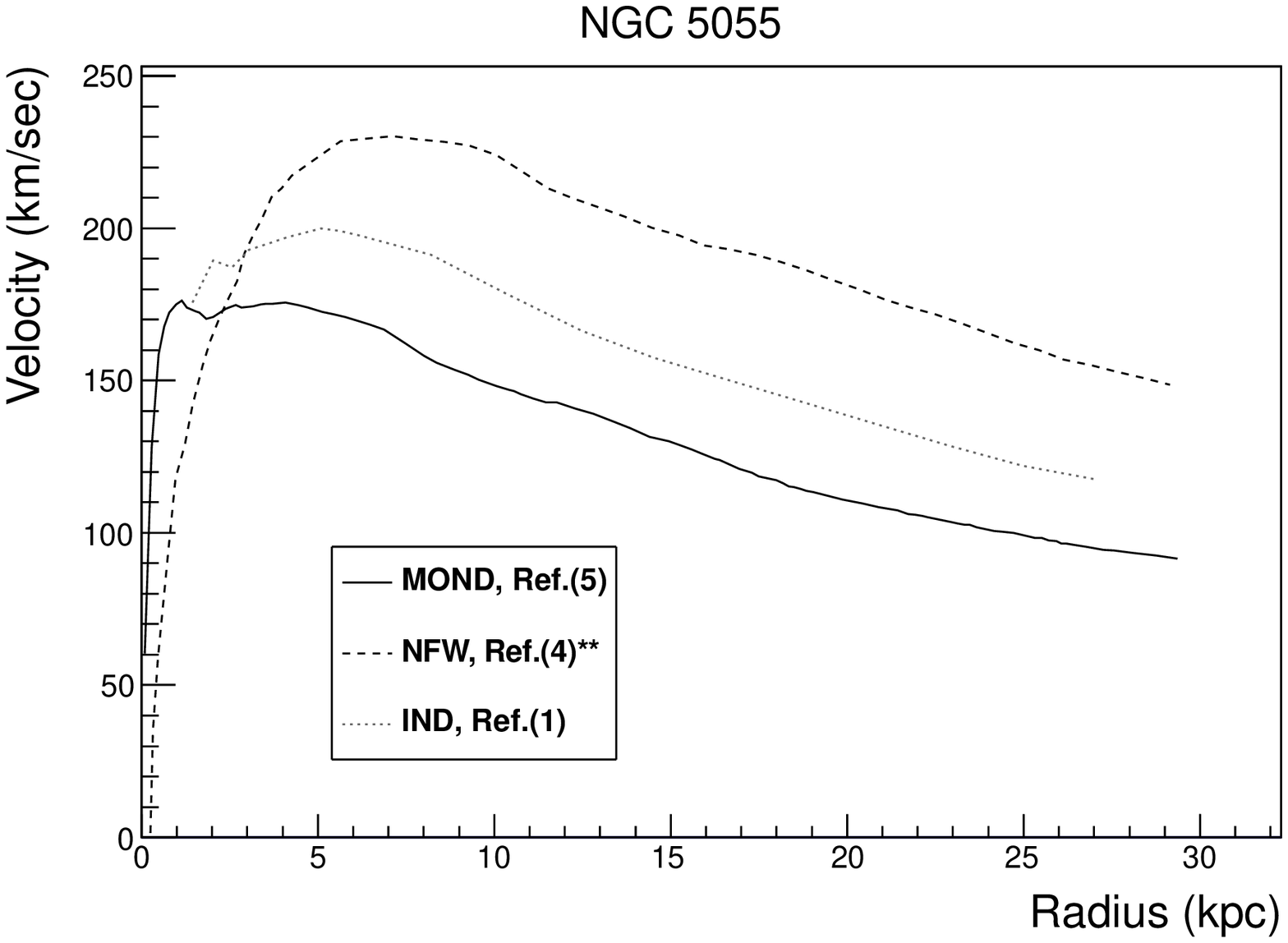}
\includegraphics[width=.40\textwidth]{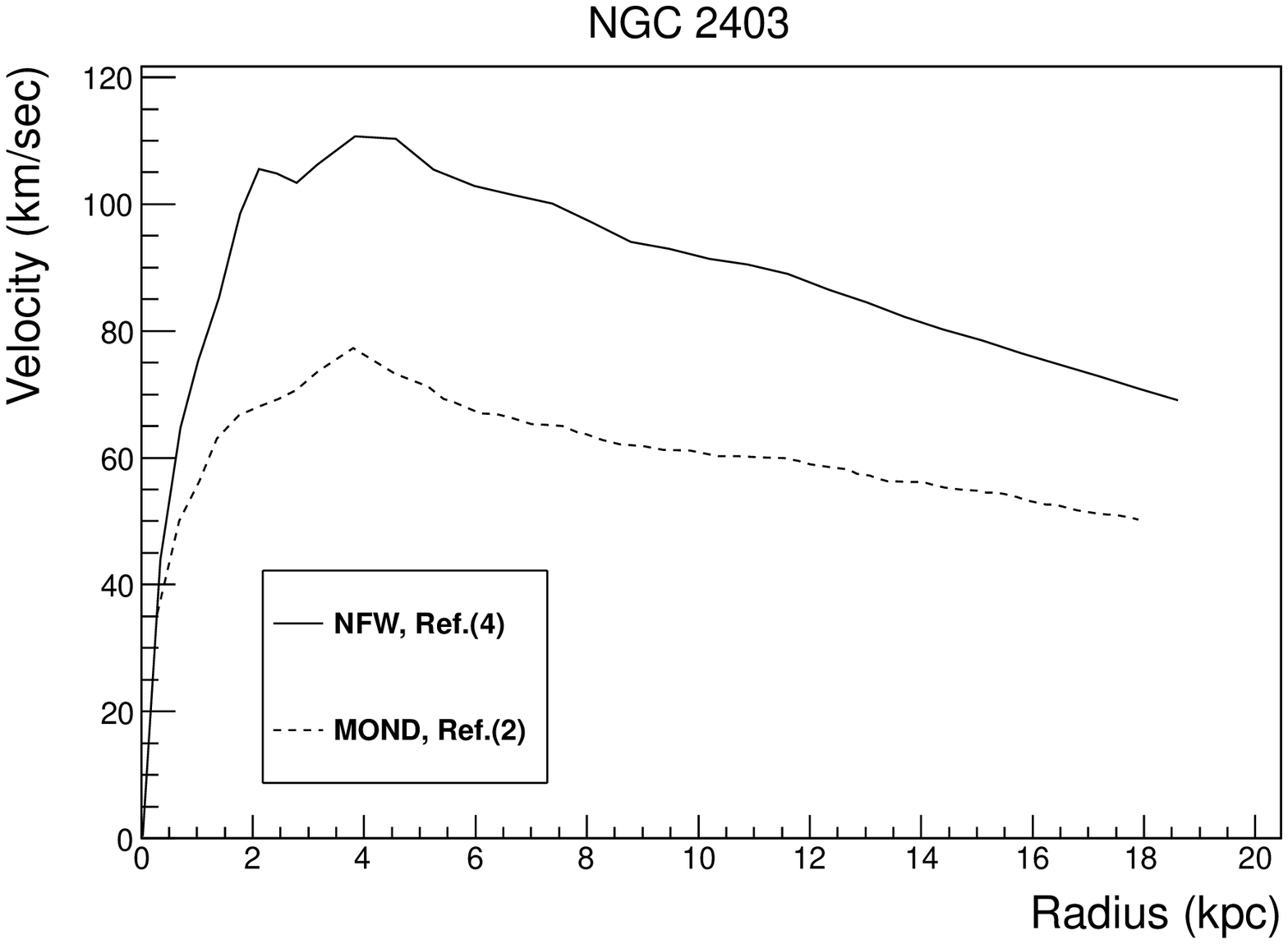}
\includegraphics[width=.40\textwidth]{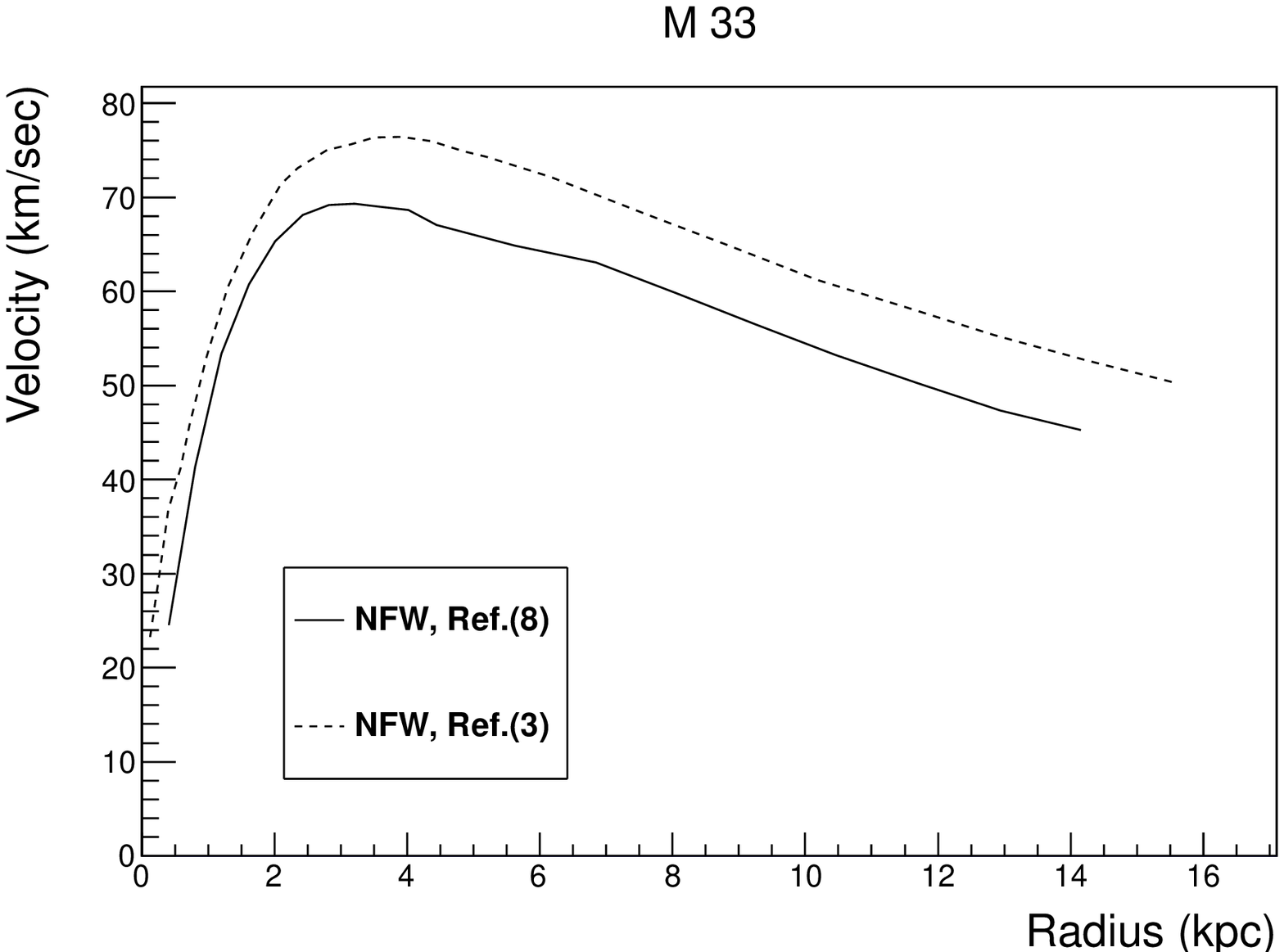}
\includegraphics[width=.40\textwidth]{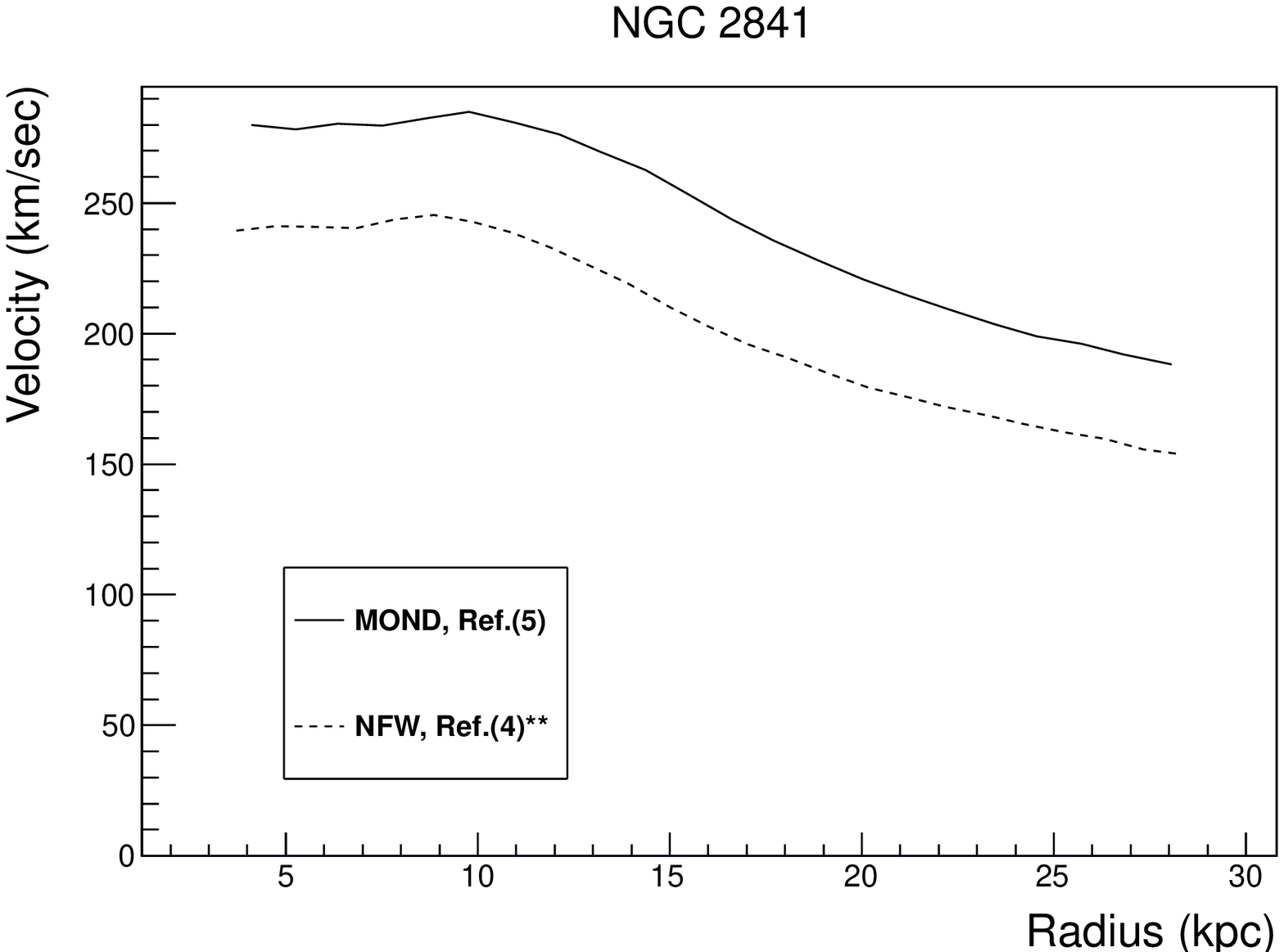}
\includegraphics[width=.40\textwidth]{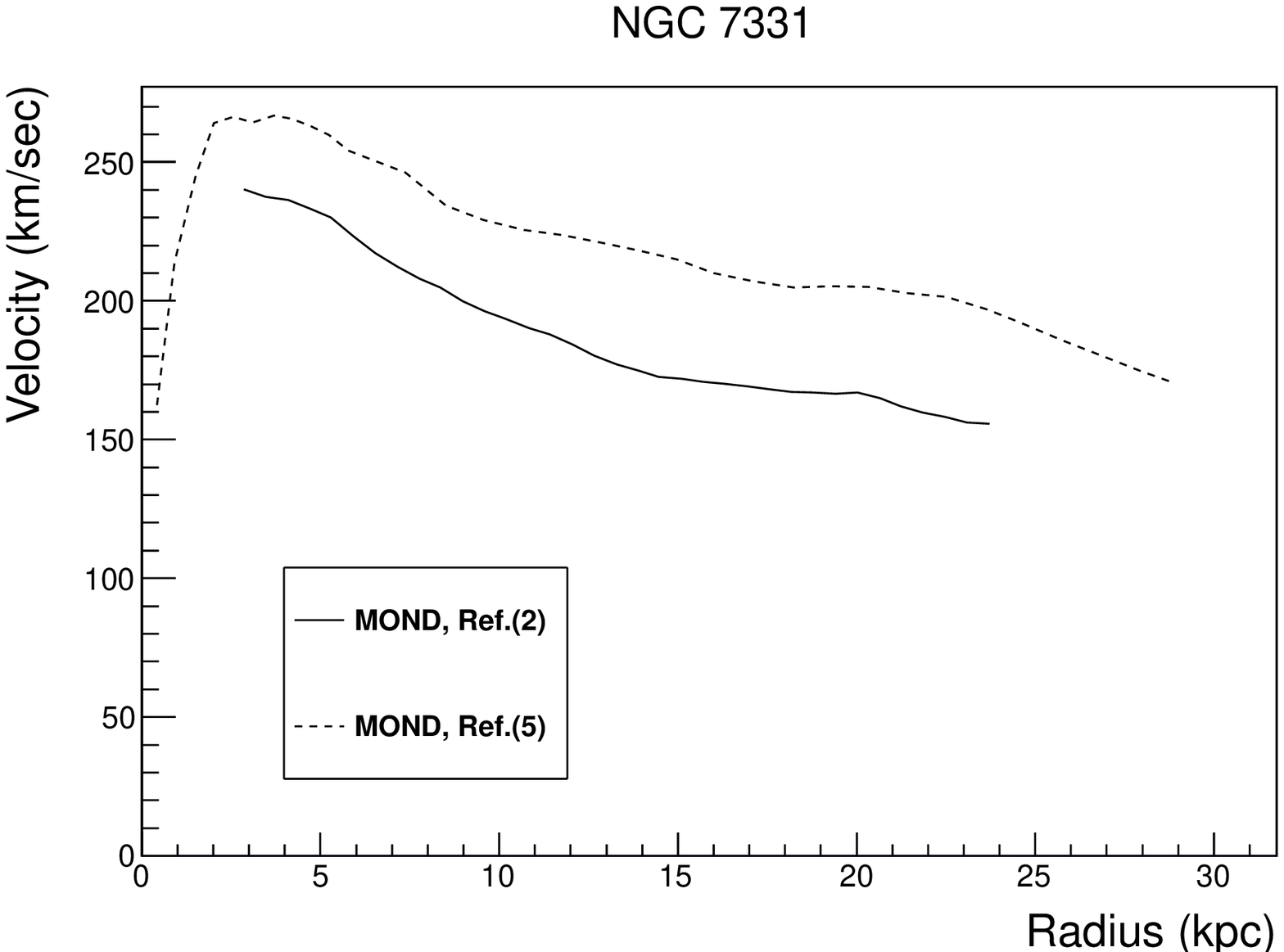}
\includegraphics[width=.40\textwidth]{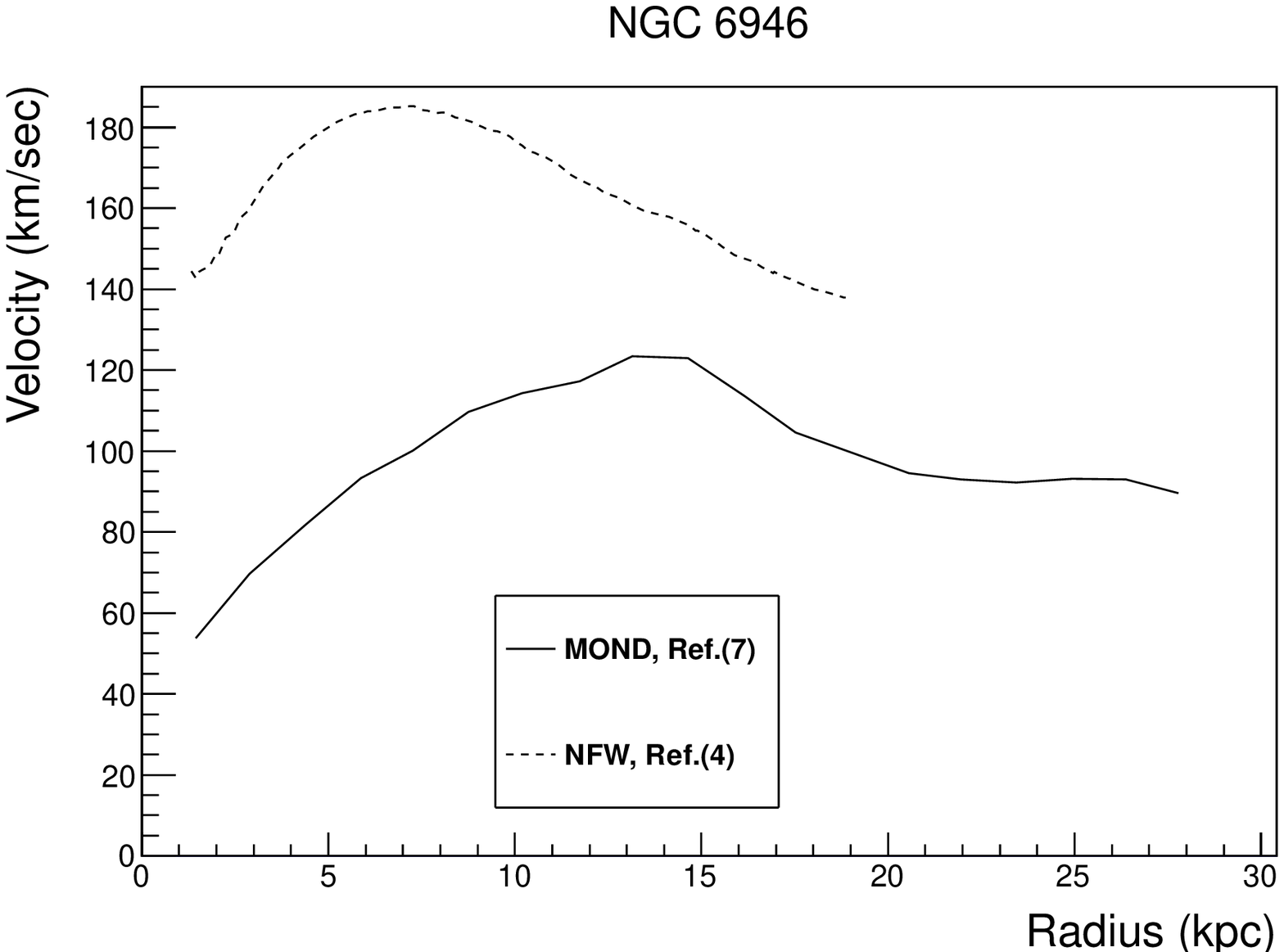}
\includegraphics[width=.40\textwidth]{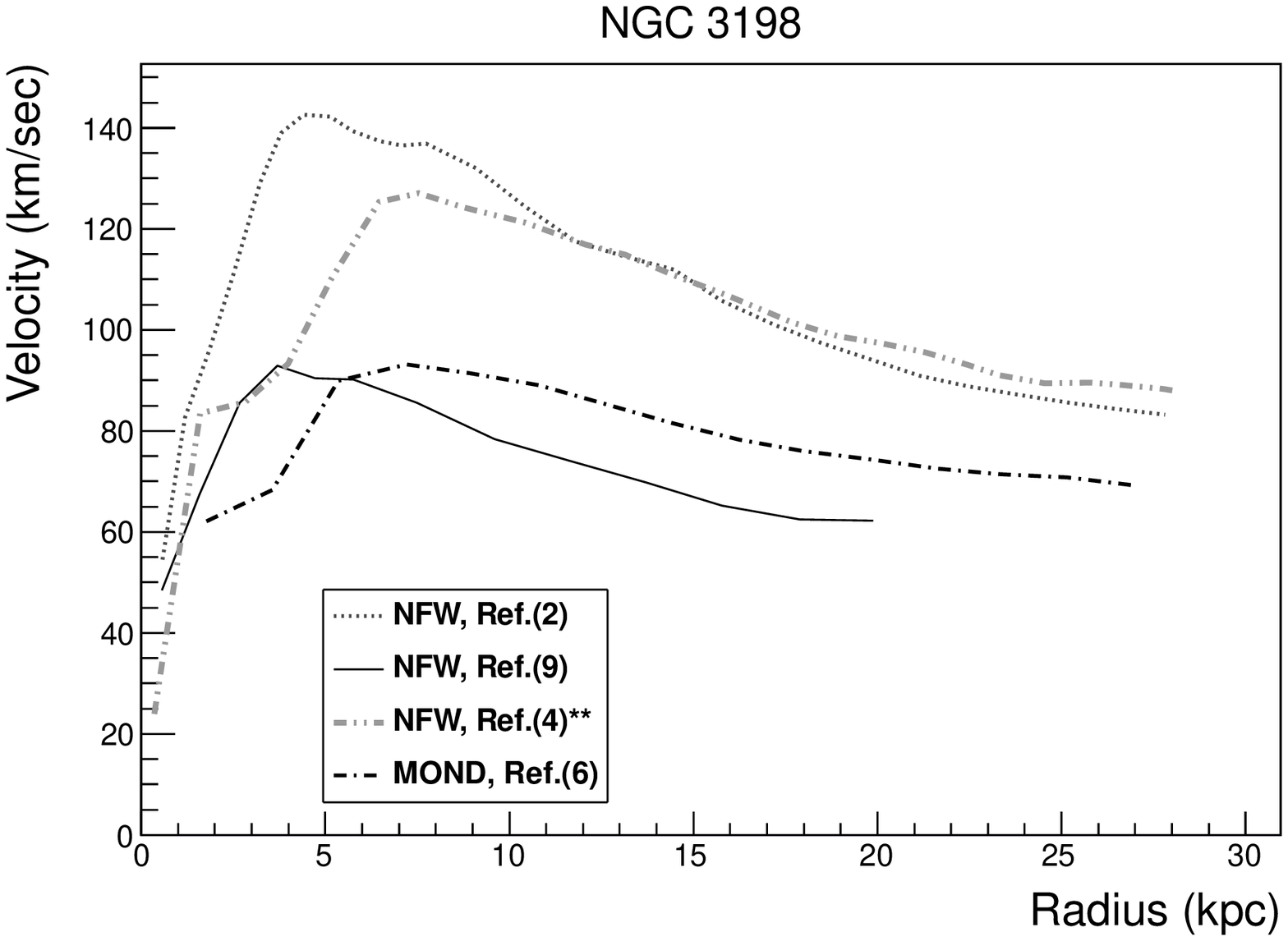}
\includegraphics[width=.40\textwidth]{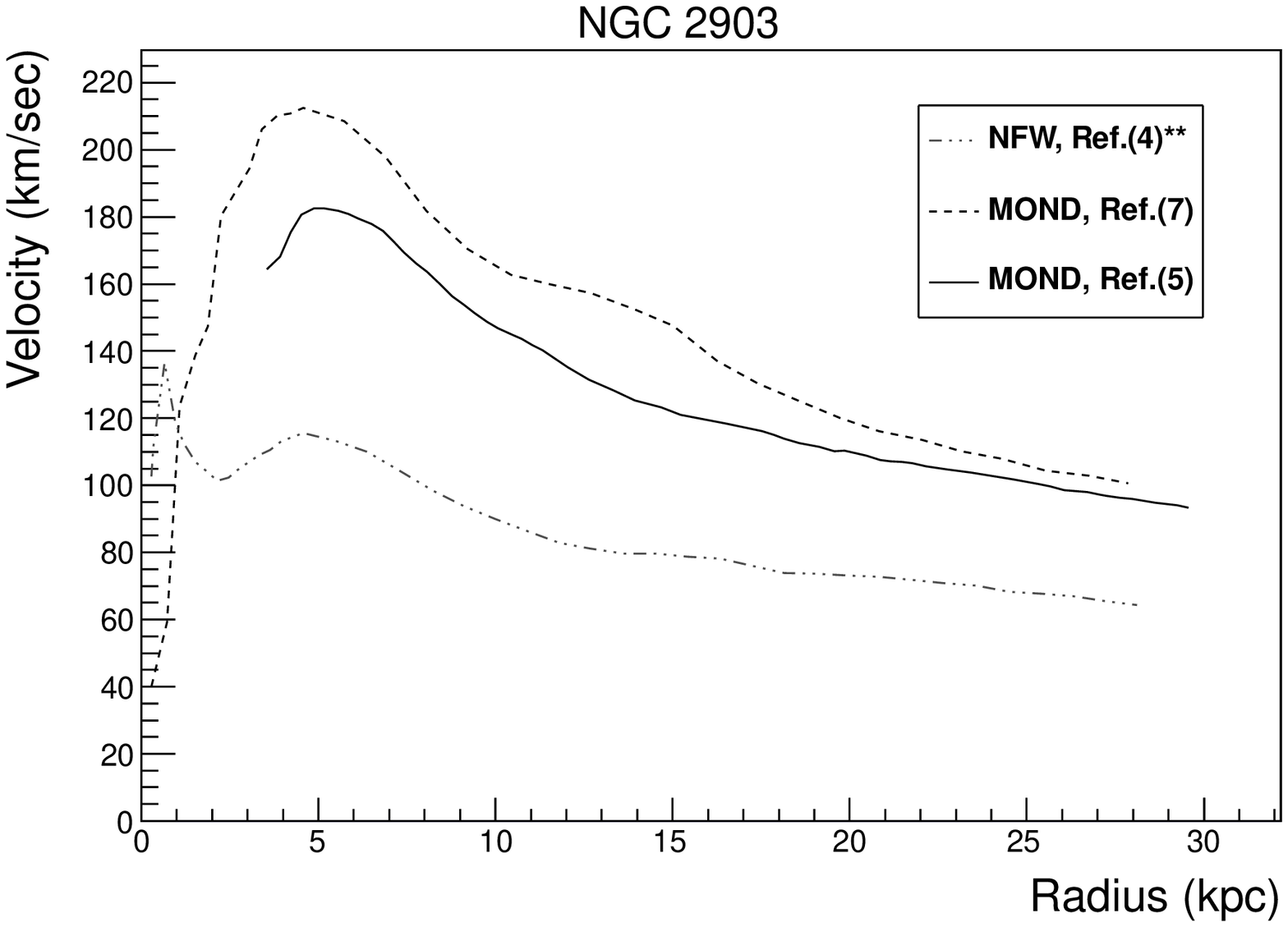}
\caption{Variations in $M_L$ reported for a different emitter galaxies.\label{fig:massmodels18}. (**) after the citation indicates    luminous profile which does not converge to an LCM fit.
Model contexts of $M_L$ are either NFW,  MOND, or  IND (model independent). 
 References are as in Table.~\ref{tab:referenceDATAs}.}
\end{figure*}

The luminous mass $M_L$ is  treated as an 
  adjustable parameter in models  such as NFW or MOND, 
as can be seen  in Fig.~\ref{fig:massmodels18} 
for eight galaxies where we have multiple data sets.  The variability in the reported $M_L$ distribution 
 for a single galaxy is due to the underconstrained  nature of luminous-matter modeling \citep{Conroy,JNav}.   The first LCM  fitting parameter  $\zeta$ 
allows model flexibility, as a dimensionless scaling of the luminous matter profiles given in the context of another model.\\
\clearpage	 
\begin{deluxetable}{l  l  l  l  l }
\tabletypesize{\scriptsize}
\tablecaption{Luminous mass-to-light Ratios\label{tab:referenceDATAs}}
\tablewidth{0pt}
\tablehead{
\colhead{Galaxy}& \colhead{Band\tablenotemark{a}} & \colhead{$M_L/L$\tablenotemark{b}}  & \colhead{Model\tablenotemark{c}}  & \colhead{Reference\tablenotemark{d}}
}
\startdata
NGC 3198					& 	B		&$ 1.1  $					   	&NFW		& 2\\
NGC 3198					&        B		&$3.8$						&NFW		&9\\	
NGC 3198**				& $3.6\mu m$	&$1.00_{\rm{bulge}}$,$0.64_{\rm{disk}}$&NFW		&4\\
NGC 3198 					&        B		&$0.48_{\rm{disk}}$				&MOND		&6\\ 	 		
\tableline
M 33						& $2.6mm$   	&$ 1.0 $						&NFW		&3\\ 		
M 33						& $3.6\mu m $	&$1.25$						&NFW		&8\\ 	 		
\tableline
NGC 5055					& F	 		&$ 3.6_{\rm{disk}}  $				&NFW		&1 \\
NGC 5055**				& $ 3.6\mu m $  &$1.0_{\rm{disk}} $				&NFW		&4\\ 		
NGC 5055					& $ 3.6\mu m  $	&$0.56_{\rm{bulge}}$, $0.55_{\rm{disk}} $&MOND	&5\\
\tableline
NGC 2403					& B    		&$1.6 $						&MOND		&2\\	
NGC 2403					& $3.6\mu m $	&$0.41 $						&NFW		&4\\ 	 		
 \tableline	 		
NGC 3521					&$3.6 \mu m 	$&$  0.71_{\rm{disk}}$			&MOND		&5 \\	 
 \tableline	 		
NGC 2841**				&$3.6 \mu m$ 	&$0.89_{\rm{bulge}} $, $  1.26_{\rm{disk}}$&NFW	&4\\
NGC 2841					& $3.6\mu m $	&$1.04_{\rm{bulge}}$,$  0.89_{\rm{disk}}$	&MOND	&5 \\ 		
\tableline	 		
NGC 7814					&$3.6 \mu m $	&$  0.71_{\rm{bulge}} $, $  0.68_{\rm{disk}}$&NFW	&11\\		
\tableline		 		
NGC 7331					&B			&$ 1.8_{\rm{bulge}} $	, $  2.0_{\rm{disk}}$	&MOND	&2\\
NGC 7331					& $3.6\mu m $	&$1.22_{\rm{bulge}} $, $0.40_{\rm{disk}}$	&MOND	&5\\ 
\tableline
 NGC 891					& $3.6\mu m $	&$ 1.63_{\rm{bulge}} $, $ 0.77_{\rm{disk}}$&IND	&11\\ 		
\tableline
M 31						& B			&$ 2.8-6.5$					&IND		&10\\
\tableline
NGC 5533					&B			&$3.4$						&MOND		&7\\
\tableline 
UGC 6973					&B, R 		&$2.7,0.4$						&MOND		&7\\
\tableline 
NGC 4088					&B, R 		&$1.0, 0.7$						&MOND		&13\\
\tableline
NGC 3992 					&B,R			&$4.9, 2.2$					&MOND		&7\\
\tableline 
NGC 4138					&B,R			&$3.5,1.0$						&MOND		&13\\
\tableline
NGC 6946					&B			&$ 0.5  $						&MOND		&7 \\
NGC 6946					& $3.6\mu m $	&$1.002_{\rm{bulge}} $, $0.64_{\rm{disk}}$&NFW	&4\\
\tableline 
NGC 3953					&B,R			&$2.7, 0.9$					&MOND		&7\\
\tableline		 			
NGC 2903					& B, R	 	&$3.6, 2.6 $				   	&MOND		&7\\
NGC 2903**				& $ 3.6\mu m $  &$0.61_{\rm{disk}} $,	$1.30_{\rm{bulge}}$&NFW		&4 \\ 	
NGC 2903					&$ 3.6\mu m  $	 &$1.71_{\rm{disk}} $				&MOND		&5\\
\tableline 
NGC 5907					&B,R			&$3.9,2.0$						&MOND		&13 \\
\tableline 
NGC 3726					&B,R			&$1.1 , 0.6$					&MOND		&12\\
\tableline 
F 563-1					&R			&$6.3$						&NFW		&9\\
\tableline 
NGC 925					&$3.6\mu m$	&$0.65$						&NFW		&4	\\
\tableline  
NGC 7793					&B  			&$2.8-6.5$						&MOND		&5  \\
\enddata
\tablecomments{}
\tablenotetext{a}{Wavelength band  for observations of total light.}
\tablenotetext{b}{Reported stellar mass-to-light ratios $M_L/L$, in units of  $(M_\odot/L_\odot)$.}
\tablenotetext{c}{Model context of $M_L$: NFW, MOND, or IND (model independent).}
\tablenotetext{d}{{\bf References.}
1:~\citet{Batt},  2:~\citet{Bot},  3:~\citet{Cor03},
4:~\citet{Blok},  5:~\citet{Gent},  6:~\citet{Maria},  7:~\citet{SanMcGa}, 
 8:~\citet{Seigar11},  9:~\citet{JNav}, 10:~\citet{Car}, 11:~\citet{Frat}, 12:~\citet{San96},  13:~\citet{Ver98}.}
\end{deluxetable}

\clearpage

The second parameter in Eq.~(\ref{eq:zople1})$,  \alpha$, is presupposed to embody the  relationship between the luminous convolution and dark matter 
  \begin{equation}
v_{DM}^2= \alpha v_{LC}^2.
\label{eq:name}
\end{equation}
 This parameter, unlike those of MOND or NFW, does not have a physical interpretation, but 
rather is a dimensionless number which characterizes the ratio of the dark matter mass to the luminous convolution term.  The substitution  of $\alpha v_{LC}^2$ for $v_{DM}^2$ in Eq.~(\ref{eq:zople1}) does not  correspond to   models such as MOND, however it does
add another perspective to  the  work on a  'universal rotation curve'
by \citet{Persic,PSS,Rub}.     The universal rotation curve phenomenon shows  the  distribution of luminous matter `traces' the dark matter by a relationship with a  median value  at the luminosity of the Milky Way.   Our preliminary investigation into $\alpha$  are shown in Sect.\ref{alphResults},  and seem to substantiate   the special status of the Milky Way's luminous mass as a critical point in the relationship between dark and luminous matter.\\
 
\subsection{Curvature Contributions}
\label{sec:curvature}
The LCM approach arises from two assumptions. The first is
that even in a curved spacetime we can always define a local Lorentz frame,
which allows us to locally define the energy $E$ of a photon \citep{Hartle} 
as
\begin{equation}
E= -{\bf u}\cdot {\bf k} = \hbar \omega_o
= -(u^tk_t + \vec{k}\cdot\vec{u}),
\label{eq:willtest}
\end{equation}
where ${\bf u}$ is the 4-velocity of the local observer, ${\bf k}$ is the
photon 4-momentum\footnote{$\hbar=1$ for the remainder of this work},
$\vec{u}$ is the spatial $3-$vector of ${\bf u}=(u^t,\vec{u})$,
and $\vec{k}$ is the wave $3-$vector of ${\bf k}=(k^t,\vec{k})$.\\

The second aspect is that the photon can be propagated out to asymptotic
infinity, enduring only negligible bending of its ray path. This argument
allows us to use the eikonal approximation \citep{Born} to solve a wave equation
for the effective index of refraction. One can not always neglect 
deviations of light geodesics from straight lines. Therefore, in general
one is forced to integrate the geodesic equations numerically,
\cite{Asaoka}. However, in the eikonal limit, as justified by the highly diffuse matter distribution of spiral galaxies,
  ray optics  allows us to focus only on the magnitude of the curvature.    \\

We assume that the scalar
wave equation will capture the relevant physics in the eikonal limit.
This wave equation for a photon with a wavefunction $\Psi(x)$ is

\begin{equation}
\Box \Psi  =  \frac{1}{\sqrt{g.}} \frac{\partial}{\partial x^\mu}
\left( g^{\mu\nu} \sqrt{g.} \frac{\partial}{\partial x^\nu} \right)\Psi
=0 \ ,
\label{eq:extrn}
\end{equation}
where $g^{\mu\nu}$are the contravariant metric components,
and $(g_.)$ is the determinant of the matrix of covariant components
$g_{\mu\nu}$.  In the eikonal approximation
for a photon with four vector  $k_\alpha=(k_t, k_i)$, emitted in the direction tangent to a circular orbit,
 instantaneously the spatial wave vector $\vec{k}=k^i \hat{\bf e}_{i}$ is 
\begin{equation}
\vec{k} = k\hat{\bf e}_{\varphi}
\end{equation}
where ${\bf  \hat{e}_{\varphi}} =\frac{\bf e_{\varphi}}{|\bf
e_{\varphi}|}=\frac{\bf e_{\varphi}}{\sqrt{g_{\varphi\varphi}}}$ is a
unit vector, with ${\bf e}_{\varphi}$ the covariant basis vector. 
When  all metric components are independent of $t$, the general wavefunction
at a frequency $\omega$ can be written as
$\Psi({\bf r},t)= e^{-i\omega  t} \Psi({\bf r}) + c.c..$, whereby the local
eikonal wave function may be written as 
\begin{equation}
\Psi(x)= \Psi_o \exp(-i\omega  t) \exp\left(i \int_{path}d{\bf r}\cdot
\hat{\bf e}_{\varphi}k \right)
\label{eq:cracker}
\end{equation}
where $\Psi_o$ is an amplitude, $d{\bf r}= {\bf e}_{\varphi} d\varphi$,
and, $\omega=k_t c$ is the frequency that
would be observed as emitted from $r$, by an asymptotic flat space observer at rest. \\

As discussed by  \citet{Narayan}, the effects of gravitational curvature on light can be described by an effective index of refraction $n$, which  relates the the vacuum light speed $c$ to the  coordinate light  speed $\tilde{c}$, $ n = c/\tilde{c}$.  Whereas the apparent slowing of the light speed in classical electrodynamics is due to  the  increased   path length because of interactions in the medium,  for gravitational effects it reflects the increased path length due to space-time curvature. \\

 Generalized  to  curved space-times by a  covariant  wave equation,
 Eq.~(\ref{eq:extrn}), the coordinate light speed at a given radius $r$ in a galaxy, $\tilde{c}(r)_{gal}$, is related to the 
   effective index of refraction $n(r)$ as  
\begin{equation}
 n(r) = \frac{c}{\tilde{c}(r)_{gal}}.
\label{eq:sinkq}
\end{equation}
Wave equation solutions for Schwarzschild and Kerr  are given in what follows.   Both metrics used are exterior vacuum solutions, meaning they are intended for use \emph{outside} a constant source of mass (in $\Phi_L$) and ($a$), the angular momentum per unit mass.   Our use of these exterior metrics \emph{inside} the plane of the galactic disc is justified by  
   the assumption in Sect.\ref{sec:lumGal}, regarding spherical symmetry and Newton's Shell theorem.   This  allows us to evaluate 
the luminous matter enclosed at each radii as the only contribution to the metric $g_{\mu\nu}$, by the terms $\Phi_L$ and $a$,  resulting in an 
exact solution of the Einstein equations at each radii and  foliations of solutions as we move out in the radial galactic coordinate $r$. \\

  The spherical assumption  does break   down out of the plane of the disc and close to the galactic center  (due to tidal forces),   or in the presence of  a   symmetry breaking feature  such as a bar. 
   However, in general,  the inner regions of galaxies do not demonstrate the mismatch between $M_L$ and $M'$,  and so this assumption allows
    a heuristic construction of the LCM  without loss of generality in  regions of interest, large radii.\\%
\subsection{Schwarzschild Wave Equation}
\label{sec:Schw}
The Schwarzschild  gravitational red-shift formula 
 for a photon emitted  
at $r$, 
\begin{equation}
 \frac{\omega_o }{ \omega( r) }=\left(\frac{1}{\sqrt{-g_{tt}} }\right)_r,
\label{eq:Clone}
\end{equation}
  relates the locally observed, characteristic frequency $\omega_o $, to 
the frequency received at infinity by 
a stationary observer 
$\omega( r)$, as a function of radial position $r$ in the potential well.  The frequency $\omega( r)$ reflects the change in photon energy due to curvature, indicated by 
the   Schwarzschild 
time metric coefficient, 
\begin{equation}
 g_{tt}=-\left(1- 2\frac{G M}{c^2r}\right)
 \label{eq:timeportion}
 \end{equation}
 which in the limit of  weak-field  metrics \citep{Hartle} , is 
\begin{equation}
g_{tt} \approx -1 + 2\Phi/c^2
\label{eq:weakfield}
\end{equation}
  for $\Phi$ the Newtonian gravitational potential at $r$ the emission point,  Eq.~({\ref{eq:potentialgeneral}).  All LCM calculations are made in this limit,  for $\Phi=\Phi_L$  the luminous matter Newtonian potential.   Eq.~(\ref{eq:Clone}) can    be derived  using either a wave equation or a Killing vector approach \citep{Wald,Cisn}.  However,  we focus   on the wave equation approach,  to make   connection with  the properties of the photon mentioned in Sect.\ref{sec:introLCM}. \\

To write the Schwarzschild wave equation, consider a general Schwarzschild metric, $(t,r,\varphi, \theta)$, whereby the nonzero $g_{\mu\nu}$ are
$g_{tt}, g_{\varphi \varphi }, g_{rr },$ and $g_{\theta
\theta} $,  which are independent of $\varphi$ and $t$.
 The components of the metric  $g$ are 
 are independent of time and  
have no cross terms $g_{oi}$,  we can therefore
write immediately
\begin{equation}
g^{00} \frac{\partial}{\partial t^2}\Psi -\frac{1}{\sqrt{g.}} \frac{\partial}{\partial x^i}
\left( g^{ij} \sqrt{g.} \frac{\partial}{\partial x^j} \right)\Psi=0
\label{eq:waveeqnSCHW}
\end{equation}
where superscripts $(i j)$ denote spatial components \citep{Hartle}.  
For  $\theta=\pi/2$
in the plane of the disc,  the Schwarzschild line element becomes

\begin{equation}
ds^2=g_{tt}dt^2 + g_{rr}dr^2 + g_{\varphi \varphi }d\varphi^2.
\label{eq:metricSchwarzschild}
\end{equation}
for $g_{rr}=1/g_{tt}$, $g_{\varphi \varphi }=r^2$, and $  \varphi$ is the  azimuthal coordinate. \\

In  a small neighborhood of the angle $\varphi=\varphi_o$, for a  photon
 emitted at ${r}$,  Eq.~(\ref{eq:cracker}) reduces to
\begin{equation}
\Psi(x) =   \Psi_o \exp(-i \omega t) \exp\left(ik \sqrt{g_{\varphi \varphi}}
(\varphi-\varphi_o) \right).
\label{eq:wakey1}
\end{equation}

\noindent Inserting  Eq.~(\ref{eq:wakey1}) into  Eq.~(\ref{eq:waveeqnSCHW}) yields, 
\begin{equation}
\omega ^2 [-g^{tt}   -
n^2 g^{\varphi \varphi } g_{\varphi \varphi } ] \Psi = 0. 
\label{eq:vhc}
\end{equation}
The general solution of  Eq.~(\ref{eq:vhc}) is
\begin{equation}
\frac{1}{\sqrt{ -g_{tt}  }} =n({r}),
\label{eq:vhcsolution} 
\end{equation}
for $n({r})$ the effective index of refraction,  defined as  Eq.~(\ref{eq:sinkq}).\\

In the Schwarzschild context, curvature affects only the energy (e.g. frequency)  of 
the photon \citep{Wald},
\begin{equation}
n({r})=\frac{\omega_o  }{ \omega} ,
\label{eq:nyc2}
\end{equation}
so that  Eq.~(\ref{eq:Clone}) equals
 Eq.~(\ref{eq:vhcsolution})
 \begin{equation}
 n(r)=\frac{c}{\tilde{c}_{gal}}=  \frac{\omega_o  }{ \omega} .
\label{eq:nyc3}
\end{equation}

Since it is that  frequency $\omega$ which gives information at infinity regarding the 
 curvature at $r$, for some   enclosed luminous mass $M(r)_L$,   it is labeled $\omega_{gal}$ or $\omega_{mw} $ in what follows, to reflect said gravitational source which it describes. \\

Alternately,  the identification of the frequencies  can be demonstrated by  construction of  a local transformation from a Schwarzschild metric to the tangent Lorentz frame.  The photon which is locally observed with an   emission frequency  $\omega_o$,     is also measured by the external observer as coming from an emitter embedded in a curvature  as indicated by $\omega$.   To connect these two observations, 
 the change of coordinates from $(t, r, \phi)\to(\hat{t}, \hat{r}, \hat{\phi})$  is
\begin{eqnarray}
dt=\frac{d\hat{t}}{\sqrt{-g_{tt}(r)}}&& \nonumber \\
dr= \frac{d\hat{r}}{\sqrt{-g_{rr}(r)}} &&\nonumber \\
{r} d\phi=\hat{{r}}d\hat{\phi}, &&\nonumber \\
\end{eqnarray}
such that  Eq.~(\ref{eq:metricSchwarzschild})  becomes the local Lorentz  frame 
\begin{equation}
ds^2=  d\hat{t}^2 +   d\hat{r}^2 +   \hat{{r} }^2 d\hat{\phi}^2.
\label{eq:coordTrans}
\end{equation}
   Starting at the emission point $({r}, \varphi_o)$, the observed frequency and wavenumber $\omega_o, k_o$  in the Lorentz frame of  Eq.~(\ref{eq:coordTrans})  must be related to the coordinate frequency  and wavenumber $\omega, k_{r}$ measured by the observer who sees the space-time as   Eq.~(\ref{eq:metricSchwarzschild}).  \\

The transformation of the  time dependent portion of the eikonal  wave function
 Eq.~(\ref{eq:cracker}), yields
\begin{equation}
  \exp(-i \omega t)  =  \exp\left(-i \omega \frac{\hat{t}}{\sqrt{-g_{tt}}}\right)
\label{eq:butter1}.
\end{equation}
The  local Lorentz  flat  frame observer, for whom 
$\omega =\omega_o $  and $g_{tt}\to\eta_{tt}=-1 $, 
\begin{equation}
  \exp(-i \omega t)=   \exp\left(-i \omega_o  \hat{t} \right) 
   \label{eq:butter2}
\end{equation}
The second observer, who sees   Eq.~(\ref{eq:metricSchwarzschild}),  finds $\omega  =\omega_{gal}$ and $g_{tt}=g_{tt}({r})$, would 
measure
\begin{equation}
 \exp\left(-i \omega\frac{\hat{t}}{\sqrt{-(g_{tt})_{r}}}\right) \equiv  \exp\left(-i \omega_o  \hat{t} \right).
   \label{eq:butter}
\end{equation}
Setting these two measurements of the same wave packet equal returns the identification of the 
frequencies with the gravitational red-shift
\begin{equation}
  \omega \frac{1}{\sqrt{-(g_{tt})_{r}}}  =
     \omega_o .
   \label{eq:squash}
\end{equation}
 \subsection{The Kerr Wave Equation}
 \label{sec:kerr}

 The Kerr wave equation is constructed from the covariant wave operator in  Eq.~(\ref{eq:extrn}).
 We consider a general Kerr-type metric in Boyer-Lindquist
coordinates, $(t,r,\varphi, \theta)$, whereby the nonzero $g_{\mu\nu}$ are
$g_{tt}, g_{t\varphi},g_{\varphi \varphi }, g_{rr },$ and $g_{\theta
\theta} $,  and are independent of $\varphi$ and $t$.
In Boyer-Lindquist coordinates, the exterior Kerr metric coefficients are \citep{chandra,oneill},

\begin{eqnarray}
g_{tt} &=& -(1-2Mr/\Sigma ) \ , \nonumber \\
g_{\varphi \varphi } &=& ((r^2 +a^2 )^2 - a^2 \Delta \sin^{2} \theta )
\sin^{2} \theta /\Sigma \nonumber \\
g_{\theta \theta } &=& \Sigma \ , \nonumber \\
g_{rr} &=&\Sigma / \Delta \ , \nonumber \\
g_{t\varphi } &=& g_{\varphi t} = -2Mar\sin^{2} \theta / \Sigma 
\end{eqnarray}
 
\noindent where

\begin{eqnarray}
\Sigma &=& r^2 + a^2 \cos^{2} \theta \ ,\nonumber \\
\Delta&=&r^2 + a^2 - 2Mr \ .
\end{eqnarray}

for $a=J/M$ the angular momentum per unit mass, and $M$  the enclosed mass at some radius $r$.\\

Writing  the eikonal approximation, as in the previous case
  Eq.~(\ref{eq:wakey1}), cross terms in time and space, $\omega k_{i}$,  are dropped as second order contributions.  In the equatorial plane of the galaxy, $\theta = \pi /2$, the Kerr wave equation yields
 \begin{equation}
\omega^2 [-g^{tt} + 2ng^{t\varphi} \sqrt{g_{\varphi \varphi } } -
n^2 g^{\varphi \varphi } g_{\varphi \varphi } ] \Psi = 0.
\label{eq:vhca}
\end{equation}
The general solution of  Eq.~(\ref{eq:vhca}) is
\begin{equation}
n(r) = \frac{ -g_{t\varphi} \pm \sqrt{(g_{t\varphi})^2-
(g_{\varphi\varphi})(g_{tt})}}{ g_{tt}\sqrt{g_{\varphi\varphi}} }.
\label{eq:index_yayaya}
\end{equation}
Note that the denominator is negative,
since $g_{tt} \approx -1 + 2\Phi/c^2$ in weak-field Kerr-type metrics,
where $\Phi$ is the Newtonian gravity potential. Therefore we
chose the $(-)$ sign preceding the square root in order to obtain
a positive $n$. \\

Interpreting the effective index of refraction  Eq.~(\ref{eq:index_yayaya}) as a ratio of 
frequencies to first order, as in the Schwarzschild case
  Eq.~(\ref{eq:nyc3}),  the Kerr effective index of refraction is:
\begin{equation}
n(r)\approx \frac{\omega_o}{\omega_{gal}(r)}.
\label{eq:index}
\end{equation}

\subsection{Parametrization of metric terms in the gravitational red-shifts}
\label{sec:gauge}  
The construction of the LCM rotation curve prediction, Eq.~(\ref{eq:zople1}), requires parametrizing the metric 
coefficients $g_{\mu\nu}$ in either    Eq.~(\ref{eq:vhcsolution}) or  Eq.~(\ref{eq:vhca}).  
The Kerr metric best mimics the physical symmetries of  spiral galaxies \citep{Hartle} ,  but the assumption  in Eq.~(\ref{eq:index}) 
 is not experimentally verified. The Schwarzschild gravitational redshift result, Eq.~(\ref{eq:nyc3}),  on the other hand, has been  experimentally confirmed \citep{Hartle,Wald}. 
As such,  the LCM function $v_{LC}^2$ will be parametrized 
 with Schwarzschild metric coefficients in all of the following fits.  We have verified that the LCM function works equally well with either  the Schwarzschild or  Kerr metric, but only the Schwarzschild-metric results will be presented here.\\

Finally, the Schwarzschild-metric coefficient  associated with time, $g_{tt}$ in Eq.~(\ref{eq:weakfield}), is parametrized with 
the   Newtonian gravitational potential  $\Phi_L$,  as defined in  Eq.~(\ref{eq:potentialgeneral}). 
The integration constant $\Phi_o$  is generally fixed so that as   
$r\to \infty$
   the potential  $ \Phi(r)\to 0 $.
\\

 This integration constant $\Phi_o$   is  the original  gauge freedom in classical gravity.  Two galaxies, with arbitrarily different luminous mass  distributions will naturally have different gauges in order to meet this  constraint.
However,  to enforce  continuity and energy conservation for a photon traveling  between two galaxies  the two
 galaxy gauges,  $\Phi_{o,1}$ and $\Phi_{o,2}$,  must be set to a common value. Consistent with Weyl's statement that,
``in the physical sense, only the ratios of the  $g_{ij}$ have an immediate tangible meaning" \citep[see][p.204]{Weyl},   we arbitrarily set both gauges $\Phi_{o,1}=\Phi_{o,2}=0$. \\

Evaluation of the Kerr metric terms requires one additional parameter: the  angular momentum per unit mass, $a $.  In the Newtonian limit, $a$ is    defined to be:
\begin{equation}	
 a(r) =	\frac{J ( r) = 4\pi \int_{r1}^{r2}  \rho \Omega r^4 dr}{ M ( r ) = r \int_{r1}^{r2}  F(r) dr/G },
\label{eq:angmom}
\end{equation} 
 where $F(r)$ is the Newtonian force in  Eq.~(\ref{eq:dynamical}),  $\Omega=  v_{lum}/r$ is
 the associated angular rotation frequency,  and  $\rho=M_L/( 4\pi r^3/3 )$ is the 
 luminous matter density. \\

\section{Validation}
\label{sec:LCM}
To validate the LCM model we perform fits of $v_{Lmod}$ (Eq.~\ref{eq:zople1}) to the observed data, $v_{obs}$.  The LCM takes the luminous mass models of both the emitter galaxy and the receiver galaxy, the Milky Way, as its inputs.  For the Milky Way (MW), we use the three mass models  shown  in Fig.~\ref{fig:bMWlum};  though the three profiles are significantly different from one another,  these differences have, in general,  little impact on the LCM fit results. The emitter-galaxy data are listed in Table~\ref{tab:referenceDATAs}. \\
 \begin{figure}
 \epsscale{.9}
\plotone{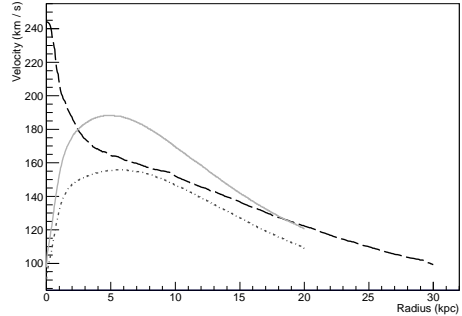}
\caption{The  MW luminous mass models used in this work are: \citet{Klypin}, model A dotted-dashed line   and  model B solid line ,  \citet{Sof81} dashed line.\label{fig:bMWlum}}
\end{figure}

\subsection{Fitting Procedure and Results}

The fits between the LCM model and the $v_{obs}$ data are calculated using the MINUIT minimization software as implemented in the ROOT data-analysis package~\citep{ROOT},  with one fit being performed for every emitter/receiver galaxy pair.  The fits are accomplished in two steps:  The first iteration of the fit yields an initial value of $\zeta$, denoted $\zeta_o$,  which is then used to rescale the luminous matter distribution $M_L$.  The second iteration of the fit is performed using the rescaled $M_L$ as the input luminous mass profile to calculate $v_{Lmod}$, resulting in the final values for $\zeta$ and $\alpha$.   

All of the fit results are reported in Table.~\ref{tab:resultsPARAMS}.  The fits for selected emitter/receiver galaxy pairs are shown in Figs~\ref{fig:resultsA}, \ref{fig:resultsB}, and~\ref{fig:resultsC}.
Fig.~\ref{fig:betaDistrib1} shows the distributions of $\zeta_o$ and $\zeta$ for all of the emitter/receiver galaxy pairs.   For most of the galaxies the second iteration results in a better fit as well as convergence of $\zeta_o$ to a median value of $\zeta=1.05\pm0.07$.\\

In a limited number of cases (four out of 35 emitter-galaxy data sets), the fits of the reported luminous mass profiles from one particular source failed to converge.   The four data sets are from \citet{Blok}:  NGC 2903, NGC 3198, NGC 2841 and NGC 5055.  The resulting parameter values are reported in Table.~\ref{tab:resultsPARAMS} and denoted with a double asterisk ($**$), but these data sets are not included in the $\alpha$ and $\zeta$ distributions or in the subsequent  fits reported in Figs.\ref{fig:resultsA}-\ref{fig:resultsC}.  For these same galaxies, alternate luminous mass profiles   are successful in LCM fits.  This suggests that the LCM can potentially constrain luminous matter modeling.\\
\begin{figure}
\epsscale{0.9}
\plotone{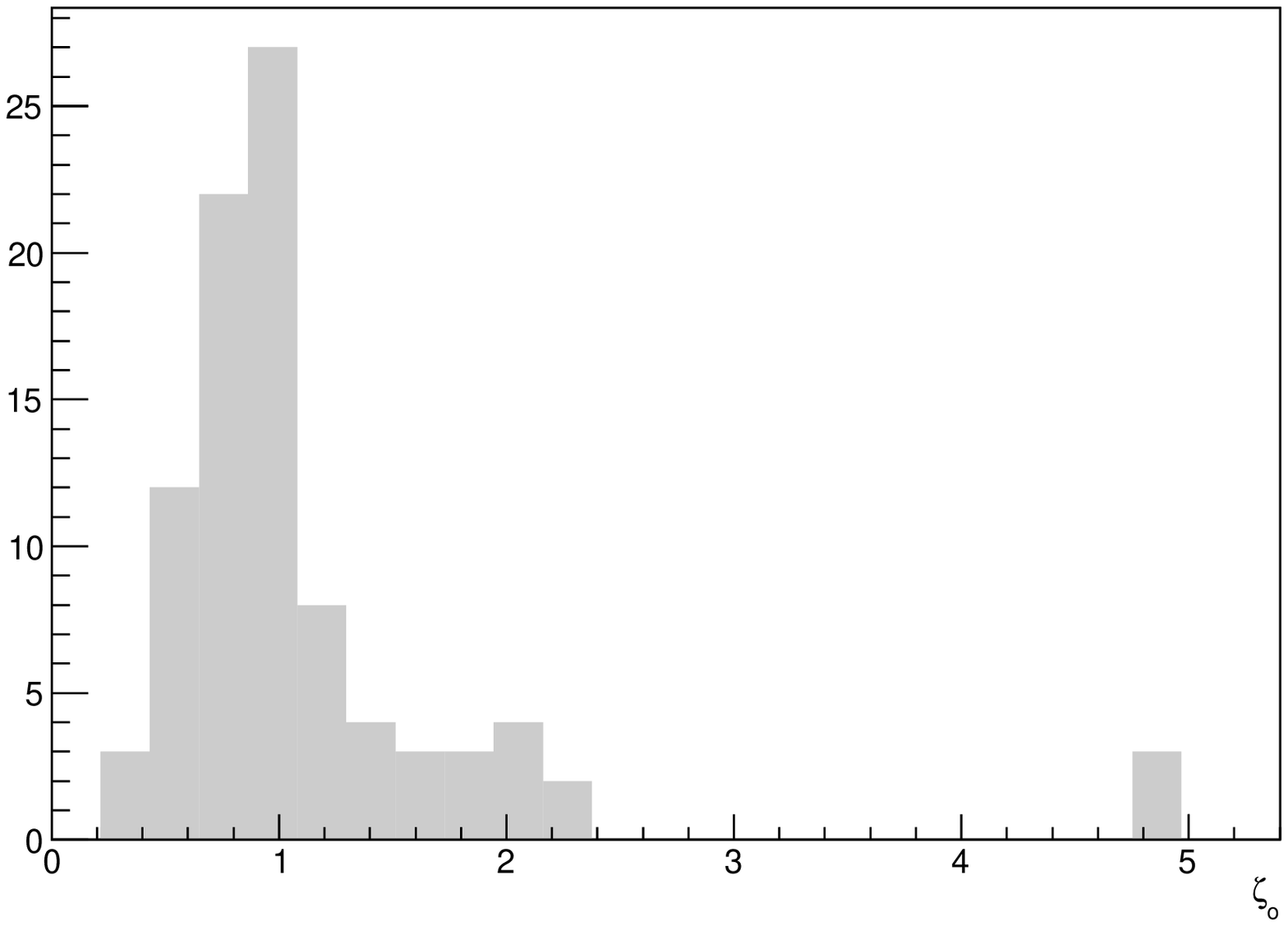}
\plotone{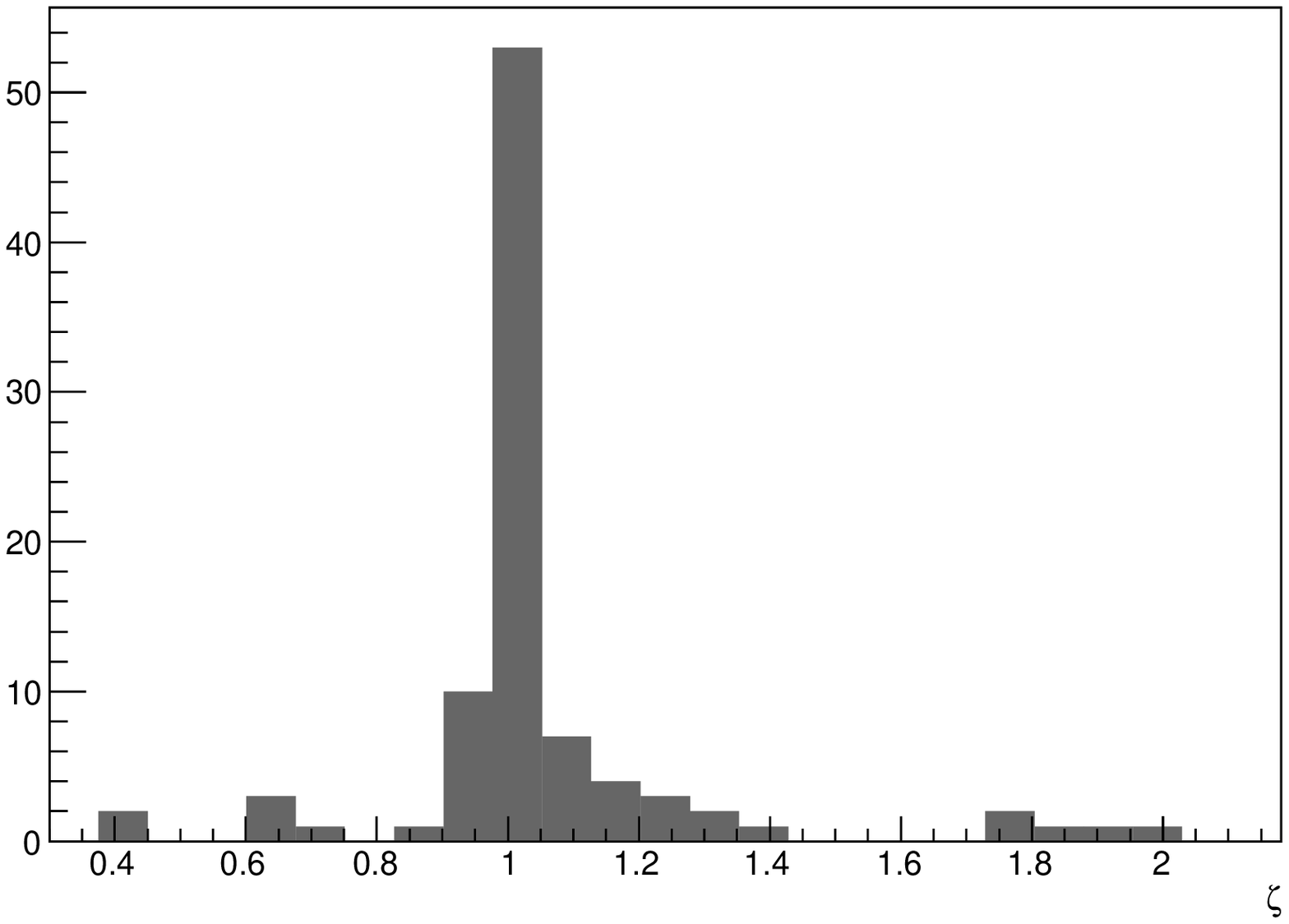}
\caption{$ \zeta$ distributions: initial (top) and after scaling (bottom).\label{fig:betaDistrib1}} 
\end{figure} 

\subsection{Results on $\alpha$ }
\label{alphResults}
Fig.~\ref{fig:alphaDistrib1} shows the values of $\alpha$ from the LCM fits as a function of the terminal value of $\kappa$.  The terminal value of $\kappa$, denoted as $\kappa_\tau$,  is the value of $\kappa$ at the largest radius available in the data for either the emitter or the receiver galaxy, in a given galaxy pair.   For the majority of emitter/receiver galaxy pairs in this analysis,  $\kappa$  approaches a constant value at large radii,  as the luminous matter has been  entirely enclosed. \\

The  $\alpha$ distribution has two interesting features.
The first is that the sign  of $\alpha$ appears to be related to the astrophysical red-shift parameter, $z$ (Eq.~(\ref{eq:zParam})). 
For values of $\kappa_\tau < 1$, $\alpha$ is negative (with the exception of 
 NGC 925), and vice versa for $\kappa_\tau > 1$.  Physically, emitter galaxies for which $\kappa_\tau < 1$ are less massive than the MW; 
 they sit in a shallower gravitational potential well.  Therefore, light coming from those galaxies will be blue shifted when observed 
 in the MW.  Similarly, emitter galaxies for which $\kappa_\tau > 1$ sit in a deeper gravitation potential well than the MW, and therefore light  emitted there will be red shifted when observed in the MW.      The exception to this pattern is NGC 925, for which   $\kappa$ has not yet approached a constant value at the largest extent of the data set.\\
 
 The second feature of interest for the $\alpha$ distribution is 
 an apparent functional relationship between $\alpha$ and the $\kappa_\tau$.   As is shown in Fig.~\ref{fig:alphaDistrib1}, 
 the apparent inflection point in the distribution for the initial $\alpha_o$ values falls slightly to the left of $\kappa_\tau = 1$, but after the 
 scaling iteration, 
 the inflection point of the distribution moves to 
$\kappa_\tau = 1$.    The fact that there appears to be a functional relationship between $\alpha$ and $\kappa_\tau$ is important because $\alpha$ parametrizes the hypothesized relationship between the dark and luminous matter, Eq.(\ref{eq:name}).  
 \begin{figure*}
 \epsscale{0.9}
\plotone{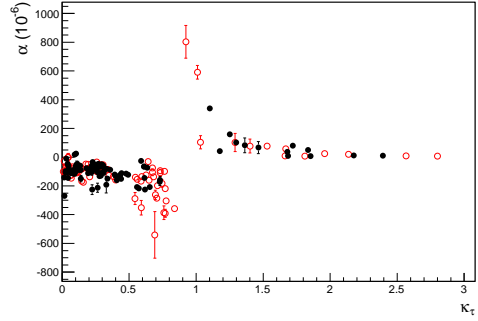} 
\caption{  The LCM  $\alpha$ fit results  vs. $\kappa_{\tau}$.  Each dot represents one 
 emitter/receiver galaxy pair) Red points represent initial $\alpha_o$ values, and black points represent  $\alpha$ values after the scaling iteration.  The error bars are  only   statistical uncertainties.\label{fig:alphaDistrib1}} 
\end{figure*} 

\begin{figure*}
\centering
\subfloat[][NGC3198~Ref.(2), $\kappa_{\tau}=.10$]{
	\includegraphics[width=0.3\textwidth]{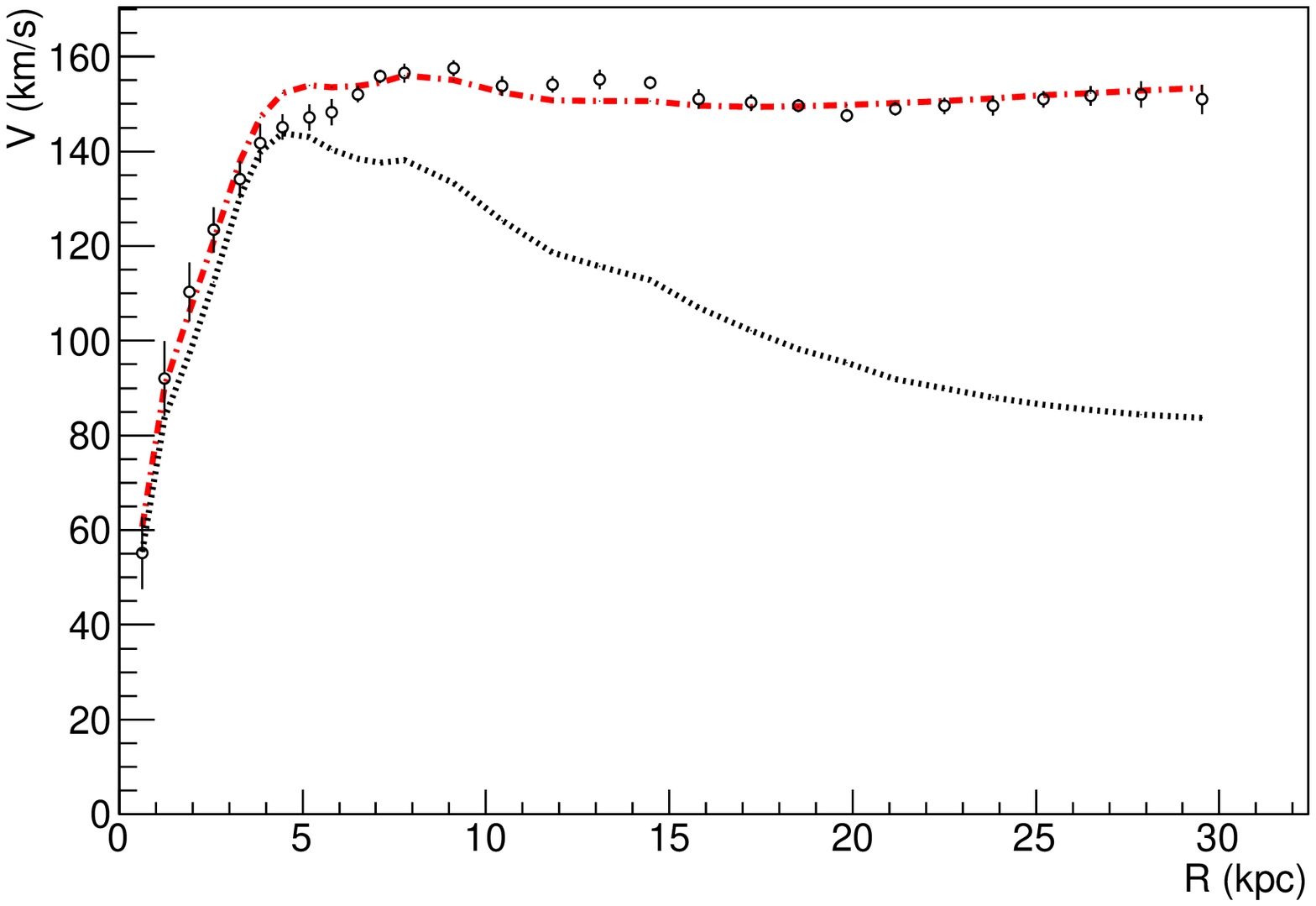}
}
\subfloat[][NGC3198~Ref.(9), $\kappa_{\tau}=.11$]{
	\includegraphics[width=0.3\textwidth]{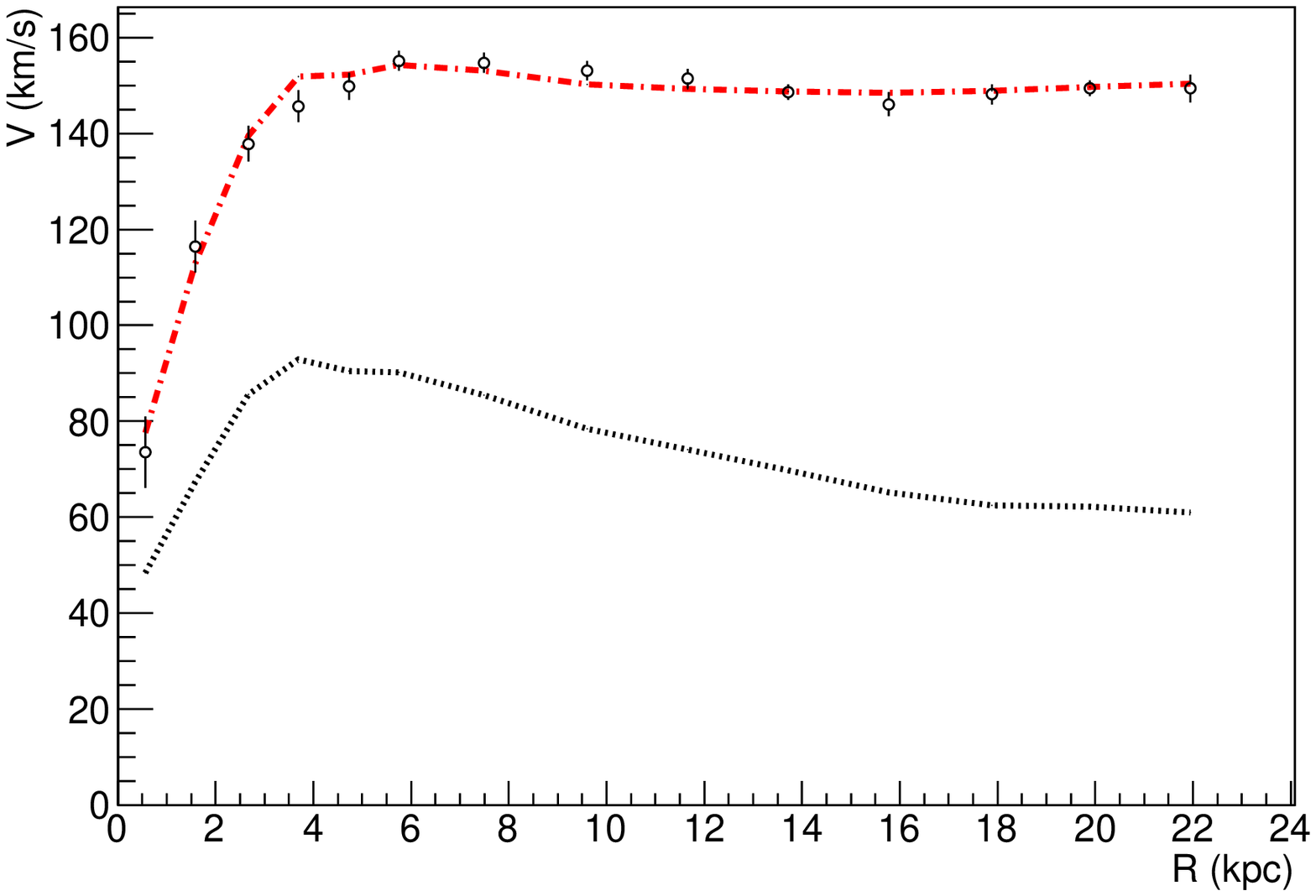}
}
\subfloat[][NGC3198~Ref.(6), $\kappa_{\tau}=0.05$]{
	\includegraphics[width=0.3\textwidth]{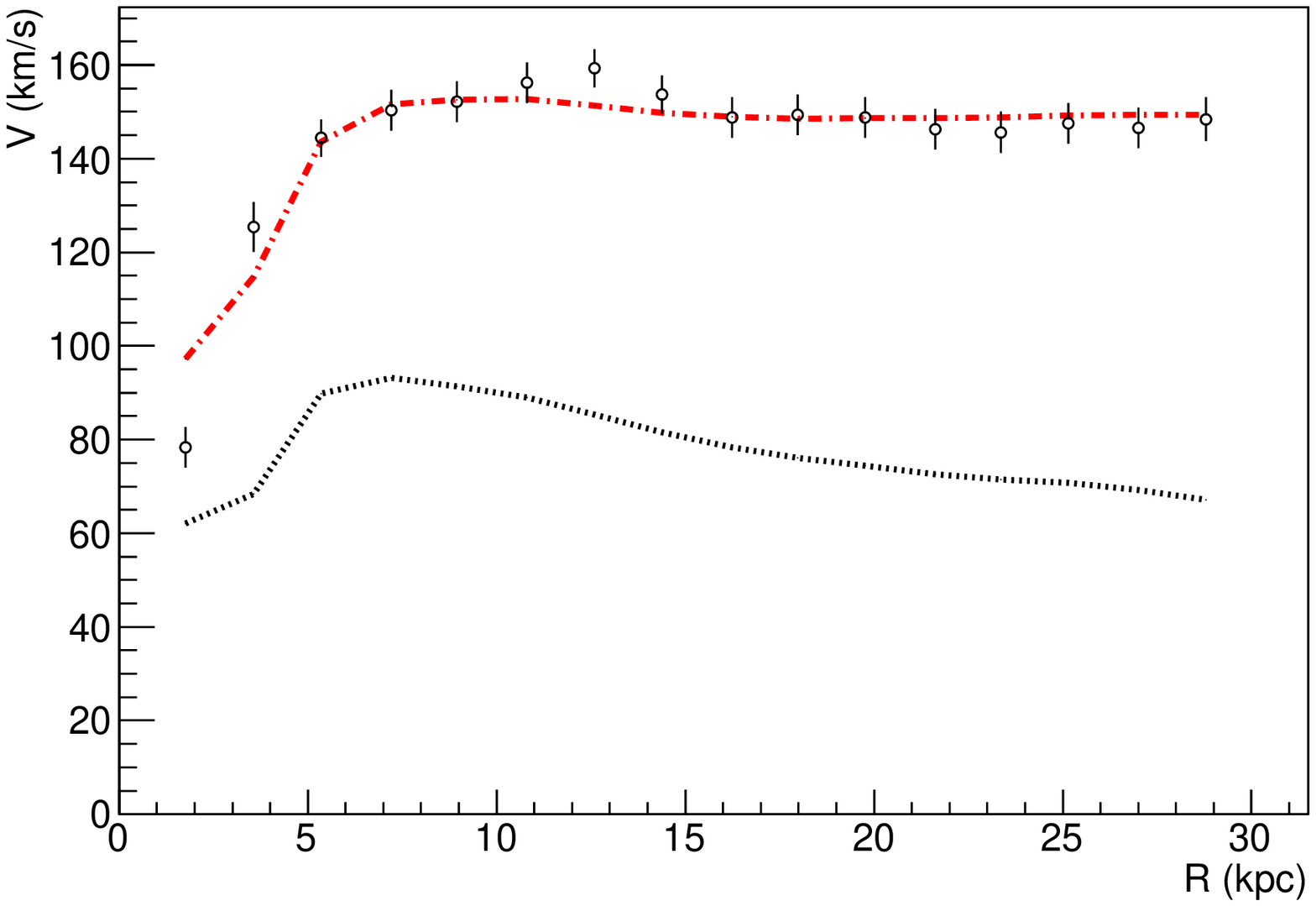}
}
\vspace{0.5cm}
\subfloat[][M33~Ref.(3), $\kappa_{\tau}=0.03$]{
	\includegraphics[width=0.3\textwidth]{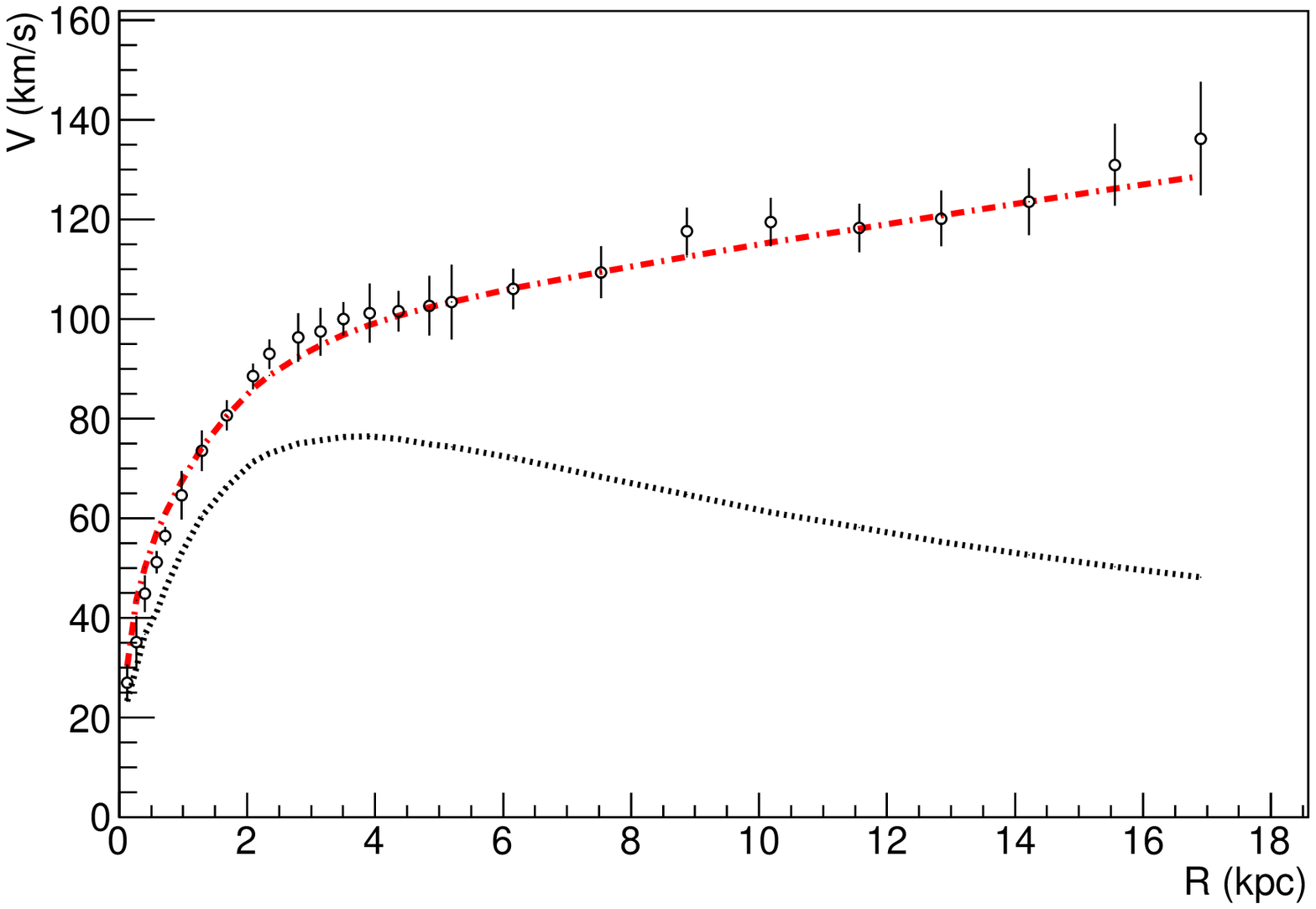}
}
\subfloat[][M33~Ref.(8), $\kappa_{\tau}=0.03$]{
	\includegraphics[width=0.3\textwidth]{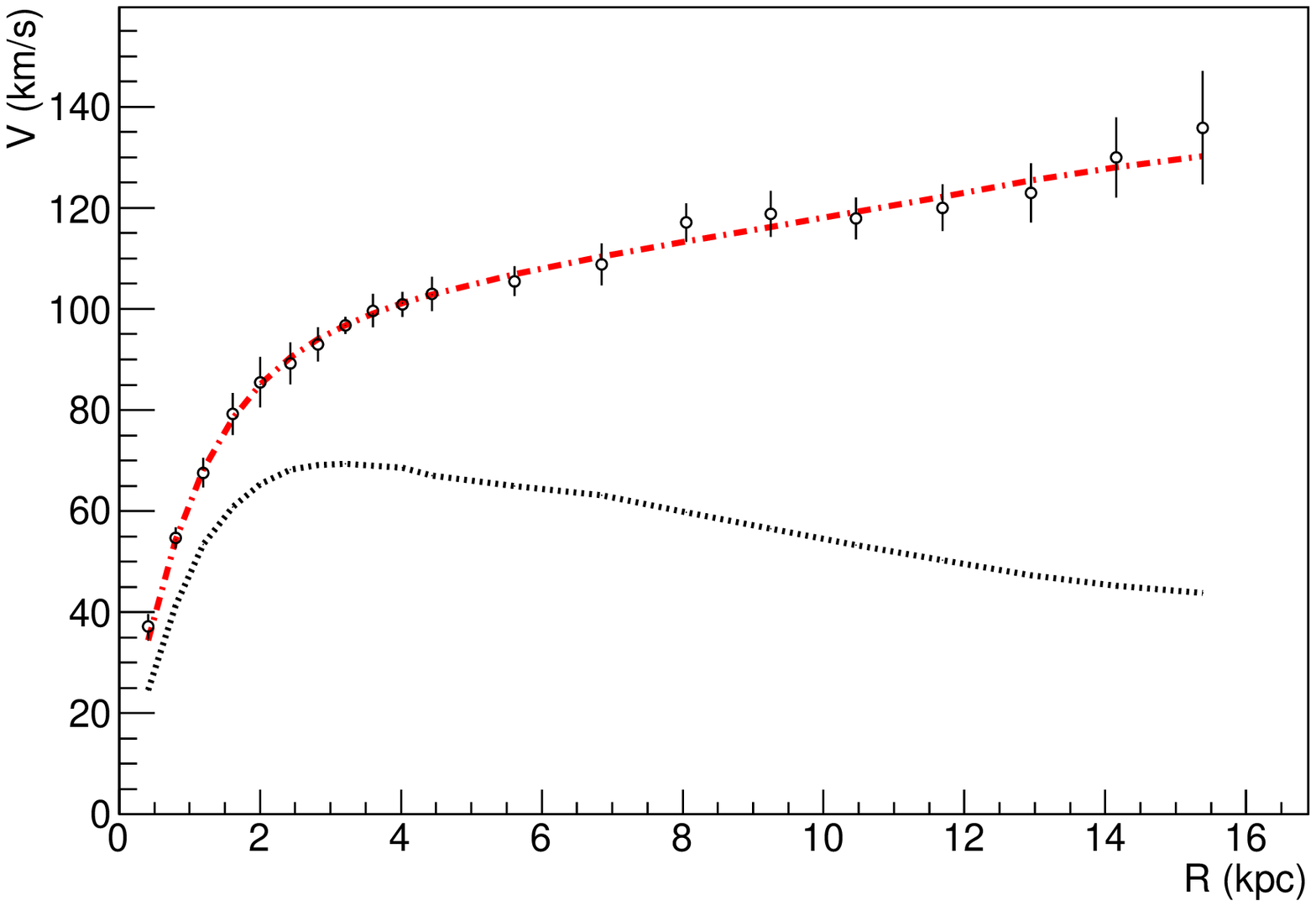}
}
\subfloat[][NGC3521~Ref.(5), $\kappa_{\tau}=0.31$]{
	\includegraphics[width=0.3\textwidth]{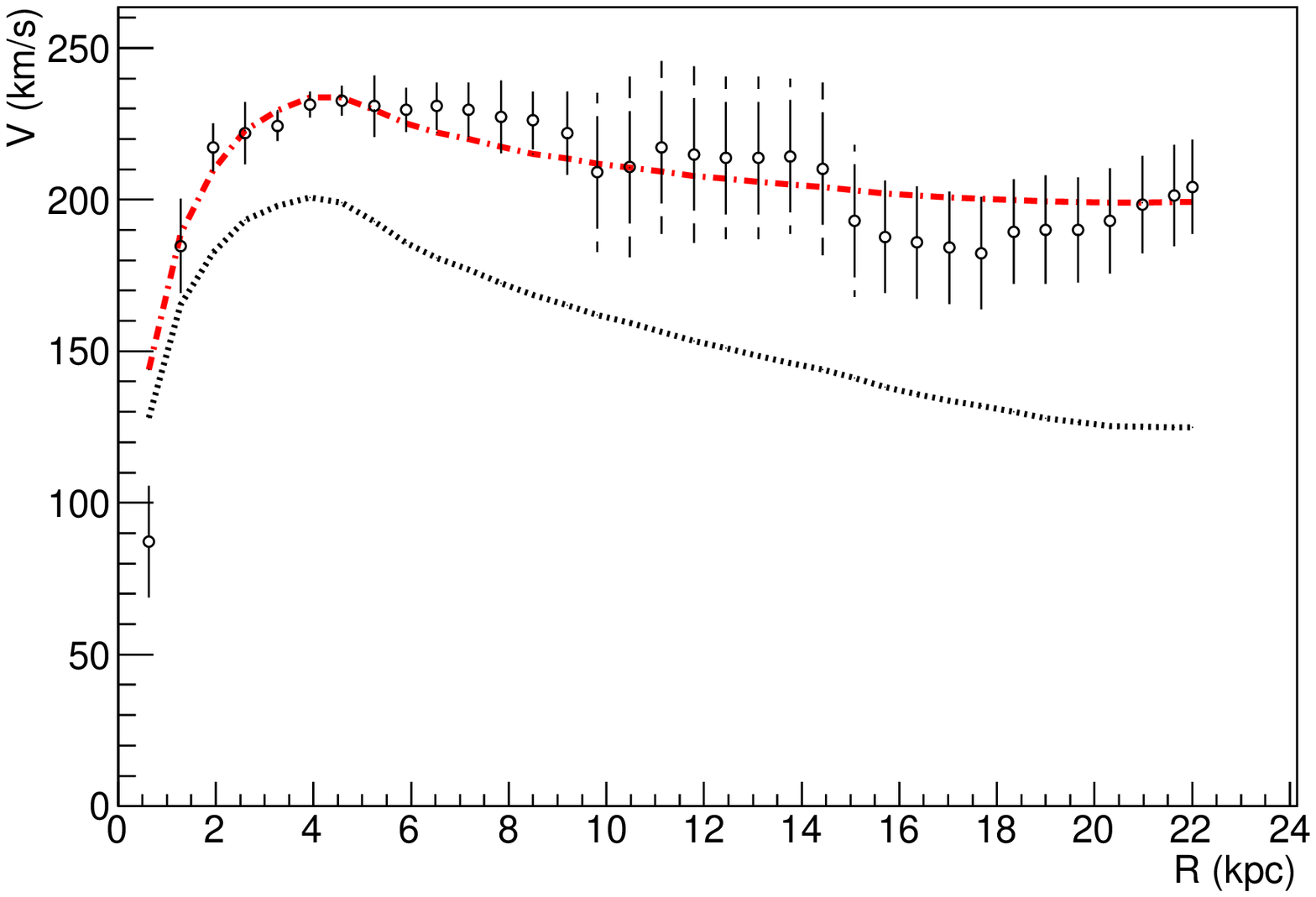}
}
\vspace{0.5cm}
\subfloat[][NGC5055~Ref.(1), $\kappa_{\tau}=0.25$]{
	\includegraphics[width=0.3\textwidth]{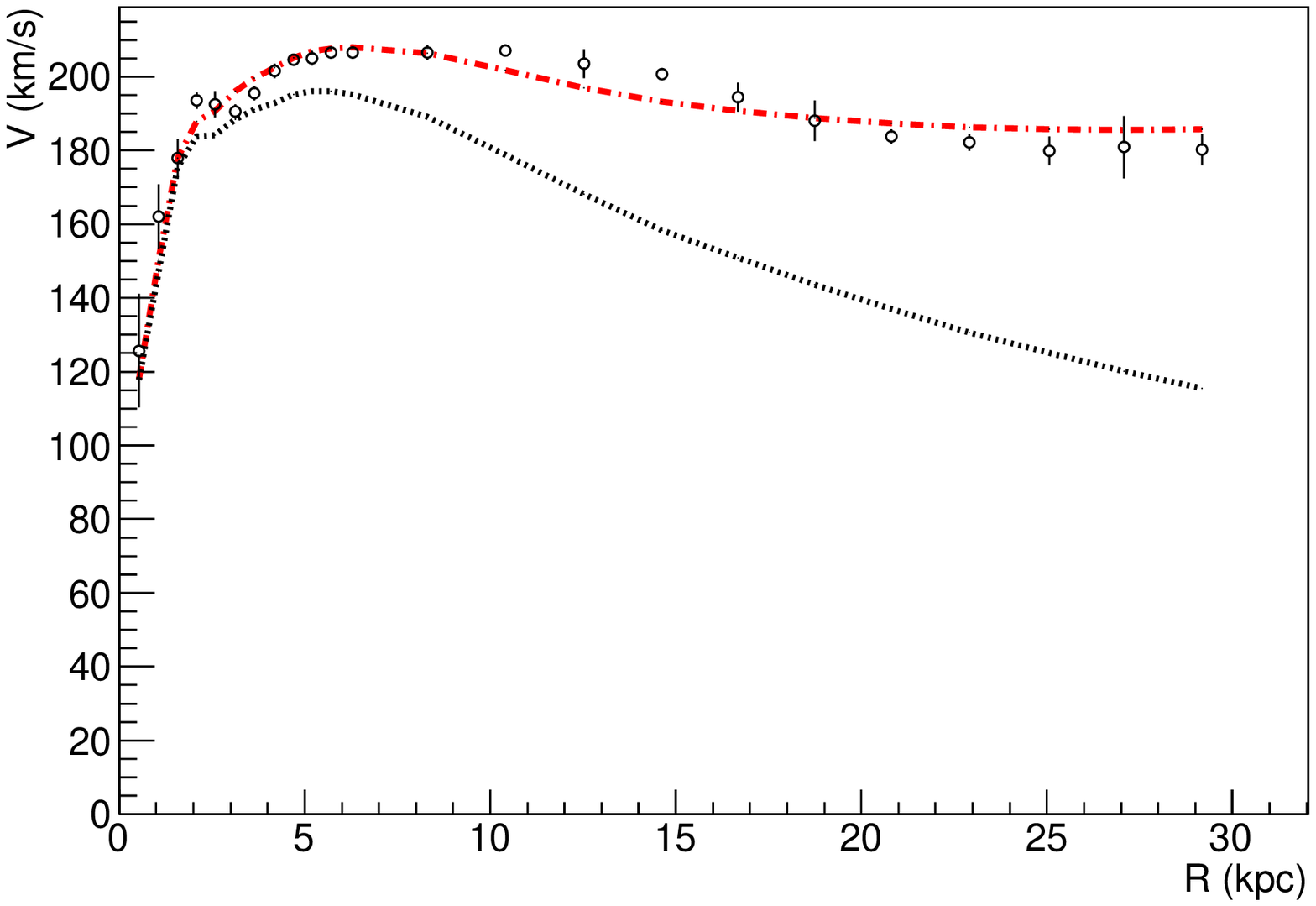}
}
\subfloat[][NGC5055~Ref.(5), $\kappa_{\tau}=0.30$]{
	\includegraphics[width=0.3\textwidth]{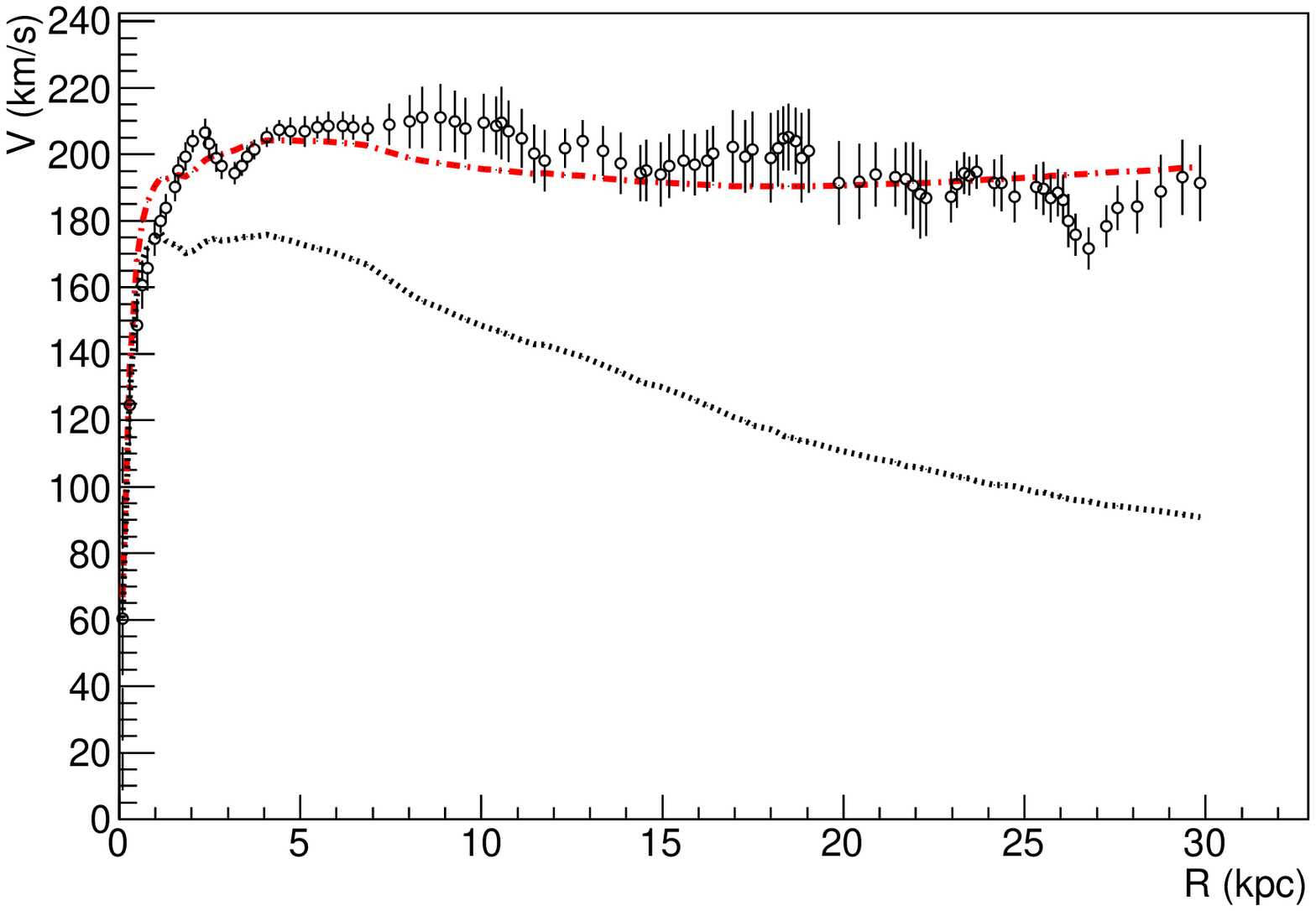}
}
\subfloat[][NGC2841~Ref.(5), $\kappa_{\tau}=0.41$]{
	\includegraphics[width=0.3\textwidth]{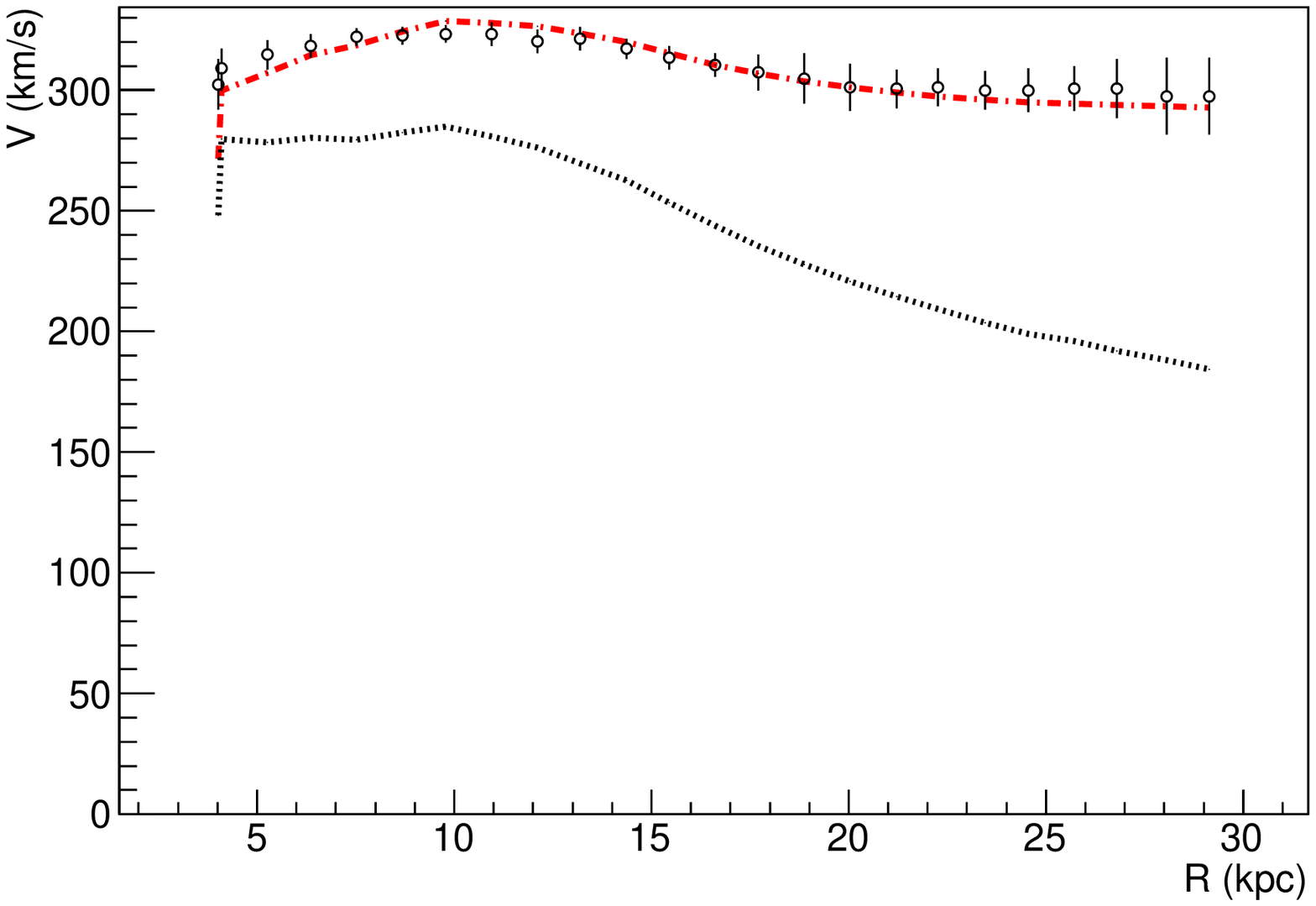}
}
\vspace{0.5cm}
\subfloat[][NGC2403~Ref.(2), $\kappa_{\tau}=0.05$]{
	\includegraphics[width=0.3\textwidth]{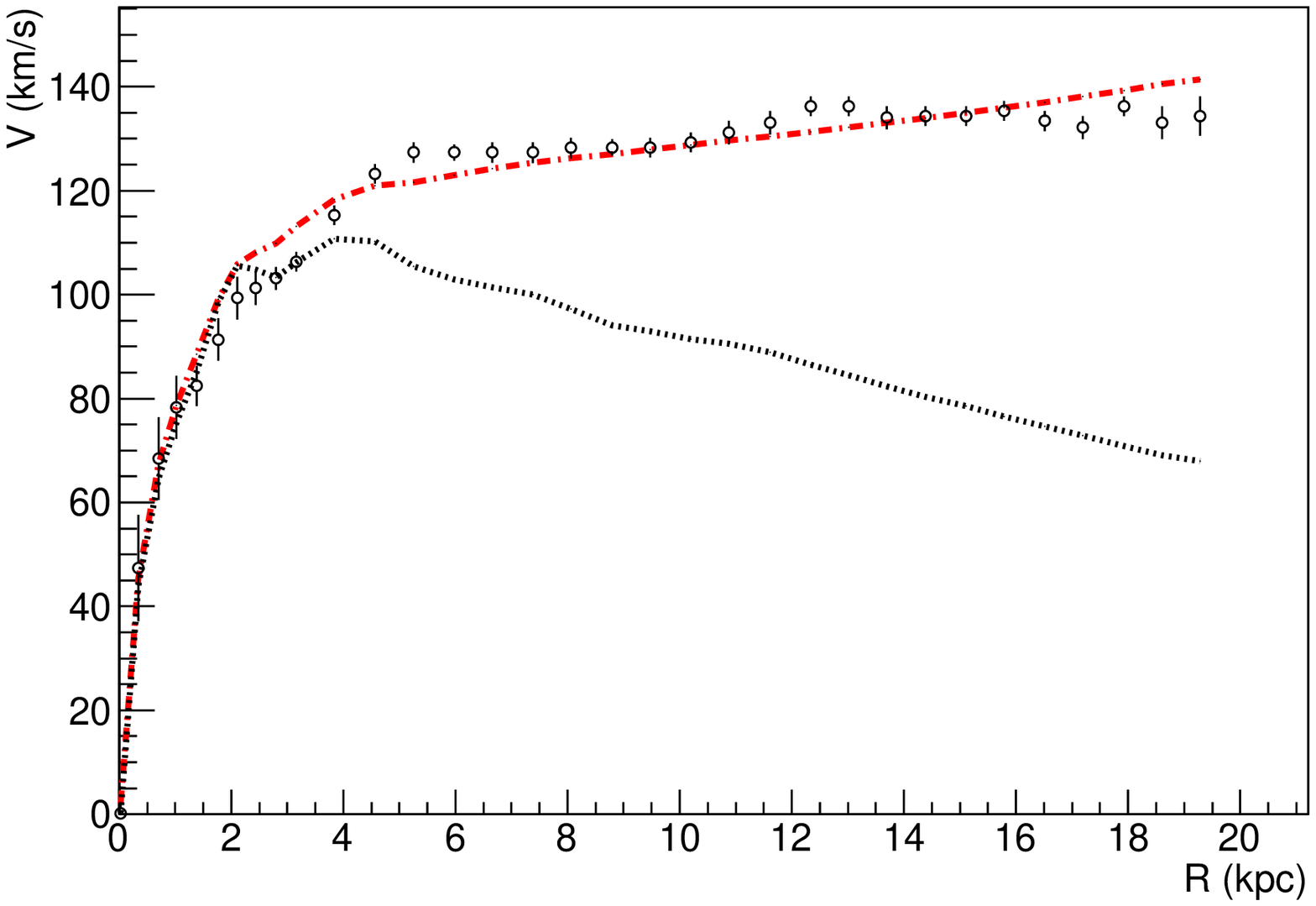}
}
\subfloat[][NGC2403~Ref.(4), $\kappa_{\tau}= 0.03$]{
	\includegraphics[width=0.3\textwidth]{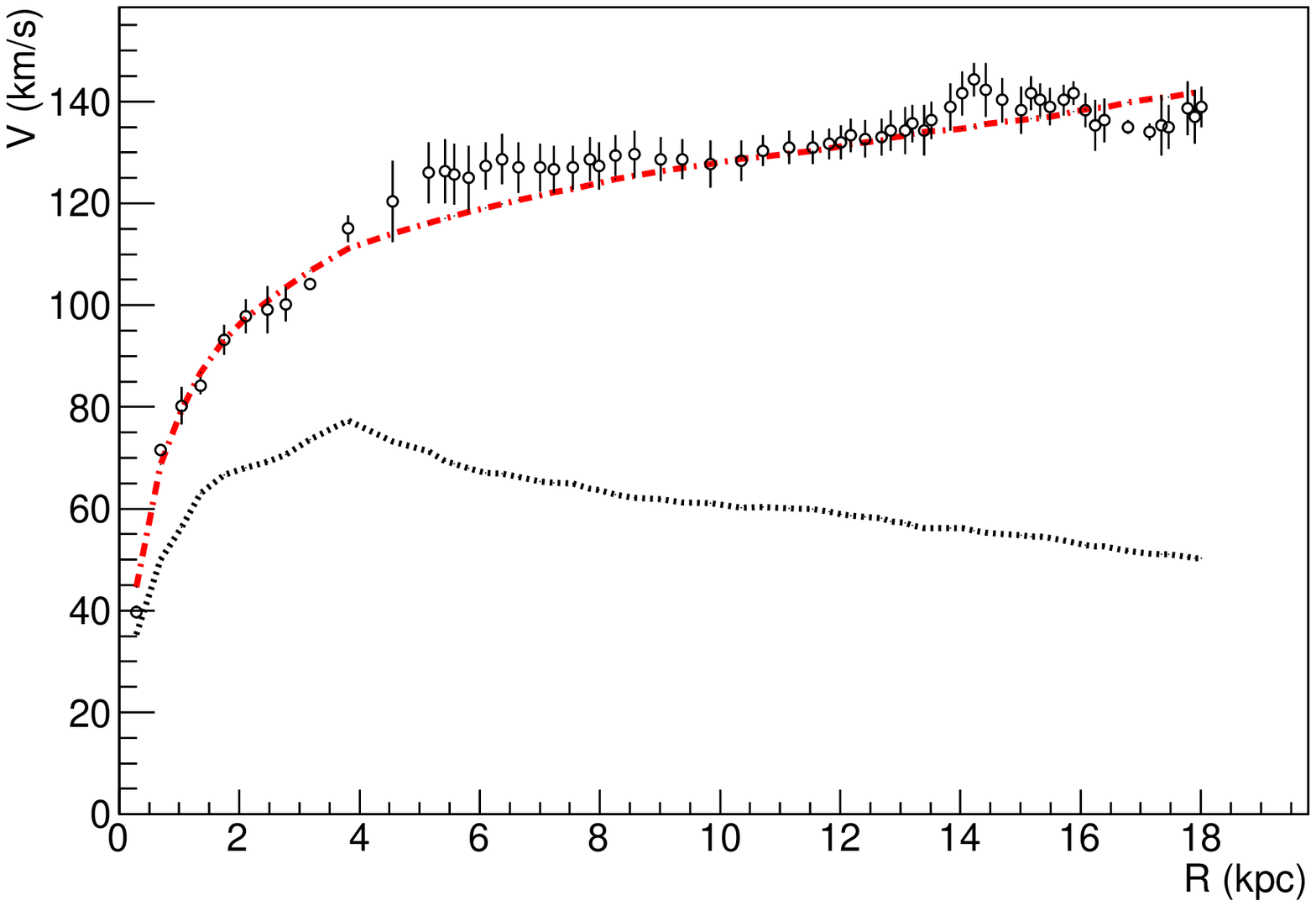}
}
\subfloat[][NGC7814~Ref.(11), $\kappa_{\tau}=0.61$]{
	\includegraphics[width=0.3\textwidth]{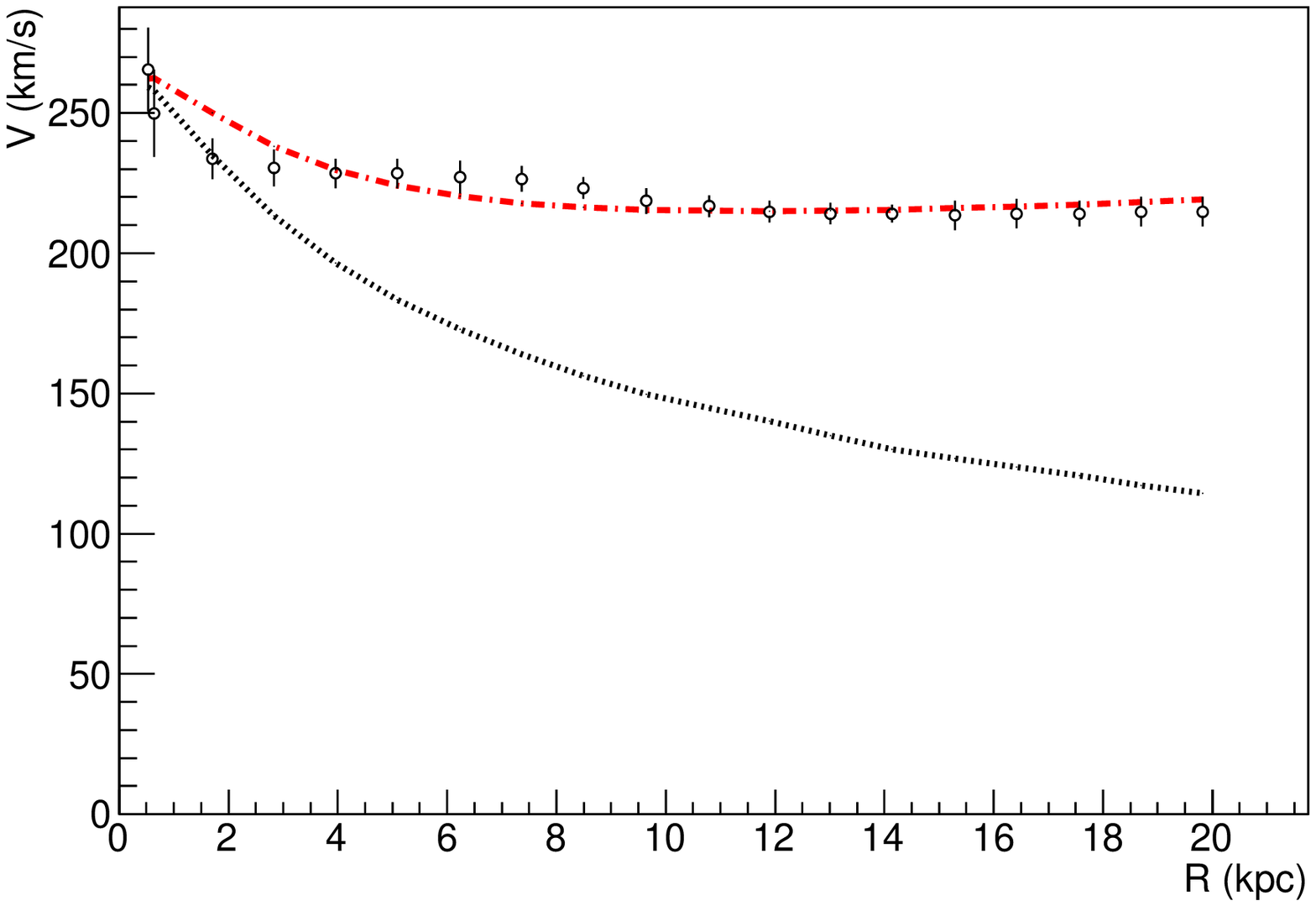}
}
\caption{LCM fits to rotation curves of spiral galaxies with published data. Emitter-galaxies are paired to MW~\citet{Sof81}.  In all panels the black circles represent the observed  rotation velocities  and the thin bars represent the reported uncertainties.  The dotted curve shows the  Newtonian rotation curve of the luminous  contributions.  The LCM best-fit is shown as a  red  dotted-dashed  line. References  are as in Table~\ref{tab:referenceDATAs}.\label{fig:resultsA}}
\end{figure*} 
 
\begin{figure*}
\centering
\subfloat[][NGC7331~Ref.(2), $\kappa_{\tau}=1.25$]{
\includegraphics[width=0.3\textwidth]{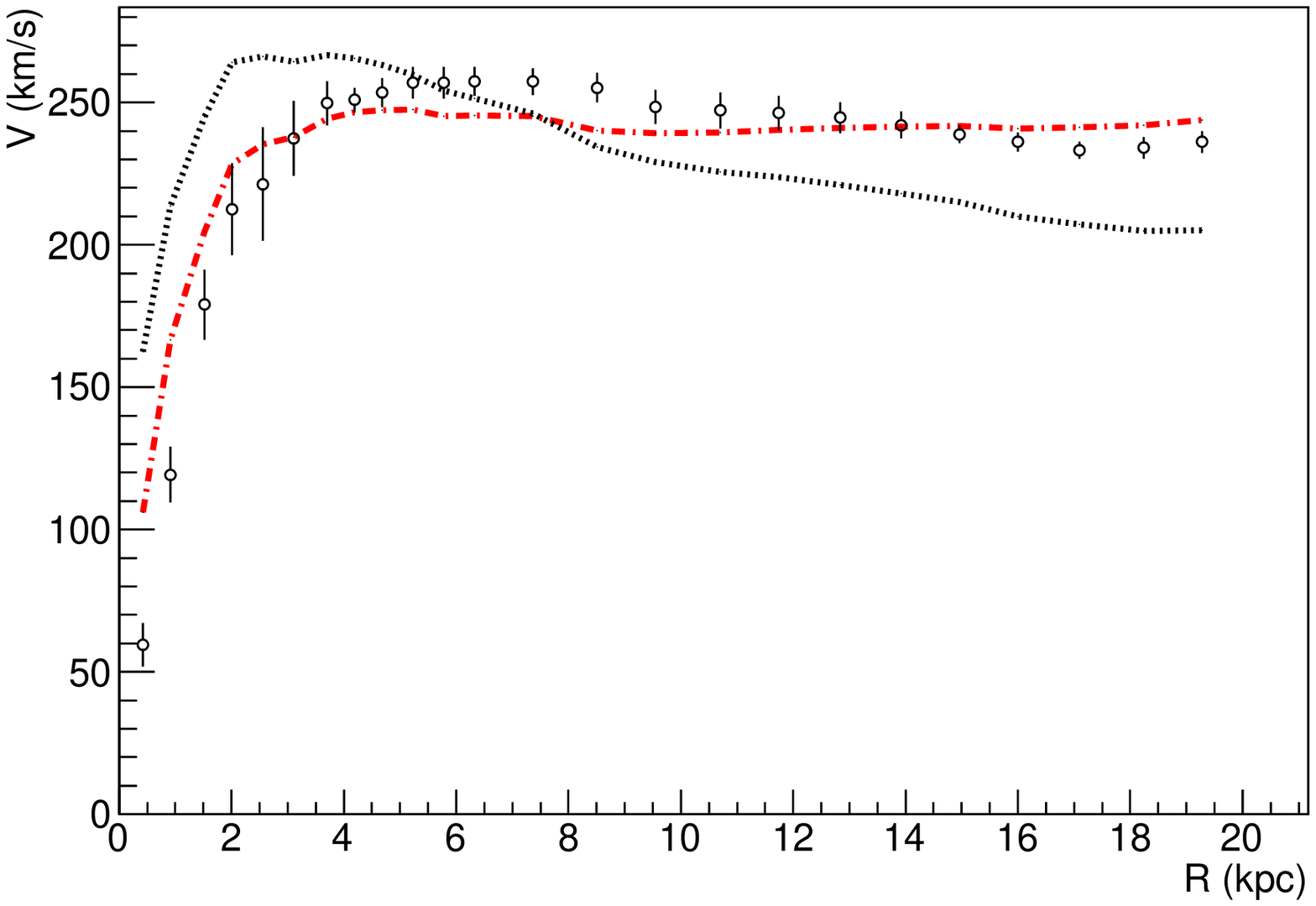}
}
\subfloat[][NGC7331~Ref.(5), $\kappa_{\tau}=0.57$]{
\includegraphics[width=0.3\textwidth]{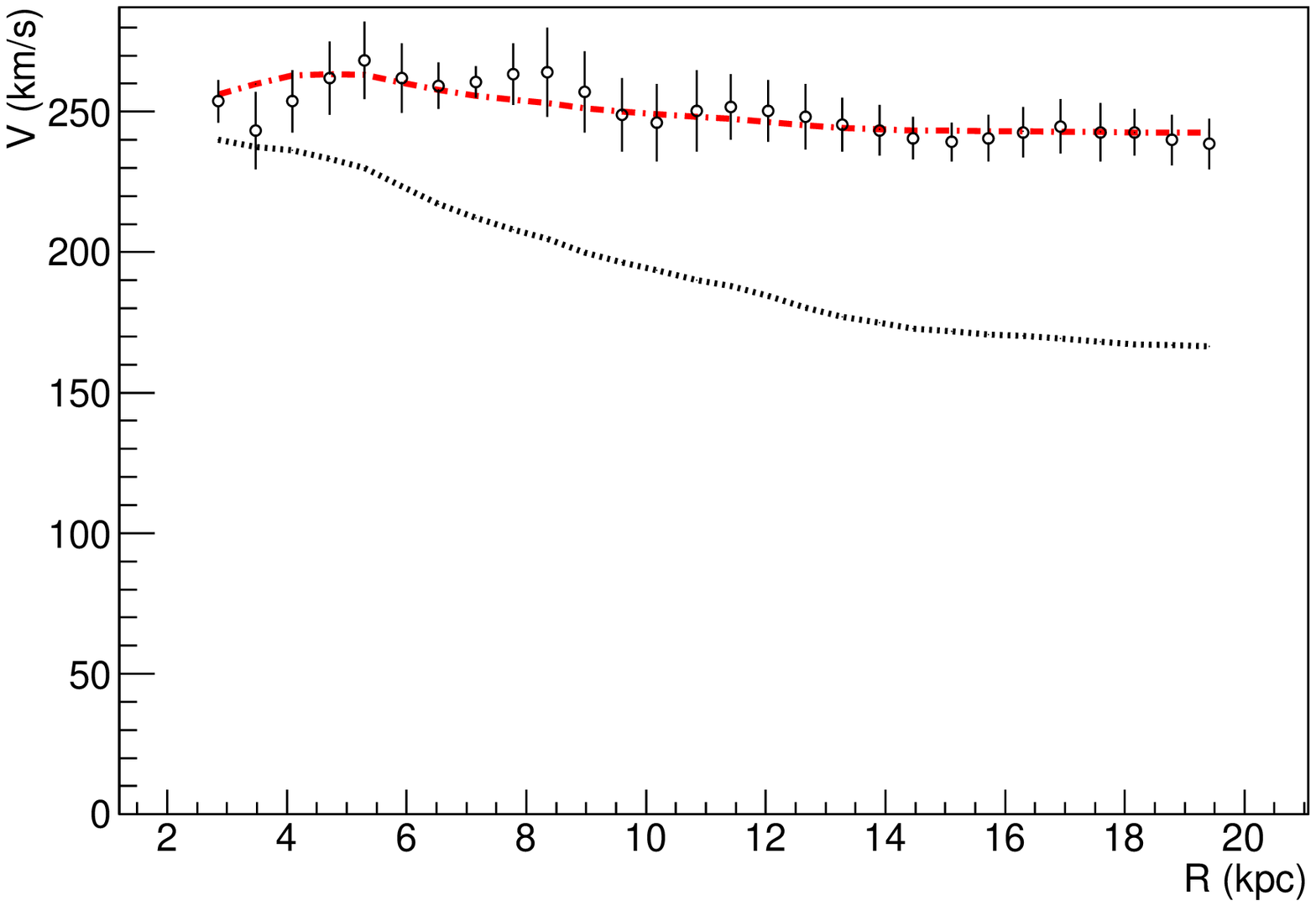}
}
\subfloat[][NGC891 ~Ref.(11), $\kappa_{\tau}=1.85$]{
\includegraphics[width=0.3\textwidth]{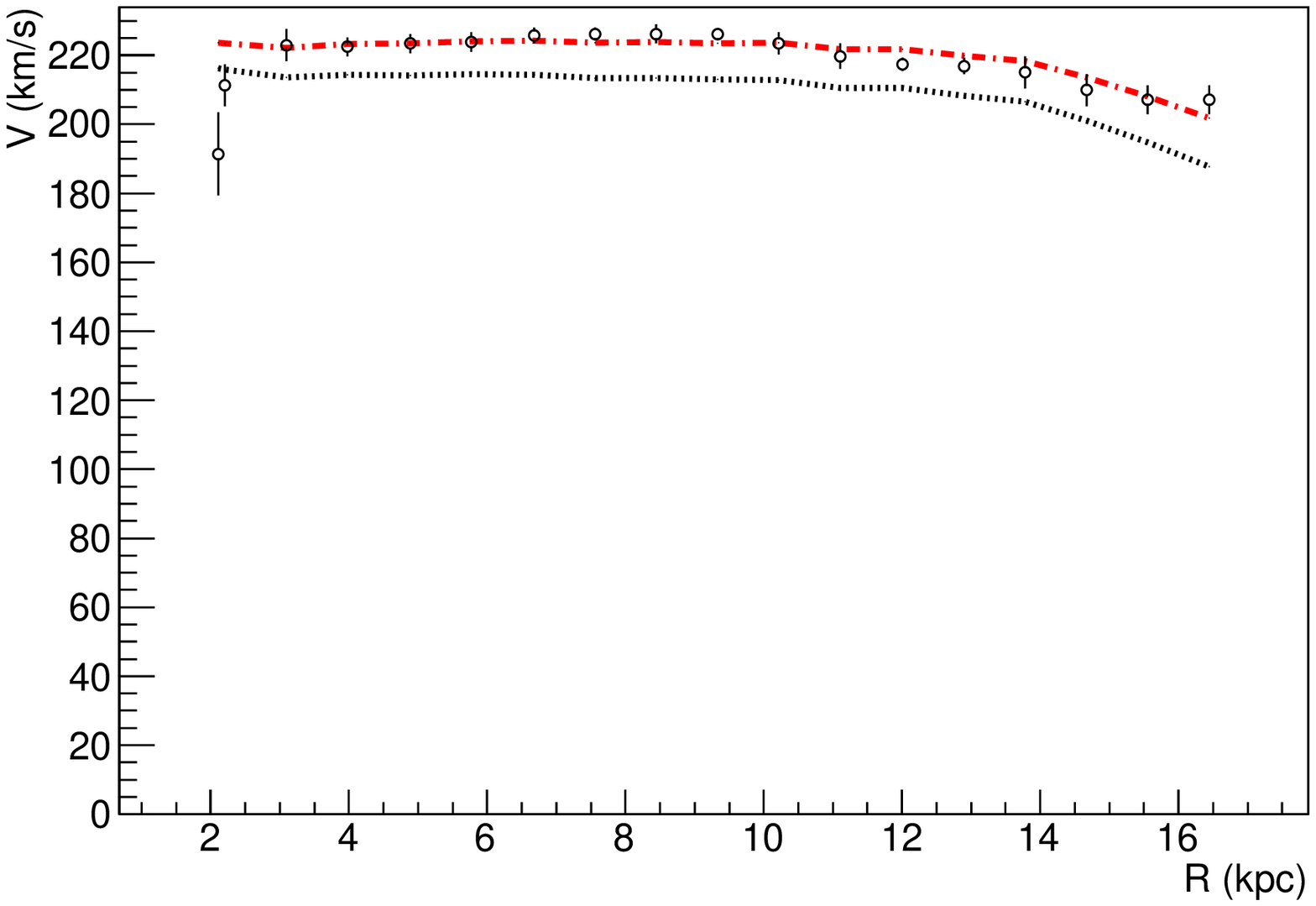}
}
\vspace{0.5cm}
\subfloat[][M31 ~Ref.(10), $\kappa_{\tau}=2.39$]{
\includegraphics[width=0.3\textwidth]{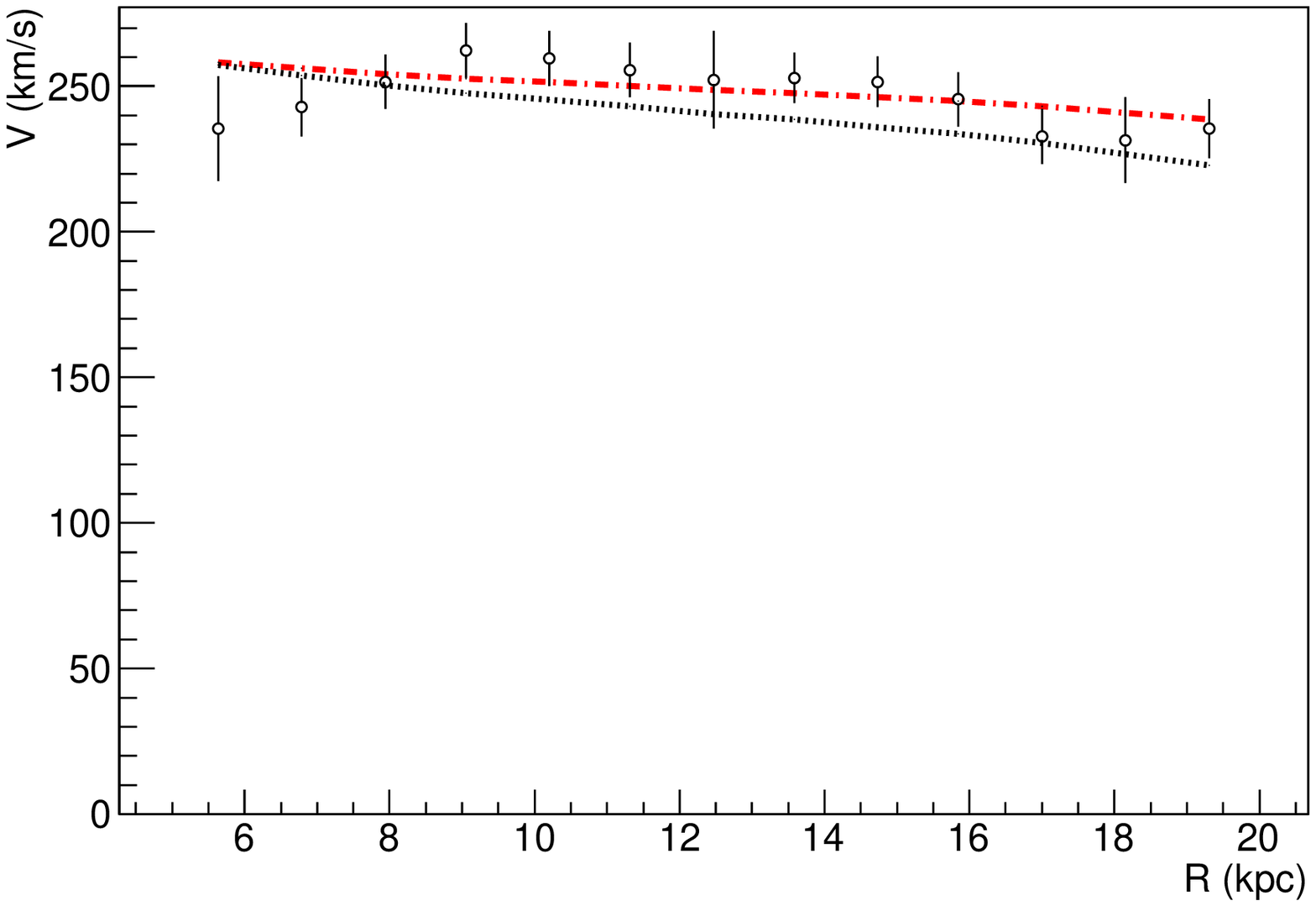}
}
\subfloat[][NGC5533~Ref.(7), $\kappa_{\tau}=1.46$]{
\includegraphics[width=0.3\textwidth]{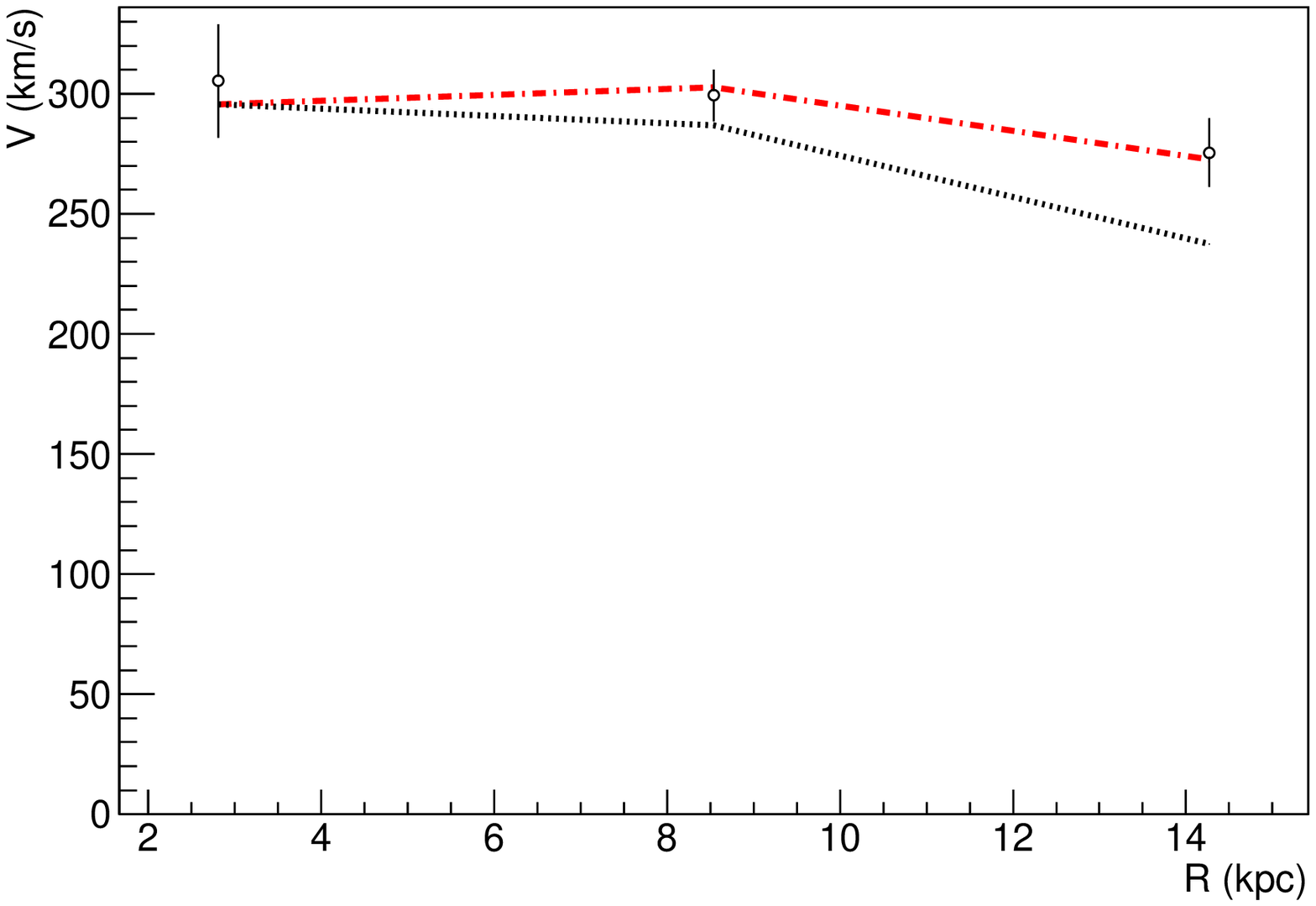}
}
\subfloat[][UGC6973 ~Ref.(7), $\kappa_{\tau}=0.28$]{
\includegraphics[width=0.3\textwidth]{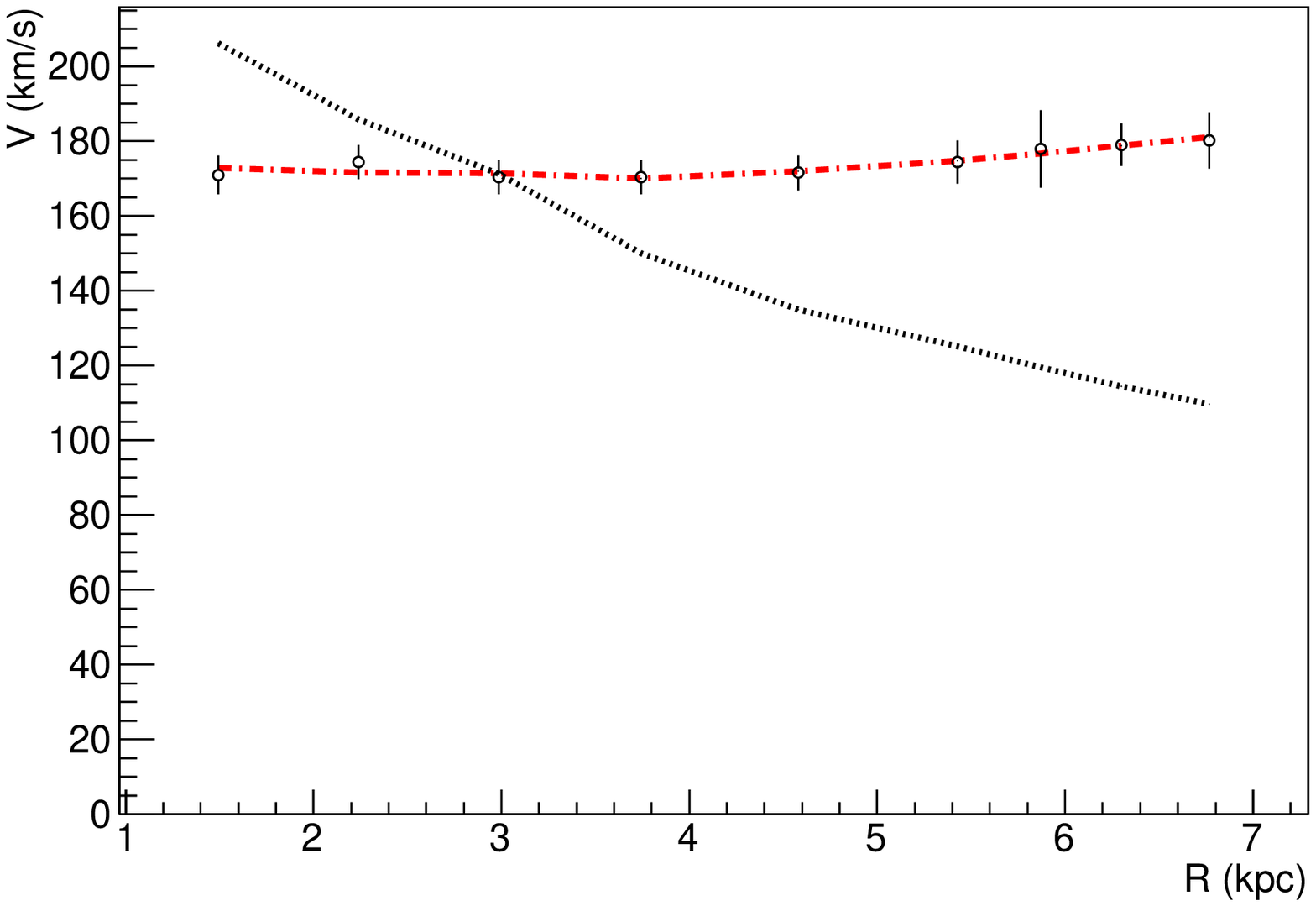}
}
\vspace{0.5cm}
\subfloat[][NGC4088~Ref.(13), $\kappa_{\tau}=0.36$]{
\includegraphics[width=0.3\textwidth]{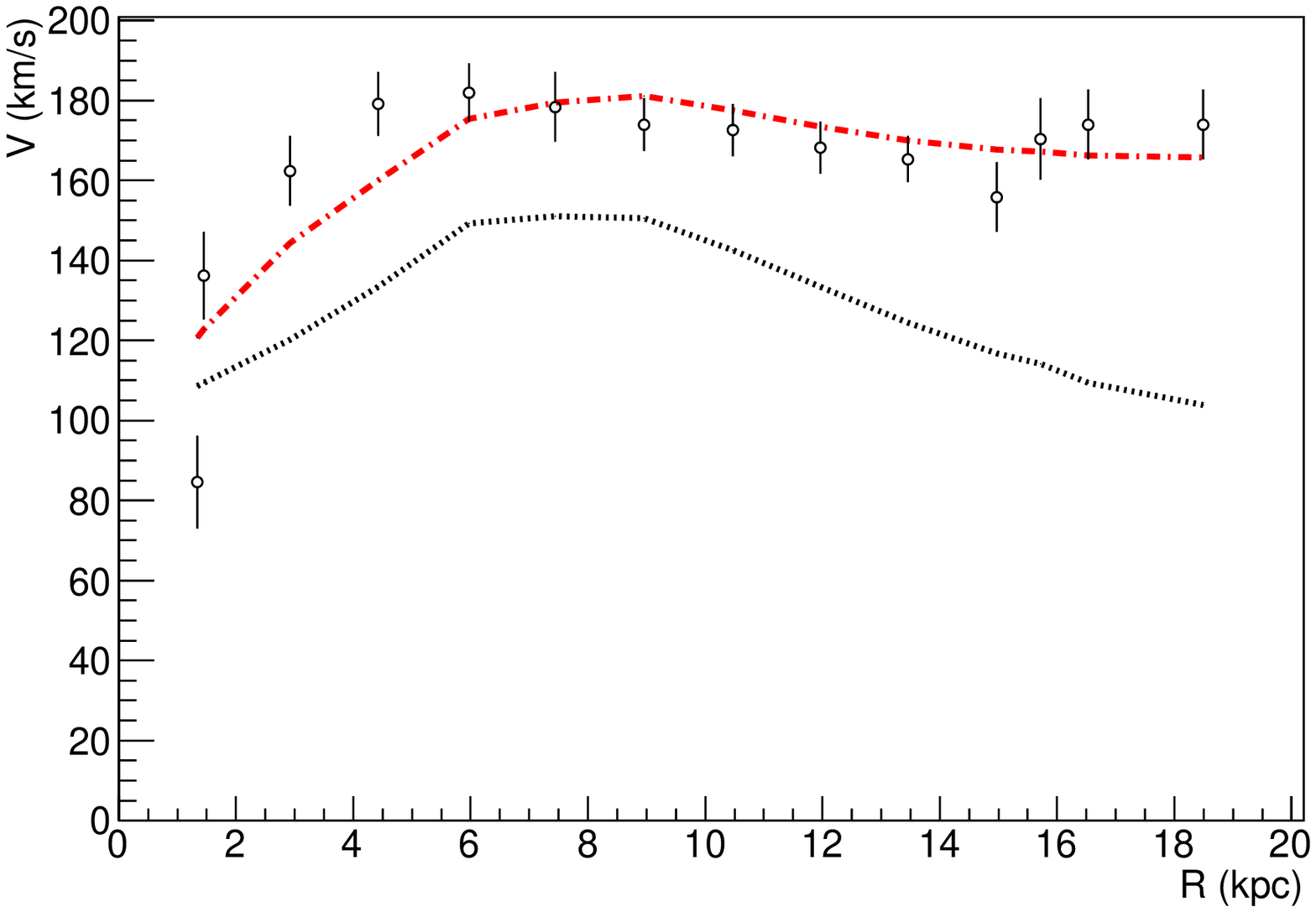}
}
\subfloat[][NGC3992~Ref.(7), $\kappa_{\tau}=0.33$]{
\includegraphics[width=0.3\textwidth]{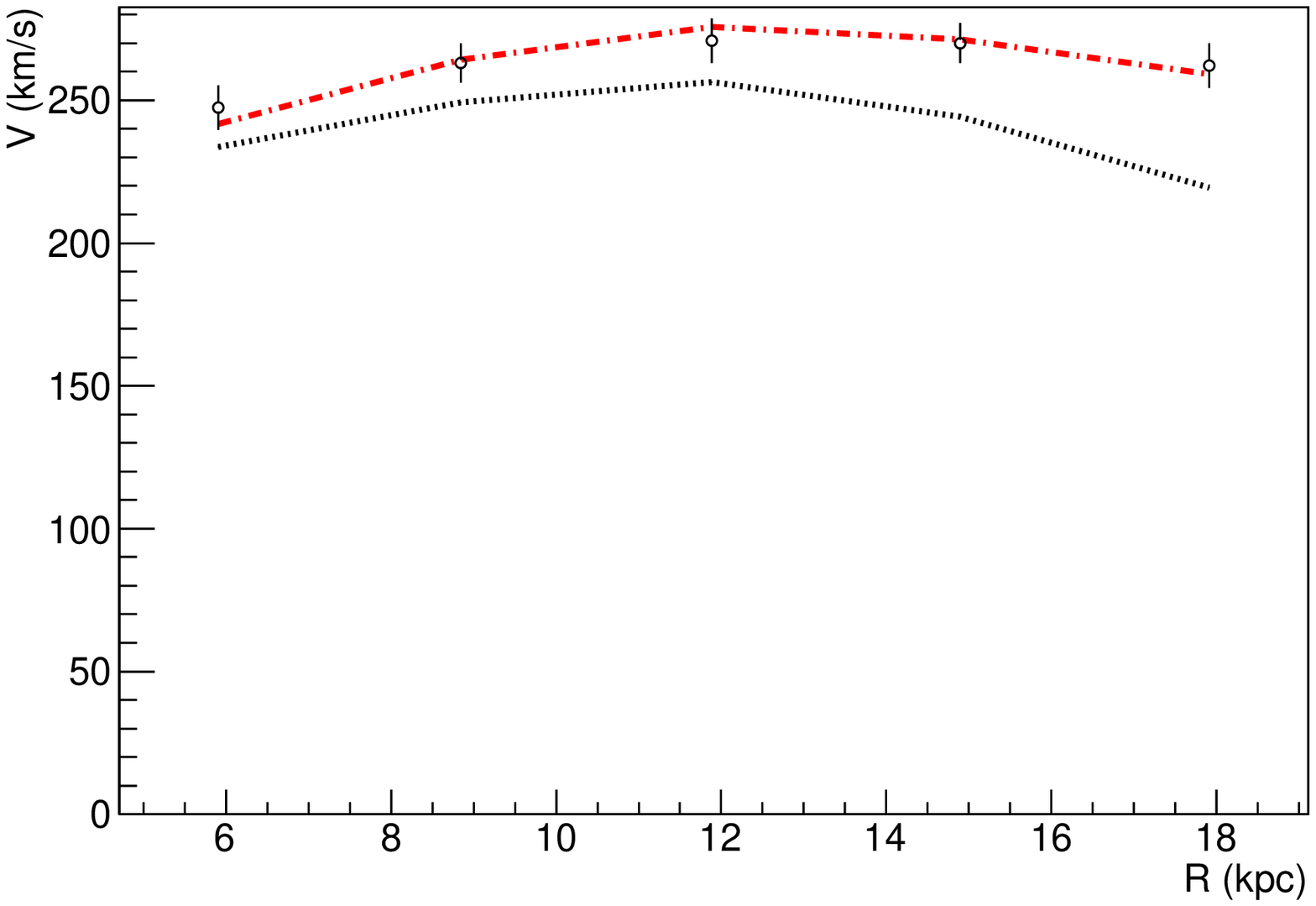}
}
\subfloat[][NGC4138~Ref.(13), $\kappa_{\tau}=0.64$]{
\includegraphics[width=0.3\textwidth]{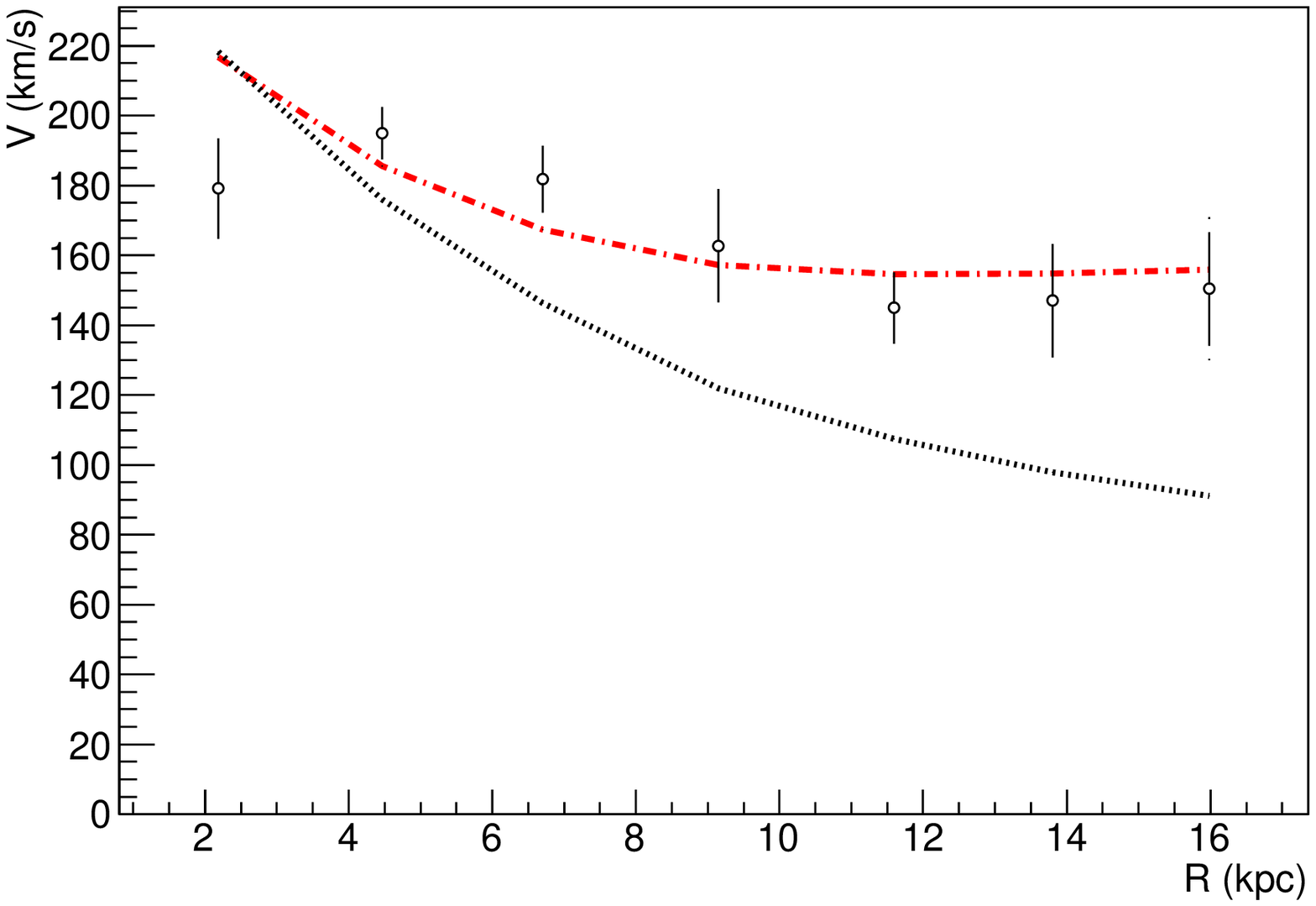}
}
\caption{LCM fits to  rotation curves of spiral galaxies with published data.  Emitter-galaxies paired  to MW \citet[Model A]{Klypin}. Symbols and curves are as in Figure~\ref{fig:resultsA}, and references are as in Table.\ref{tab:referenceDATAs}.\label{fig:resultsB}}
\end{figure*}

\begin{figure*}
\centering
\subfloat[][NGC6946~Ref.(7), $\kappa_{\tau}=0.21$]{
\includegraphics[width=0.3\textwidth]{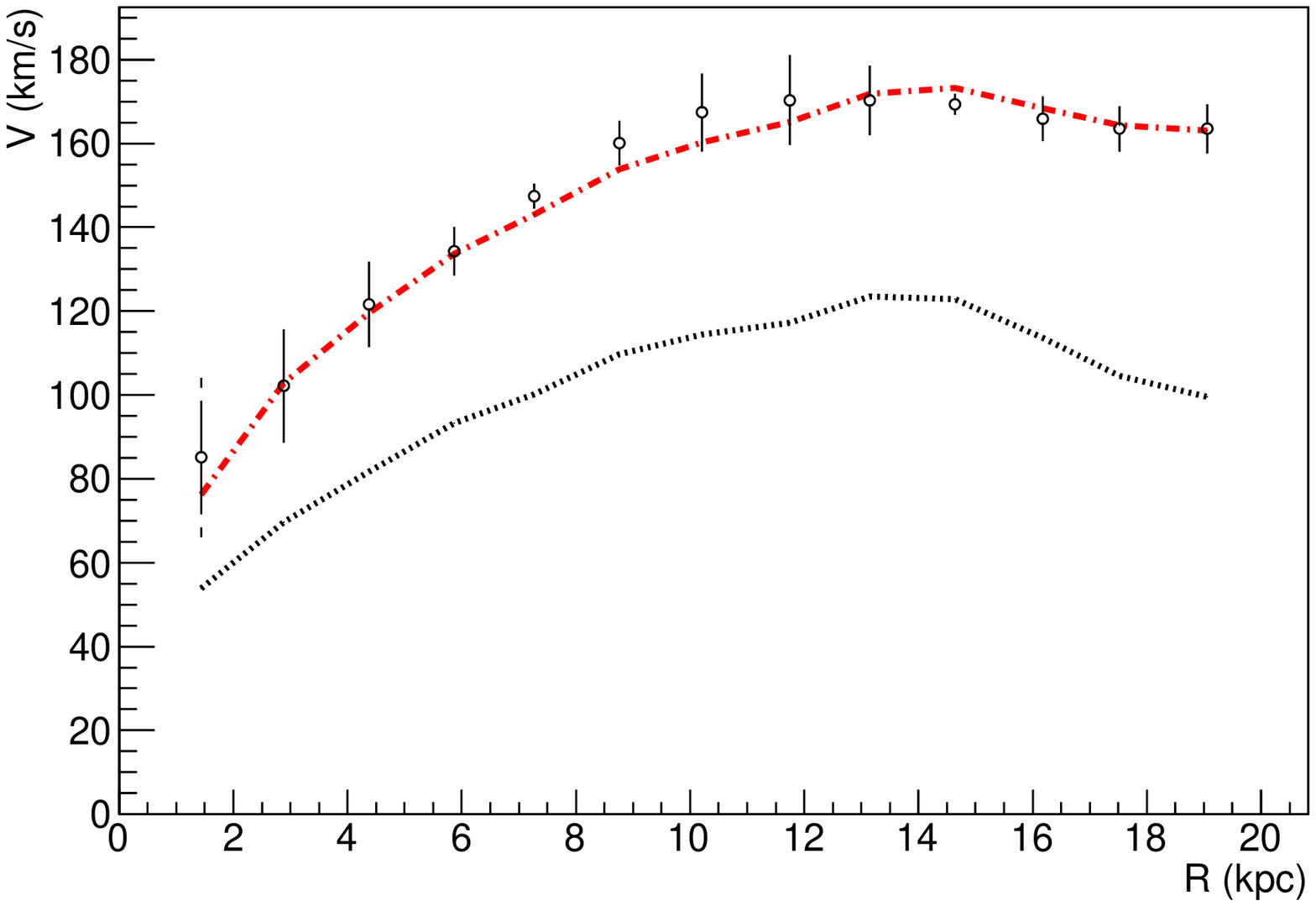}
}
\subfloat[][NGC6946~Ref.(4), $\kappa_{\tau}=0.56$]{
\includegraphics[width=0.3\textwidth]{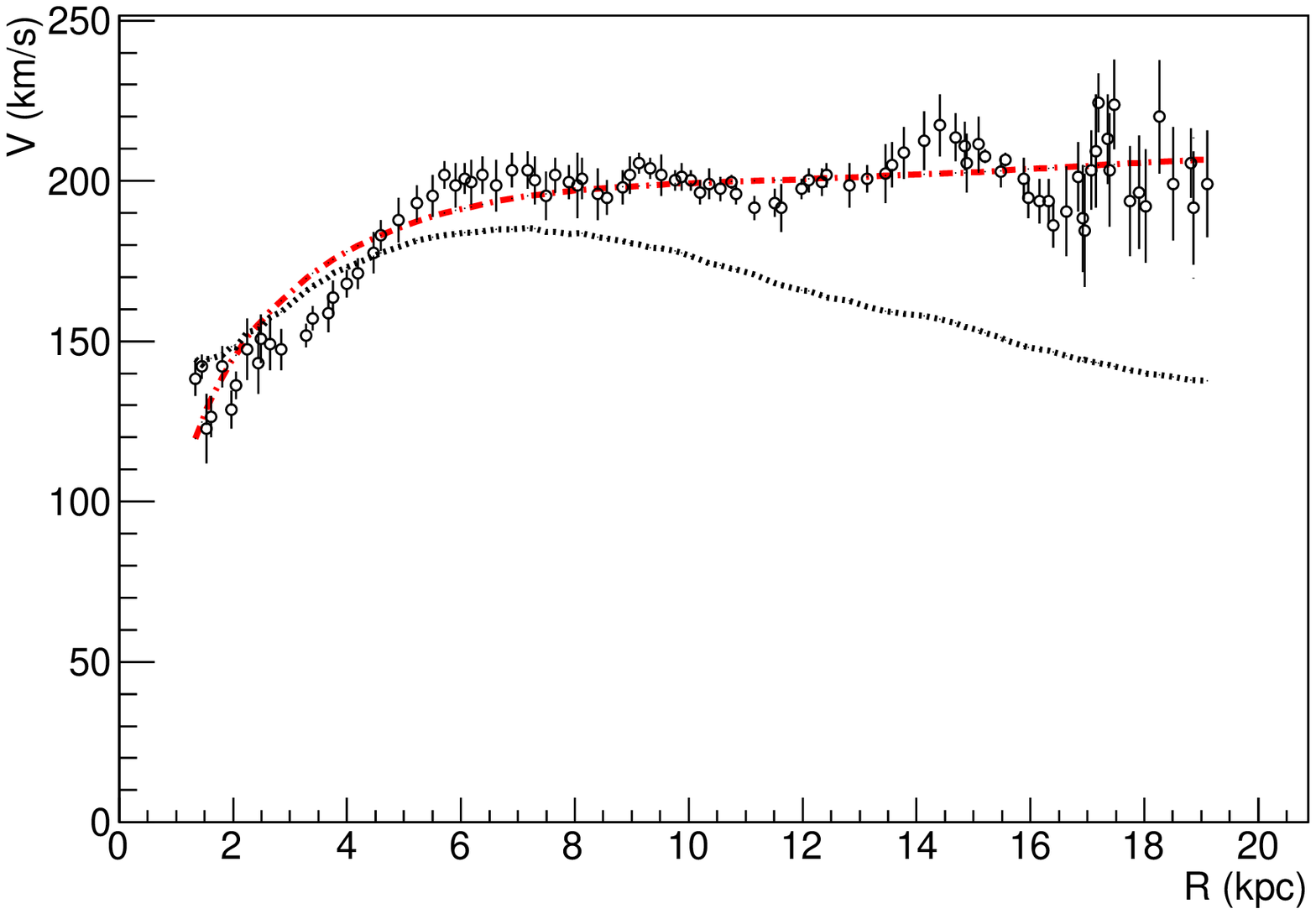}
}
\subfloat[][NGC3953~Ref.(7), $\kappa_{\tau}=0.27$]{
\includegraphics[width=0.3\textwidth]{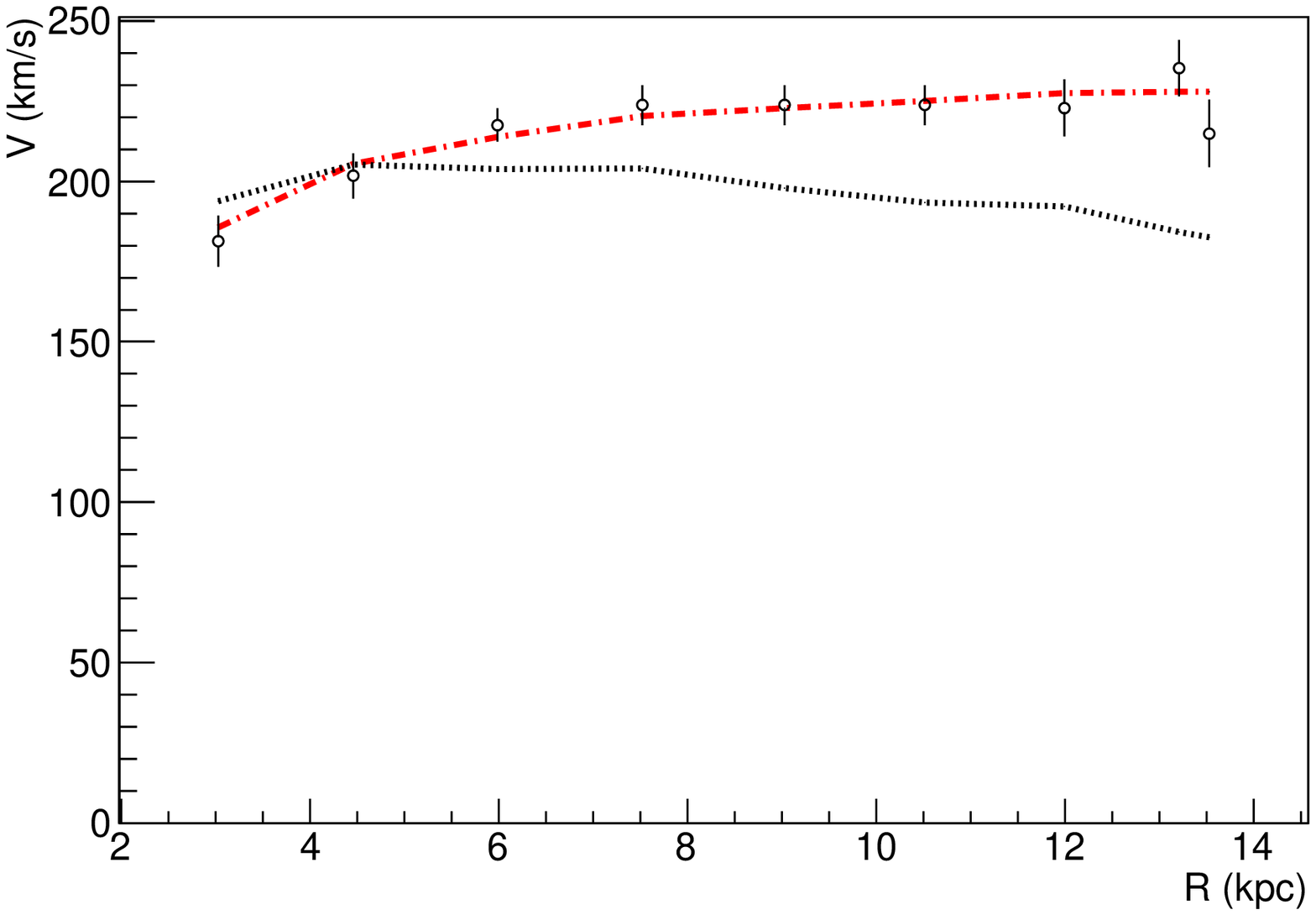}
}
\vspace{0.5cm}
\subfloat[][NGC2903~Ref.(7), $\kappa_{\tau}=0.48$]{
\includegraphics[width=0.3\textwidth]{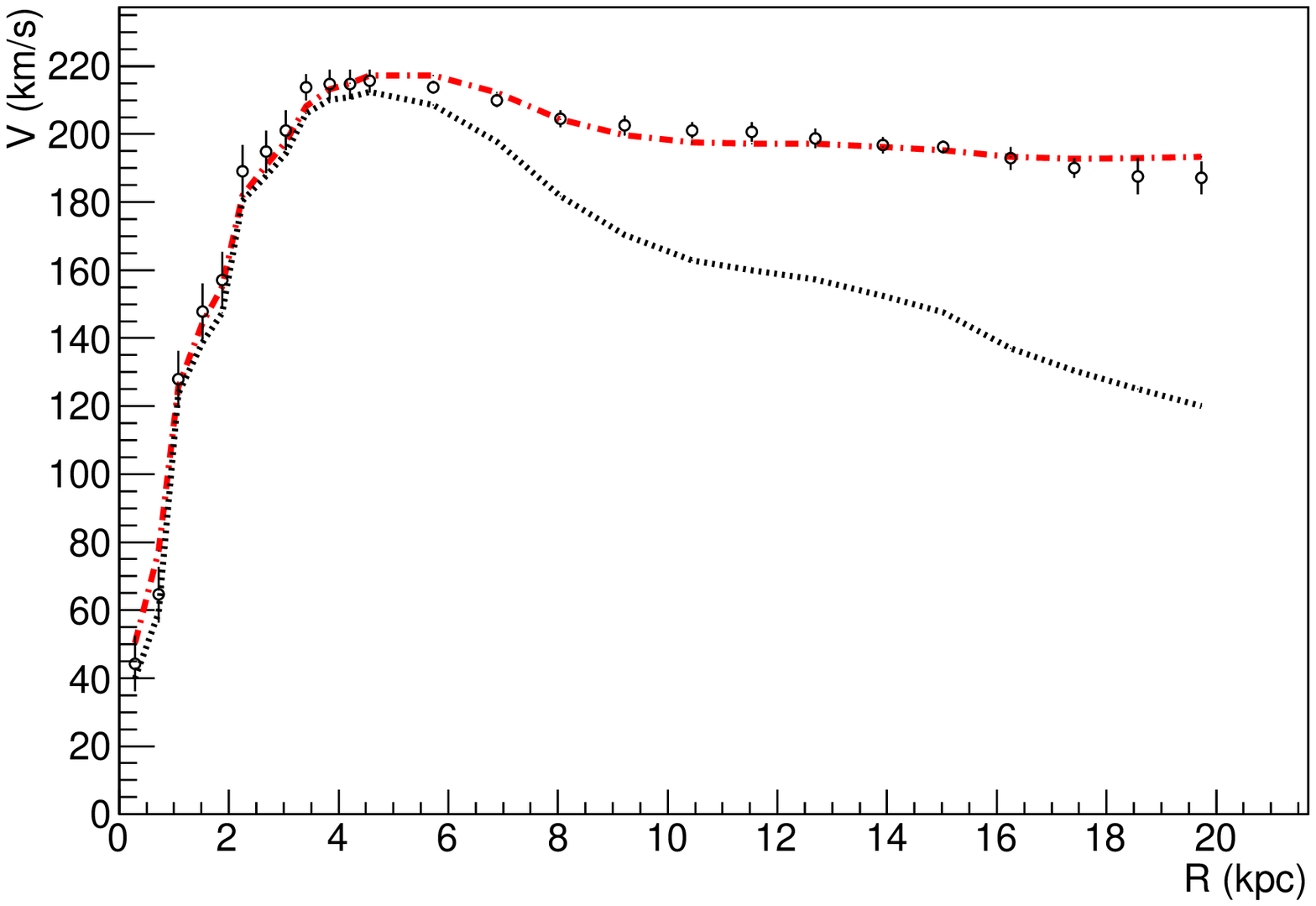}
}
\subfloat[][NGC2903~Ref.(5), $\kappa_{\tau}=0.40$]{
\includegraphics[width=0.3\textwidth]{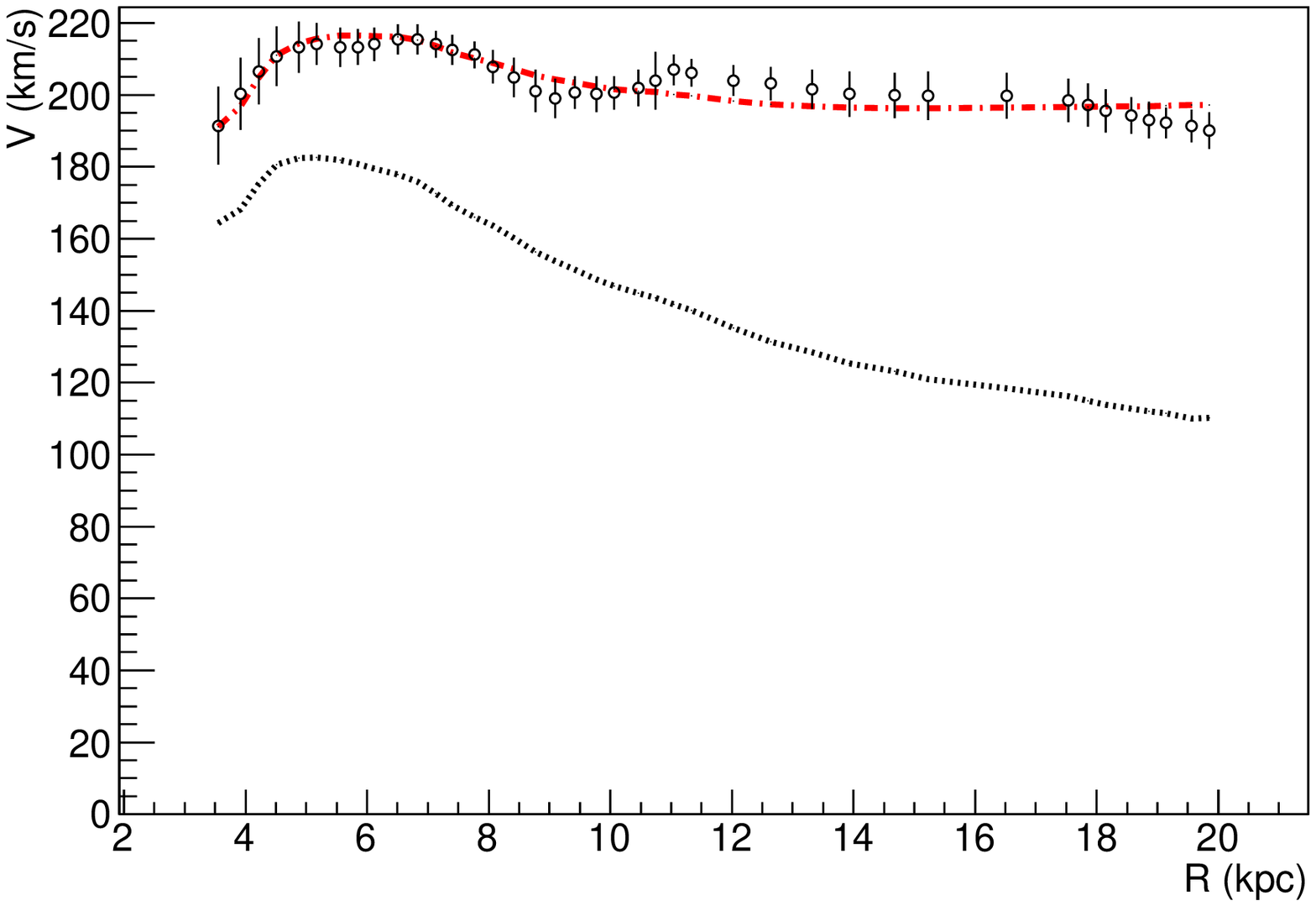}
}
\subfloat[][NGC5907~Ref.(13), $\kappa_{\tau}=0.44$]{
\includegraphics[width=0.3\textwidth]{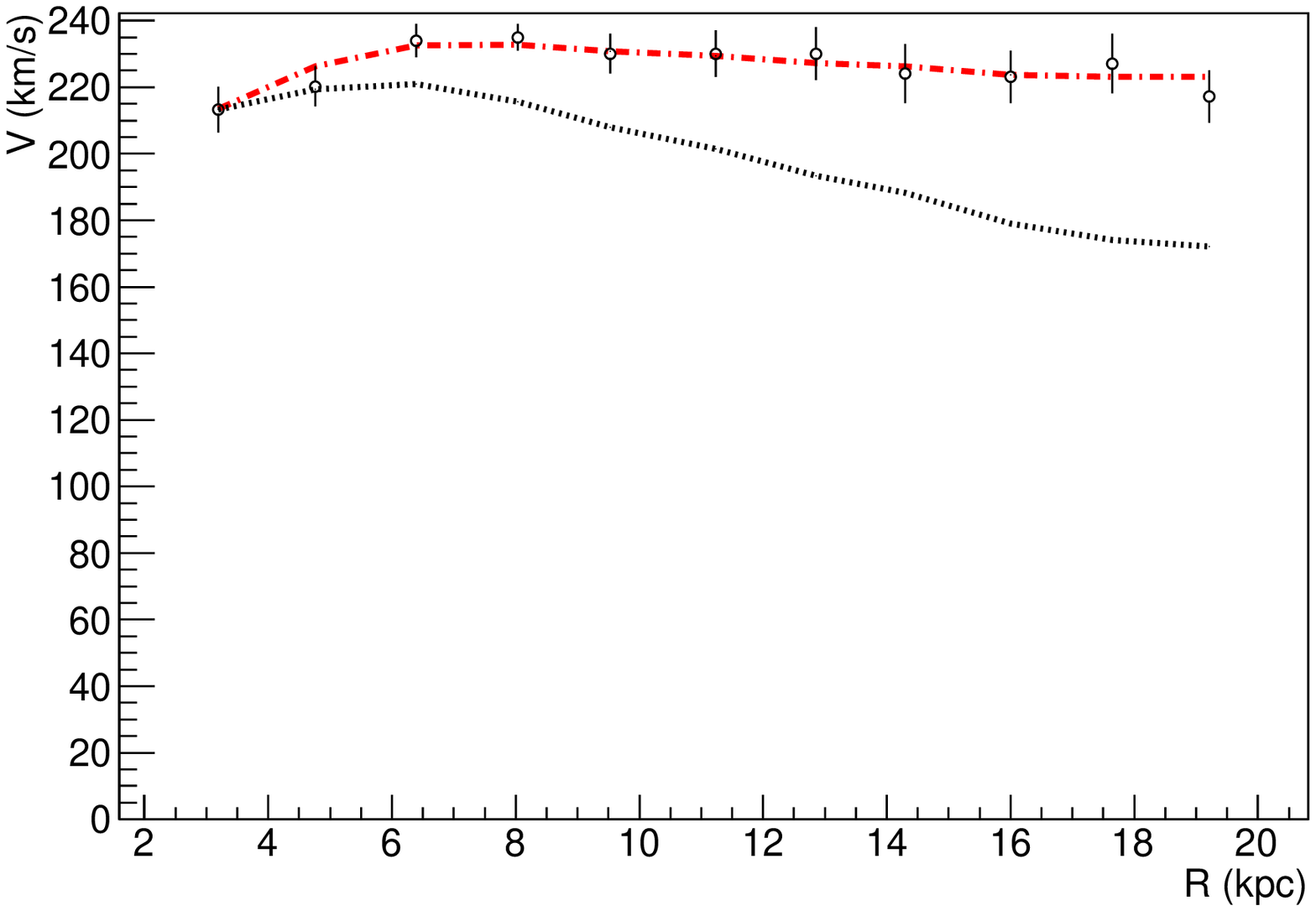}
}
\vspace{0.5cm}
\subfloat[][NGC3726~Ref.(12), $\kappa_{\tau}=0.27$]{
\includegraphics[width=0.3\textwidth]{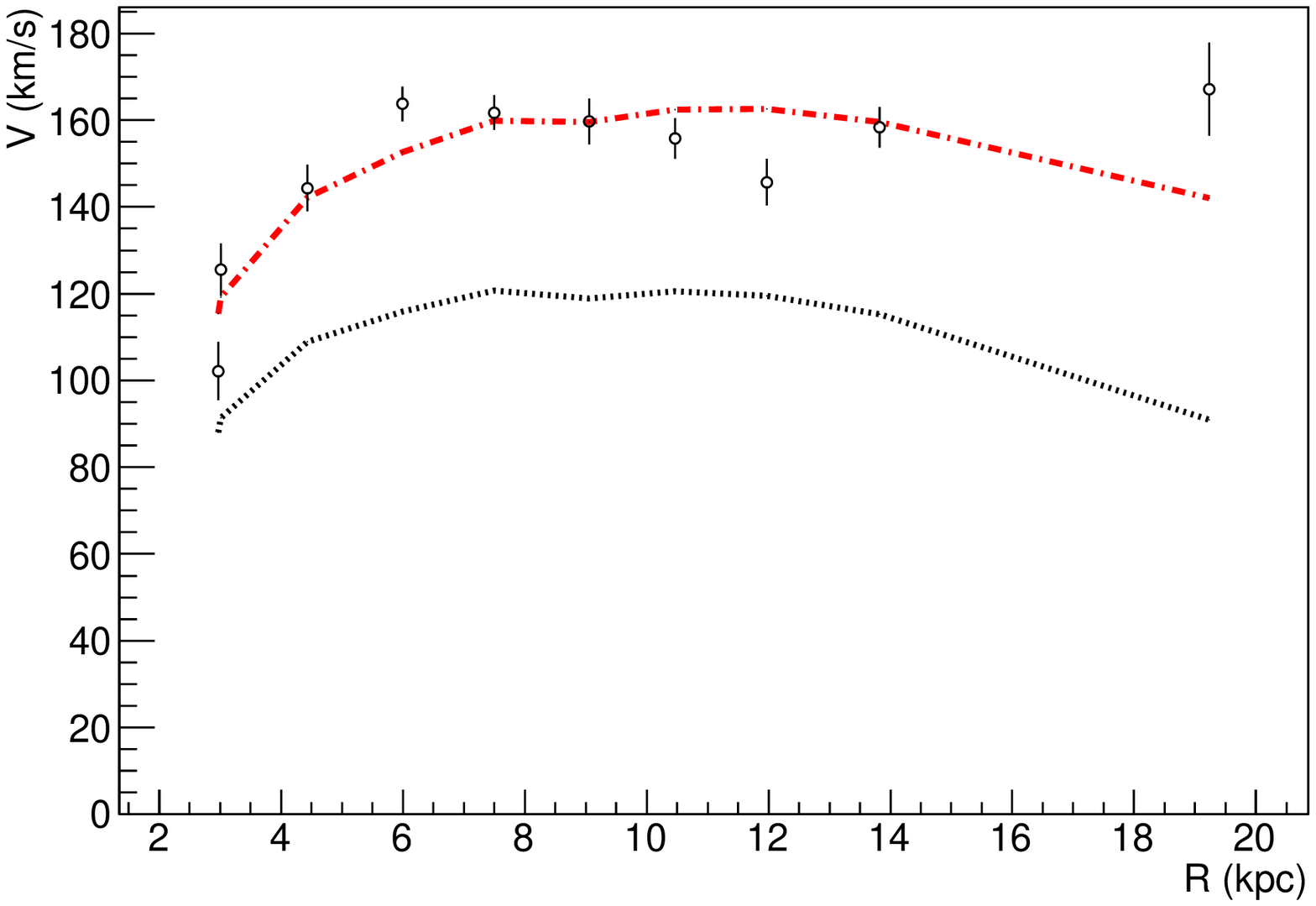}
}
\subfloat[][F563-1~Ref.(9), $\kappa_{\tau}=0.11$]{
\includegraphics[width=0.3\textwidth]{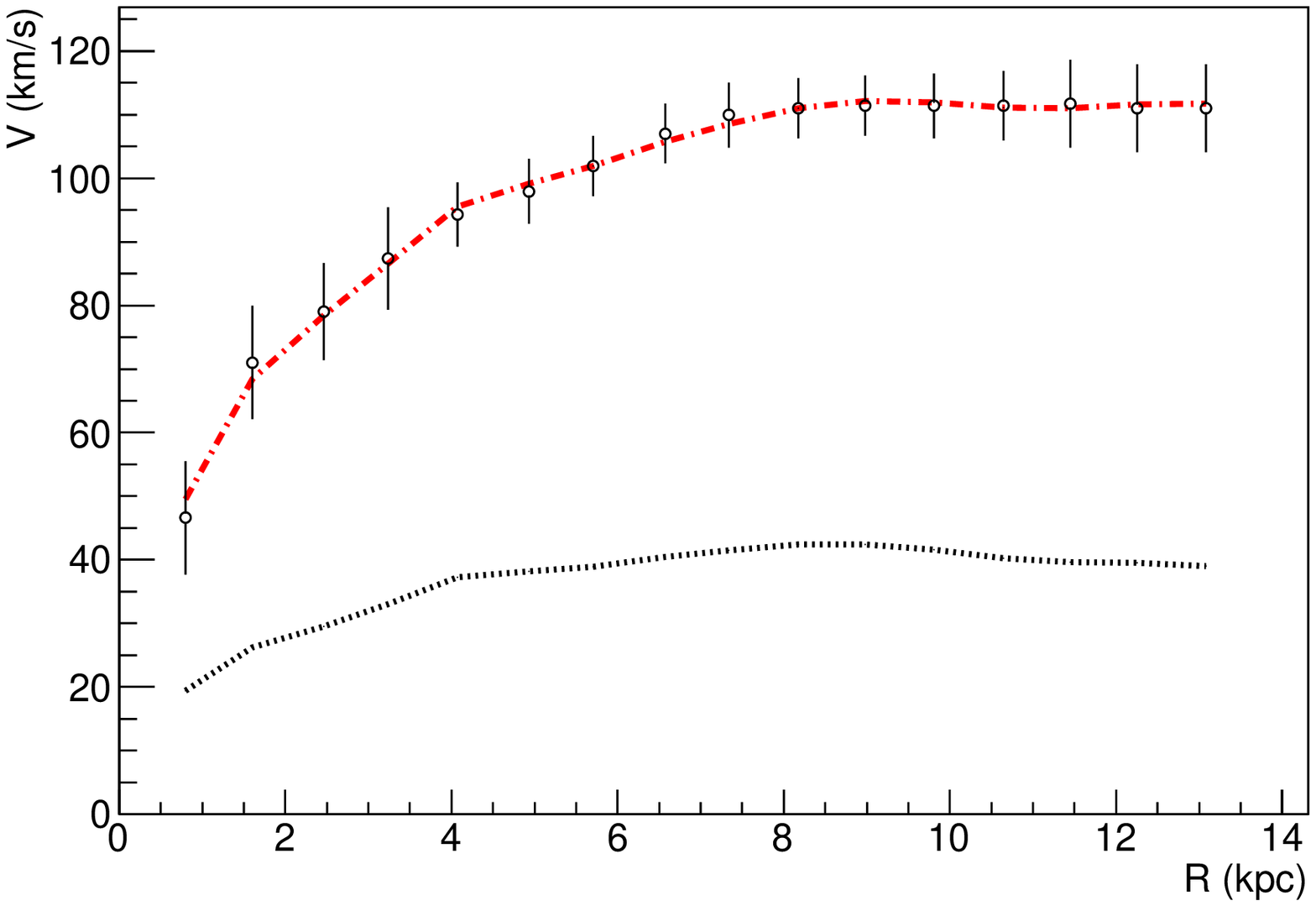}
}
\subfloat[][NGC925~ Ref.(4), $\kappa_{\tau}=0.09$]{
\includegraphics[width=0.3\textwidth]{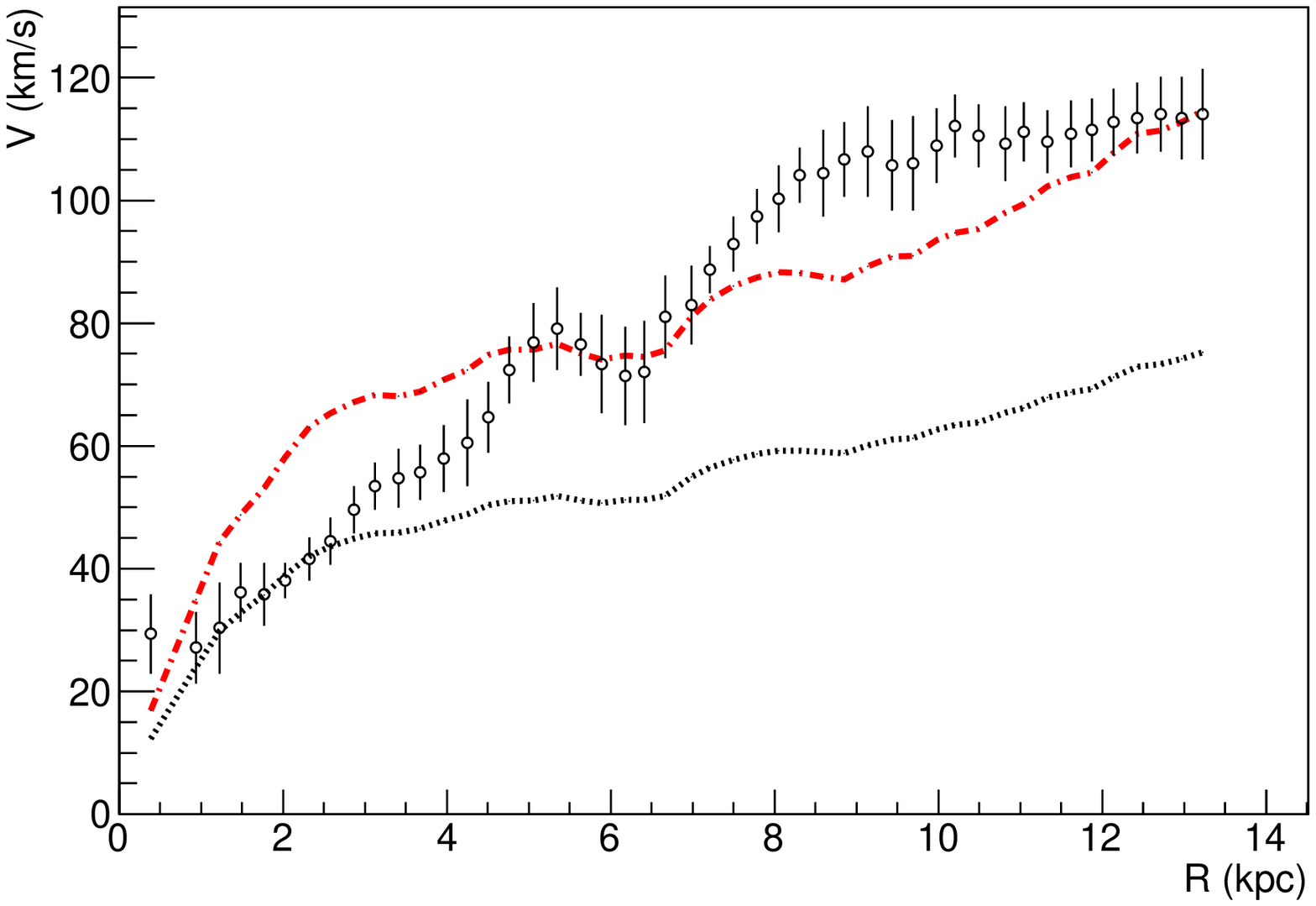}							 
}
\vspace{0.5cm}
\subfloat[][NGC7793,~Ref.(5), $\kappa_{\tau}=0.05$]{
\includegraphics[width=0.3\textwidth]{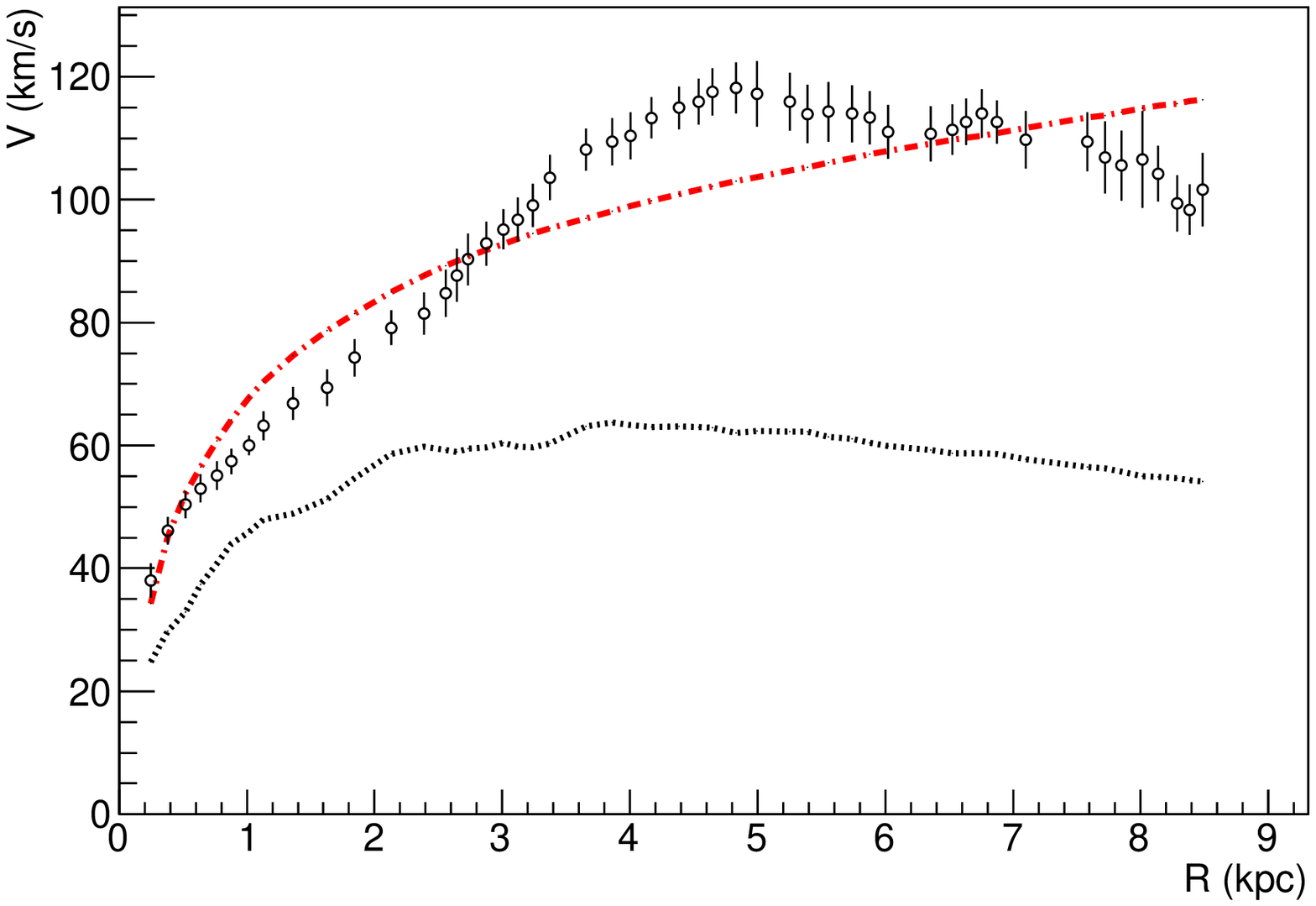}
}
\caption{LCM fits to  rotation curves of spiral galaxies with published data. Emitter-galaxies paired  to MW \citet[Model B]{Klypin}. Symbols and curves are as in Figure~\ref{fig:resultsA}, and references are as in Table.\ref{tab:referenceDATAs}.\label{fig:resultsC}}
\end{figure*}

 \clearpage
 
 \begin{deluxetable}{l l c c c r c c c}
 \rotate
\tablecaption{Results from LCM fits\label{tab:resultsPARAMS}}
\tablewidth{0pt}
\tablehead{
\colhead{Galaxy\tablenotemark{a}}  & 
\colhead{$\kappa_{\tau}$}  & 
\colhead{$\alpha$\tablenotemark{b} ($10^{-6}$)}  &  
\colhead{$\zeta$\tablenotemark{c}}  &  
\colhead{$ \zeta_o$}	&
\colhead{LCM Fit $\frac{\chi^2}{\rm{dof}} $}  &
\colhead{Other Model $\frac{\chi^2}{\rm{dof}}$\tablenotemark{d}}  &
\colhead{Reference\tablenotemark{e}}  & 
\colhead{Reference\tablenotemark{f}}
}
\startdata
NGC3198 				 &$0.10$	&$-59.70\pm0.93$ 	&$1.00\pm 0.01$	&$0.90$		&$47.48/26=1.83$ 				& $1.34_{\rm{N}}$  					&2	  		& 16  \\
					&$0.26$	&$-79.04\pm 1.84$ 	&$1.01\pm0.01$		&$0.84$  		&$42.30/19=2.23$ 				&       								&  			&14\\
 					&$0.24$	&$-76.28\pm 1.77$ 	&$1.01\pm0.01$		&$0.85$  		&$41.99/19=2.21$ 				&    						 		&  			&15 \\

NGC3198				&$0.11$	&$-0.54 \pm 1.69$ 	 &$0.99 \pm0.01$	&$2.12$		&$10.32/12=0.86$ 				&  $0.67_{\rm{N}}$  		 			&9   			& 16\\
					&$0.31$	&$-70.64\pm2.33$ 	&$0.97\pm0.01$		&$2.09$		&$10.81/11=0.98$				&       								&  			&14\\
 					&$0.28$	&$-67.61\pm2.23$ 	&$0.98 \pm0.01$	&$2.10$		&$10.76/11=0.98$				&       								& 			 &15	 \\	
 
NGC3198**			&$0.06$	& $106.01\pm 2.76$	&$1.02\pm 0.02$	&$0.65$		& $451.04/24=18.79$ 				& $0.84_{\rm{N}}$  					&4   			& 16 \\
					&$0.09$	&$-180.89\pm 8.30$	&$1.02\pm 0.05$	&$0.43$		&$420.93/14=30.07$ 				&    								&  			&14\\
 					&$0.09$	&$ -179.37\pm 8.21$	&$1.03\pm 0.05$	&$0.43$		&$417.12/14=29.79$ 				&    								&  			&15 	 \\		
 
NGC3198 				&$0.05$	&$-58.55\pm4.20$ 	& $0.99\pm0.02$	&$0.99$		&$29.81/14=2.13$ 				&  $4.48_{\rm{M}}$  					&6  			& 16\\
			 		&$0.23$	&$-89.43\pm 9.18$	&$0.97\pm 0.04$	&$1.87$		&$25.18/9=2.80$ 				&  						 		&  			&14 \\
 					&$0.21$	&$-86.58\pm 8.89$	&$ 0.97\pm 0.04$	&$1.88$		&$25.31/9=2.81$ 				&   								&			&15  \\	
\tableline
 
M33					&$0.03$	&$-97.91\pm5.67$  	&$1.01\pm 0.05$	&$0.68$		&$  25.86/24=1.08$ 				& $0.29_{\rm{N}}$					&3  			& 16\\
					& $0.07$	&$-101.94\pm5.93$	&$1.02\pm 0.05$	&$0.66$		& $30.19/24=1.26$				&       								&			&14\\
 					&$0.07$	&$-100.56\pm5.85$ 	&$1.01\pm0.05$		&$0.67$		&$29.81/24=1.24$				&     						 		&  			&15  \\		
 						
M33					&$0.03$	& $-106.00\pm5.22$	&$1.00\pm 0.04$	&$0.91$		& $4.03/18=0.22$				&$0.16_{\rm{N}}$ 					&8  			& 16 \\
					&$0.07$	&$-111.40\pm5.50$	&$1.06\pm 0.04$	&$0.89$		&$3.67/18=0.20$ 				&								&  			&14 \\
 					&$0.07$	&$-110.00\pm5.45$	&$ 1.00\pm 0.04$	&$0.90$		&$3.65/18=0.20$ 				&    								& 			&15 \\	
\tableline
 
NGC5055 				&$0.25$	&$-52.97\pm 1.48$ 	&$ 1.01\pm 0.01$	&$0.87$		&$ 70.24/21=3.34$ 				&$3.96_{\rm{N}} $ 					&1  			&16\\
					&$0.31$	&$-109.96\pm 4.54$ 	&$ 2.03\pm 0.01$	&$0.40$		& $30.23/16=1.89$				&     						 		&  			&14 \\
 					&$0.44$	&$-111.50\pm4.60$ 	&$1.26\pm 0.04$	&$0.62$		&$28.93/21=1.81$				&      						 		&  			&15 \\			
 
NGC5055**			&$0.30$	& $-39.78\pm 3.47$	&$1.01\pm  0.01$	&$0.86$		&$ 4915.35/51=96.38$ 			&$75.78_{\rm{N}}$  					&4			& 16\\
					&NA		& NA 		    	&NA				&$10^{-11}$	&NA 							&     								&  			&14	 \\
 					&NA	&NA			&NA				&$10^{-10}$	&NA 							&    								& 			&15 \\
		 								 
NGC5055				&$0.30$	& $-53.46\pm 1.12$	& $ 1.00\pm 0.01$	&$0.97$		& $171.09/92=1.86$				&$1.05_{\rm{M}}$ 					&5  			& 16 \\
	 				&$0.39$	&$-120.16\pm 4.13$	&$1.80\pm 0.02$	&$0.47$		&$80.01/63=1.27$ 				&     								&  			&14 \\
 					&$0.61$	&$-144.63\pm 5.04$	&$1.06 \pm 0.01$	&$0.80$		&$97.42/63=1.55$ 				&    								&  			&15 \\		
\tableline
 
NGC2403				&$0.05$	&$-79.90\pm1.35$  	&$ 1.01 \pm 0.01$	&$0.66$		&$ 97.63/31=3.15$ 				&$4.98 _{\rm{M}} $ 					&2	  		& 16\\
					& $0.14$	&$-88.93\pm1.50$  	&$1.03\pm 0.02$	&$0.63$		&$102.97/31=3.32$				&      								&  			&14 \\
 					&$0.13$	&$-86.90\pm 1.47$ 	&$ 1.02\pm 0.02$	&$0.65$		&$102.53/31=3.31$				&        							&  			&15 \\		
 
NGC2403				&$0.03$	& $-108.00\pm 1.41$	& $1.00\pm 0.02$	&$0.89$		& $115.71/61=1.90$				&$ 1.81_{\rm{N}}$   		 			&4  			& 16\\
					&$0.09$	&$-116.41\pm1.52$  	&$ 1.01\pm 0.02$	&$0.85$		&$117.99/61=1.93$ 				&    								& 			 &14  \\
 					 &$0.09$	&$-114.42\pm1.49$	&$1.01\pm 0.02$	&$0.87$		&$118.19/61=1.94$ 				&     								&  			&15\\	
\tableline
 NGC 3521				&$0.31$	& $-56.82\pm6.52$	&$0.99 \pm 0.02$	&$1.14$		& $21.27/32=0.66$				&$0.97_{\rm{M}}$  					&5 			 & 16\\
					&$0.73$	&$-171.00 \pm25.53$  	&$1.02\pm 0.03$	&$0.98$		&$23.76/28=0.85$ 				&    								&  			&14  \\
 					&$0.73$	&$-158.62 \pm23.77$	&$ 0.98\pm 0.03$	&$1.07$		&$24.36/28=0.87$ 				&    								&  			&15  \\	
\tableline
 			
NGC2841**			&$0.45$	&$-175.00 \pm 7.57$  	&$0.96 \pm 0.01$	&$1.28$		&$3.61/24 = 0.15$ 				&$0.57 _{\rm{N}} $					&4  			& 16\\
					& $2.97$	&$25.31\pm 1.70$  	&$0.48\pm 0.01$	&$3.22$		& $6.44/14 = 0.46$				&       								& 			&14\\
 					&NA	&NA			&NA			&$10^{-7}$	&NA				&     								& 			&15 \\	
 
NGC2841				&$0.41$	& $-160.26\pm 8.28$	&$1.00 \pm 0.01 $	&$1.02$	& $20.48 /22 = 0.93$					&$1.08_{\rm{M}}$		 			&5   			& 16 \\
					&$1.72$	&$79.99 \pm 6.27$  	&$ 0.73 \pm 0.01$	&$1.70$	&$5.95 /12 = 0.50$ 					&    								& 			&14 \\
 					&$1.84$	&$52.17 \pm 5.44$	&$0.63 \pm 0.00$	&$1.99$	&$9.53/12 = 0.79$ 					&    								&  			&15 \\
\tableline		
 
NGC7814				&$0.61$		& $-65.63\pm 2.03$	&$1.06 \pm 0.02$&$0.80$		& $ 19.38/17 = 1.14$				&$9.11_{\rm{N}}$ 				&11 			& 16\\
					&$1.68$		&$37.37 \pm 1.12$  	&$0.96 \pm 0.03$&$0.79$		&$11.61/17 =0.68$ 				&    							& 			&14	  \\
 					&$1.30$		&$102.45 \pm 2.99$	&$0.61 \pm 0.04$&$0.66$		&$30.52/17 = 1.80$ 				&   							&  			&15  \\
\tableline
 
NGC7331				&$0.32$		&$-88.92\pm 2.26$  	&$1.22 \pm 0.02$&$0.51$ 		&$270.21/32 = 8.44$ 				&$6.80 _{\rm{M}} $ 				&2			& 16	\\
					& $1.25$		&$158.33\pm 6.73$  	&$1.11 \pm 0.01$&$0.75$		& $122.78/22 = 5.58$				&   							&			& 14   \\
 					&$1.10$		&$339.91\pm 11.10$ 	&$1.28\pm 0.01$&$0.72$		&$58.58/22 =2.66 $				&     							&			& 15      \\		
 
NGC7331				&$0.29$		& $-110.00\pm 5.47$	&$1.00 \pm 0.02$&$1.02$		& $12.01/34 =0.35$				&$0.45 _{\rm{M}}$ 				&5			& 16      \\
					&$0.57$		&$-217.52\pm 15.53$  	&$ 1.29\pm 0.03$&$0.74$		&$5.821/26 = 0.22$ 				&     							&			& 14    \\
 					&$0.62$		&$-225.56\pm 16.11$	&$1.08\pm 0.03$&$0.88$		&$5.91/26 = 0.23$ 				&     							&			& 15  \\	
\tableline
 NGC891				&$0.59$		& $-25.22\pm 11.24$	&$ 1.02 \pm 0.04$	&$0.92$	& $27.74/16 =  1.73$				&   IND						&11			& 16    \\
					&$1.85$		&$6.54\pm 3.11$  	&$1.00\pm 0.02$	&$1.02$	&$28.42/16 = 1.78$ 				&     							&			& 14    \\
 					 &$1.69$		&$8.66\pm 4.92$	&$ 1.00 \pm 0.02$	&$1.01$	&$29.74/16 = 1.86$ 				&    							&			& 15    \\	
 \tableline
M 31					&$1.18$		& $42.14\pm 19.02$	&$ 0.88 \pm 0.02$	&$1.14$	& $7.93/21 = 0.38$				&  IND						&10			& 16	     \\
					&$2.39$		&$10.19\pm 6.76$  	&$1.00\pm 0.08$	&$0.85$	&$8.05/11 = 0.73$ 				&      							&			& 14   \\
					 &$2.18$		&$12.30\pm 8.83$	&$1.00\pm 0.09$	&$0.85$	&$8.381/11 = 0.76$ 				&    							&			& 15    \\	
\tableline
 
NGC 5533				&$0.43$		& $-124.00\pm 18.01$	&$1.04 \pm 0.06$	&$0.79$	& $1.35/3 =0.45$				&$3.55 _{\rm{M}}$				&7			& 16     \\
					&$1.46$		&$67.00\pm 41.61$  	&$ 0.99 \pm 0.04$	&$1.04$	&$0.29/1 = 0.29$ 				&     							&			& 14    \\
 					&$1.36$		&$82.83\pm 51.61$	&$0.99\pm 0.04$	&$1.06$	&$0.31/1 =0.31$ 					&  							&			& 15      \\	
\tableline
 
UGC 6973				&$0.10$		& $-99.10\pm 5.35$	&$1.02 \pm 0.03$	&$0.54$	& $0.60/7 = 0.09$				&$23.50 _{\rm{M}}$				&7			& 16     \\
					&$0.28$		&$-132.20\pm 7.14$  	&$ 1.14 \pm 0.04$	&$0.47$	&$0.59/7 = 0.08$ 				&      							&			& 14  \\
 					&$0.29$		&$-128.56\pm 6.96$	&$1.06 \pm 0.03$	&$0.52$	&$0.71/7 = 0.10$ 				&    							&			& 15    \\	
\tableline
 
NGC 4088				&$0.13$		& $-59.70\pm 8.58$	&$1.00 \pm 0.04$	&$1.10$	& $30.70/12 =2.56$				&$ 4.41_{\rm{M}}$					&13			& 16     \\
					&$0.36$		&$-86.94\pm 12.29$  	&$ 0.99 \pm 0.04$	&$1.04$	&$28.71/12=2.39$ 				&     							&			& 14   \\
 					 &$0.33$		&$-81.84\pm 11.63$	&$0.99 \pm 0.04$	&$1.06$	&$29.28/12=2.44$ 				&     							&			& 15   \\	
\tableline
 
NGC 3992				&$0.27$		&$-104.95\pm 9.67$	&$ 1.00 \pm 0.02$	&$0.93$	& $1.81/7=0.26$					&$0.50 _{\rm{M}}$					&7			& 16	\\
					&$0.33$		&$-191.67 \pm 57.59$ &$1.88\pm 0.07$		&$0.48$	&$1.16/3=0.39$ 					&							&    			& 14	  \\
 					&$0.42$		&$-200.39\pm 60.05$	&$1.30\pm 0.07$	&$0.66$	&$1.09/3=0.36$ 				&							& 			& 15\\	
					 	
\tableline
 
NGC4138				&$0.23$		& $-34.20\pm 6.85$	&$ 1.00 \pm 0.05$	&$0.88$	& $10.65/5=2.13$				&$ 2.12_{\rm{M}}$					&13			& 16	     \\
					&$0.64$		&$-73.04 \pm 15.01$  	&$0.93\pm 0.05$	&$0.87$	&$ 11.88/5=2.38$ 				&     							&			& 14    \\
 					&$0.61$		&$-65.70\pm 13.62$	&$ 1.00 \pm 0.05$	&$0.91$	&$12.36/5=2.47$ 				&     							&			& 15    \\	
 \tableline
  
NGC 6946				&$0.09$		&$-86.74\pm 5.20$  	&$0.99\pm 0.02$	&$1.41$	&$ 12.79/18=0.71$ 				&$ 3.03_{\rm{M}}$					&7			& 16    \\
					& $0.22$		&$-103.76\pm 22.45$  &$0.97\pm 0.06$	&$1.40$	& $7.21/11=0.66$				&      							&			& 14    \\
 					&$0.21$		&$-99.61\pm 21.76$ 	&$0.97 \pm 0.06$	&$1.41$	&$7.64/11=0.69$				&       							&			& 15    \\		
NGC 6946				&$0.19$		& $-111.23\pm 2.87$	&$ 0.65\pm 0.01$	&$1.026$	& $157.841/94=1.68$				&$ 3.67_{\rm{N}}$  				 &4			& 16    \\
					&$0.65$		&$-207.90\pm 5.44$  	&$ 0.45\pm 0.01$	&$1.246$	&$201.33/94=2.14$ 				&    							 &			& 14    \\
 					&$0.56$		&$-208.00\pm 5.44$	&$0.45\pm 0.01$	&$1.15$	&$ 201.33/94=2.14$ 				&   							  &			& 15   \\	
\tableline
  
NGC3953				&$0.14$		& $-150.00\pm 22.12$	&$1.03\pm 0.05$	&$0.68$	& $3.95/7=0.56$					&$ 1.35_{\rm{M}}$					&7			& 16    \\
					&$0.22$		&$-226.06\pm 33.43$  &$1.80 \pm 0.08$	&$0.38$	&$ 3.94/7=0.56$ 				&     							&			& 14    \\
 					&$0.27$		&$-213.00\pm 31.45$	&$ 1.38 \pm 0.07$	&$0.49$	&$ 3.94/7=0.56$ 				&   							&			& 15   \\		
\tableline
 
NGC 2903				&$0.22$		& $-58.90\pm 1.09$	&$ 1.00 \pm 0.01$	&$0.92$	& $78.31/31=2.53$ 					& $8.10 _{\rm{M}}$  				&7			& 16 	     \\
					&$0.49$		& $-122.59\pm 3.57$ 	&$1.08\pm 0.01$	&$0.76$	&$20.77/23=0.90$ 					&      							&			& 14   \\
 					&$0.48$		&$-114.76\pm 3.34$ 	&$1.04\pm 0.01$	&$0.80$	&$20.66/23=0.90$ 					&  							 &			& 15   \\	
 
NGC 2903**			&$0.01$		&$-311.23 \pm 4.82$  	&$1.50\pm 0.12$	&$0.10$	&$2481.72/40=62.04$ 				&  $ 247.18_{\rm{N}}$ 			&4			& 16    \\
					& NA		&NA  			&NA			&$ 10^{-11}$	& NA						&       							&			& 14   \\
 					&NA		&NA 			&NA			&$10^{-12}$	&NA						&        						&			& 15   \\		 	
 
NGC 2903				&$0.18$		& $-75.98\pm 1.65$	& $1.00\pm 0.01$	&$1.25$	& $54.52/59=0.92$ 					& $0.53 _{\rm{M}}$   				&5			& 16	     \\
					&$0.43$		&$-138.60\pm 4.49$  	&$0.99 \pm 0.01$	&$1.13$	&$ 19.59/38=0.52$ 					&     							&			& 14   \\
 					 &$0.40$		&$-129.87\pm 4.21$	&$0.99 \pm 0.01$	&$1.14$	&$20.05/38=0.53$ 					&    							&			& 15   \\
\tableline
 
NGC5907				&$0.25$		& $-82.70\pm 5.11$	&$ 1.01 \pm 0.02$	&$0.87$	& $3.55/15=0.24$					&$0.48 _{\rm{M}}$					&13			& 16  \\
					&$0.34$		&$-149.26\pm 18.08$  &$ 1.86 \pm 0.04$	&$0.45$	&$2.45/9=0.27$ 					&							&		 	& 14      \\
 					&$0.44$		&$-151.51\pm 18.35$	&$1.28 \pm 0.04$	&$0.63$	&$2.45/9=0.27$ 					&							&		 	& 15    \\	
\tableline
 
NGC 3726				&$0.10$		&$-67.30 \pm 9.40$	&$ 0.99\pm 0.03$	&$1.35$	& $37.43/11=3.40$					&$ 7.10_{\rm{M}}$					&12 			& 16 \\
					&$0.29$		&$-49.39 \pm 19.08$  	&$0.98\pm 0.04$	&$1.52$	&$30.47/8=3.81$ 					&							&  		  	& 14\\
 					 &$0.27$		&$-46.70\pm 18.16$	&$0.98\pm 0.04$	&$1.53$	&$30.59/8=3.82$ 					&							&  			 & 15\\	
\tableline
 
F563-1				&$0.04$		&$-47.08\pm 21.88$	&$ 0.99 \pm 0.10$	&$4.94$	& $0.52/14=0.04$					&$0.05_{\rm{N}}$				&9 			& 16	\\
					&$0.12$		&$-52.98 \pm 24.59$  	&$ 0.97\pm 0.11$	&$4.91$	&$0.51/14=0.04$ 					&							&   			& 14  \\
 					&$0.11$		&$-51.92 \pm 24.11$	&$ 0.97\pm 0.11$	&$4.92$	&$0.51/14=0.04$ 					&							&  			& 15	 \\	
\tableline
 
NGC925				&$0.03$		& $-10.29\pm 19.20$	&$1.08 \pm 0.08 $	&$1.91$	& $308.24/45=6.85$					&$3.07_{\rm{N}}$				&4			& 16    \\
					&$0.10$		&$24.42\pm 16.98$  	&$1.18\pm 0.06$	&$2.23$	&$ 306.47/45=6.81$ 					&      							&			& 14   \\
 					 &$0.09$		&$18.77\pm 17.32$	&$1.16\pm 0.06$	&$2.20$	&$307.35/45=6.83$ 					&     							&			& 15 \\
\tableline
 
NGC 7793				&$0.02$		& $-141.10\pm 7.62$	&$1.04\pm 0.07$	&$0.73$	& $265.74/48=5.54$					&$4.11 _{\rm{M}}$				&5			& 16     \\
					&$0.02$		&$-269.99\pm 12.34$  &$1.16 \pm 0.17$	&$0.27$	&$ 324.48/47=6.90$ 					&      							&			& 14  \\
 					 &$0.05$		&$-144.90\pm 7.91$	&$ 1.05\pm 0.07$	&$0.71$	&$273.82/48=5.70$ 					&     							&			& 15   \\
\enddata
\tablenotetext{a}{Results marked with $**$ indicate a fit that did not converge.}
\tablenotetext{b,c}{Uncertainties are statistical only.}
\tablenotetext{d}{N=NFW; M=MOND; blank indicates model independent.}
\tablenotetext{e}{Emitter-galaxy references: as in Table~\ref{tab:referenceDATAs}}
\tablenotetext{f}{MW References: 
14. \citet{Klypin} model A (\emph{no }exchange of angular momentum),
15. \citet{Klypin} model B (\emph{with }exchange of angular momentum),
16. \citet{Sof81}.}
14. Sofue 1981, 15. Klypin 2002, model A (\emph{no }exchange of angular momentum), 16. Klypin 2002, model B 
\end{deluxetable}

\section{Conclusion: Discussion and future work}
\label{sec:conclusion}
While the derivation of the LCM is neither fundamental nor Lorentz-invariant, it does provide a working set of assumptions which can be tested against observations.  The model    is successful in 
fitting observed rotation curves with the reported luminous matter alone. It is in this sense that the LCM may provide a working constraint to stellar population synthesis models and luminous matter modeling.\\

Two major goals for future work include   extending the LCM formalism to more general geometries of dark matter, and inverting the LCM formalism  to  predict the  preferred luminous matter profile for a specific galaxy  given the observed data  $v_{obs}$.\\

The first goal, to extend the LCM to a broader category of distance scales and geometries could most easily begin with weak gravitational lensing, \citep{Narayan}, as the formalisms are parallel.   The   LCM symmetry assumptions currently only apply in the plane of the galactic disc, where spherical symmetry can be expected to approximate the functional shape of a  disc potential.  Furthermore, the  general spherical assumptions which are necessary for analytic solution of the wave equation may only generalize to galaxy and globular clusters numerically.    Extensions of the LCM formalism even to dynamics above/below the plain of the galactic disc  will require   intensive numerical modeling of the variations in the potential as well as analysis regarding the appropriate metric.    Extension of the LCM to the flat rotation curves of the Milky Way itself  will require careful  study of  how to apply the convolution when the emitter is embedded in within
the receiver frame.    
 \\

The second goal, to use the LCM as a constraint on luminous matter modeling, can be investigated upon identification of a preliminary functional form for the $\alpha$ parameter as a function of the 
 relative galaxy curvatures $\kappa_\tau$.    Such an inversion protocol  would allow the LCM to predict the luminous profile $M_L$  from the   observed data, $v_{obs}$.\\

\subsection{Acknowledgements}
The authors would like to thank V.P. Nair,  Ed Bertschinger, Janet Conrad,  Marco Inzunza, Peter Fisher, Timothy Boyer, Joel Gersten, and Vassili Papavassiliou.    S. Cisneros is supported by  the MIT Martin Luther King Jr. Fellowship, while J.\,A. Formaggio and N.\,A. Oblath are  supported by the United States Department of Energy under Grant No. DE-FG02-06ER- 41420.\\

\clearpage

\bibliography{LCM}{}
\bibliographystyle{astrolastfirstyear1}

\end{document}